\documentclass[12pt]{article}
\usepackage{amsmath,amssymb,theorem,cite,epsfig,url,psfrag,eepic,amsmath,mathtools,mathrsfs,amsbsy,dsfont,esint,braket,cancel}
\usepackage{indentfirst}
\usepackage{graphicx}
\usepackage{tikz,pgfplots}
\pgfplotsset{compat=newest}
\usepackage[bookmarks=true,hyperfigures=true,colorlinks=true,linkcolor=blue,citecolor=blue,urlcolor=blue,bookmarksnumbered]{hyperref}
\numberwithin{equation}{section}
\advance \topmargin by -\headheight
\advance \topmargin by -\headsep     
\evensidemargin \oddsidemargin
\usepackage{bm,upgreek}

\makeatletter
\def\Appendix{\appendix
  \def\@seccntformat##1{Appendix~\csname the##1\endcsname.~~}}
\makeatother

\def\XXint#1#2#3{{\setbox0=\hbox{$#1{#2#3}{\int}$}
\vcenter{\hbox{$#2#3$}}\kern-.5\wd0}}

\begin{document}
\title{\textbf{Analytical results for 
large-$N_c$ scalar \texorpdfstring{$\text{QCD}_2$}{}}\vspace*{0.3cm}}
\date{}
\author{Pavel Meshcheriakov\thanks{p.meshcheriakov@princeton.edu}
\\[\medskipamount]
\parbox[t]{0.85\textwidth}{\normalsize\it\centerline{Joseph Henry Laboratories, Princeton University, Princeton, NJ 08544, USA}}}
\maketitle
\vspace*{-0.5cm}
\begin{abstract}
    We study large-$N_c$ scalar QCD$_2$, a $1+1$-dimensional confining gauge theory with fundamental scalar quarks, whose meson spectrum is governed by a Bethe-Salpeter equation structurally parallel to the 't Hooft equation. Exploiting this structural analogy, we develop a nonperturbative analytic framework, based on integrability and inspired by the Fateev–Lukyanov–Zamolodchikov (FLZ) method, originally devised for the 't Hooft model and later extended in our previous works. Notably, the same Bethe-Salpeter equation also arises in the description of interchain mesons in the doubled Ising model coupled via a spin-spin interaction term. Within the FLZ approach, we find spectral sums and derive a systematic large-$n$ WKB expansion for the meson spectrum. The analytic results reproduce the expected behavior in key asymptotic regimes, such as the near-critical limit $m\to g/\sqrt{\pi}$ and the heavy-quark regime $m\gg g$, and are in good agreement with numerical data. Finally, by analytically continuing the mass parameter into the complex plane, we uncover two infinite families of singularities where individual mesons become massless, suggesting a hidden connection to nontrivial Conformal Field Theories.
\end{abstract}
\tableofcontents
\section{Introduction}\label{Introduction}
Understanding the mechanisms of confinement remains one of the most enduring challenges in Quantum Field Theory. Direct analytical progress in four-dimensional Yang–Mills theory and Quantum Chromodynamics (QCD) has proven extremely difficult, as these theories resist conventional perturbative techniques. A fruitful way forward is to examine simplified settings where essential features of confinement are preserved but the dynamics become more tractable. In particular, lower-dimensional gauge-theory models serve as theoretical laboratories, allowing one to probe non-perturbative structures that may shed light on the dynamics of their four-dimensional counterparts.

In this context, the 't Hooft model \cite{THOOFT1974461} stands as a $1+1$-dimensional analogue of QCD. The theory becomes exactly solvable in the double-scaling limit introduced by 't Hooft, where $N_c\to\infty$ with the 't Hooft coupling $g^2=g_{\text{YM}}^2N_c$ held fixed. In this regime, only planar diagrams without quark loops survive, while all other contributions are suppressed by powers of $1/N_c$. Remarkably, the resulting planar theory already captures several hallmarks of QCD$_4$, including the absence of deconfined quarks and the Regge-like growth of meson masses, establishing the model as an indispensable low-dimensional laboratory for confinement physics.

The designation ``exactly solvable” refers to the fact that the entire planar dynamics can be resummed into a single integral bound state equation—the ’t Hooft equation—governing the meson spectrum,
\begin{equation}\label{tHooft-eq-fermions}
    2\pi^2\lambda\;\phi(x)=\left(\frac{\alpha_1}{x}+\frac{\alpha_2}{1-x}\right)\phi(x)-\fint_0^1\limits dy\frac{\phi(y)}{(x-y)^2},
\end{equation}
where the dimensionless parameters $\alpha_{1,2}$ and eigenvalues $\lambda_n$ (corresponding to the $n$-th meson with wave function $\phi_n(x)$) are related to the quark masses $m_{1,2}$, the ’t Hooft coupling $g$, and the meson masses $M_n$ via
\begin{equation}\label{mass-notation}
    \alpha_i=\frac{\pi m_i^2}{g^2}-1,\quad M_n^2=2\pi g^2\lambda_n.
\end{equation}

In spite of the beautiful and concise form of the 't Hooft equation, it has proven difficult to solve analytically. In his original work \cite{THOOFT1974461}, 't Hooft combined numerical calculations of the meson spectrum with analytic estimates of the leading linear asymptotics \cite{THOOFT1974461}. Subsequent studies refined both aspects: numerical methods were significantly improved, leading to precise determinations of the spectrum \cite{Hanson:1976ey,Brower:1979PhysRevD,Kochergin:2024quv}, while subleading analytic corrections were also worked out \cite{Brower:1979PhysRevD}.

Nevertheless, an explicit analytic solution of the 't Hooft model, valid nonperturbatively in the mass parameters $\alpha_i$, was achieved only recently \cite{Fateev:2009jf,Litvinov:2024riz,Artemev:2025cev}.
A decisive advance came from Fateev, Lukyanov, and Zamolodchikov (FLZ) \cite{Fateev:2009jf}, who showed that for $\alpha_1=\alpha_2=0$ the 't Hooft equation can be reformulated as a Baxter-type TQ relation, enabling exact evaluation of spectral sums and systematic WKB analysis. This integrability-based approach was subsequently extended to arbitrary equal masses, $\alpha_1=\alpha_2=\alpha \geq -1$, and to the fully generic case of unequal masses, $\alpha_1\neq \alpha_2$ with $\alpha_i \geq -1$ \cite{Litvinov:2024riz,Artemev:2025cev}. It was further adapted to the two-particle approximation of the Ising Field Theory (IFT) \cite{Litvinov:2025geb}, where confinement is modeled by interpreting ``mesons’’ as bound states of two Majorana fermions in a magnetic field---an analogue of the 't Hooft integral equation \cite{Fonseca:2006au}. Moreover, the FLZ method can be naturally applied to explore the theory at complex values of the parameters. By treating spectral sums as analytic functions of the mass parameter $\alpha$, one can perform analytic continuation into the complex domain and study the resulting singularities. Such singularities indicate the appearance of massless states in the spectrum and some of them may point to a potential connection with nontrivial Conformal Field Theories. Together, these developments highlight the effectiveness of the FLZ method for probing the analytic structure of bound states in $1+1$-dimensional Quantum Field Theories, naturally suggesting their application to other low-dimensional confining models.

In this work, building on \cite{Fateev:2009jf,Litvinov:2024riz, Artemev:2025cev,Litvinov:2025geb}, we adapt the FLZ method to Yang-Mills theory coupled to $N_f$ flavors of scalar quarks (squarks) in the fundamental representation of $SU(N_c)$ in the 't Hooft (planar) limit, which we will refer to as large-$N_c$ scalar QCD$_2$. This theory is defined by the Lagrangian density
\begin{equation}
    \mathcal{L}=-\frac{1}{2}\text{tr }F_{\mu\nu}F^{\mu\nu}+\sum_{a=1}^{N_f}(D_\mu\varphi_a)^*(D^{\mu}\varphi_a)-m^2_{0a}\varphi^*_a\varphi_a.
\end{equation}
Although scalar quarks have not been observed, they naturally appear in supersymmetric extensions of the Standard Model as the scalar superpartners of quarks \cite{Nilles:1983ge}. In phenomenological studies, they are often interpreted as effective ``diquarks”, with their bound states sometimes referred to as ``tetraquarks”, providing a useful proxy for exploring the conjectured tetraquark spectrum in QCD$_4$ \cite{Grinstein:2008wm}. In this light, large-$N_c$ scalar QCD$_2$ provides a tractable arena for exotic spectroscopy and a simplified setting to probe structures relevant to supersymmetric field theories. In what follows, however, we will use the terminology ``scalar quark”, with ``mesons”\footnote{Beyond mesons, large-$N_c$ QCD$_2$–type theories also contain baryons (e.g. as topological solitons \cite{Witten:1979kh}) and exotic multiquark states suppressed at leading order in $1/N_c$. These sectors lie outside the scope of the present work and will not be considered here.} understood as color-singlet bound states of a scalar quark and antiquark. Their masses are determined by a Bethe-Salpeter equation (bound state equation), which in this model reduces to an integral equation structurally analogous to the 't Hooft equation \eqref{tHooft-eq-fermions} \cite{Halpern:1976gd,Shei:1977ci,Bardeen:1976tm,Tomaras:1978nt}:
\begin{equation}\label{tHooft-eq-bosons}
    2\pi^2\lambda\;\Phi(x)=\mathcal{H}\;\Phi(x)=\left(\frac{\alpha_1}{x}+\frac{\alpha_2}{1-x}\right)\Phi(x)-\fint_0^1\limits dy\frac{\Phi(y)}{(x-y)^2}\frac{(x+y)(2-x-y)}{4x(1-x)},
\end{equation}
with $\alpha_{1,2}$ and $\lambda$ given in \eqref{mass-notation}. Besides the evident difference on the right-hand side compared to the 't Hooft model, scalar QCD$_2$ also requires a logarithmic mass renormalization \cite{Shei:1977ci}. Accordingly, the masses $m_i$ in \eqref{mass-notation} should be interpreted as renormalized rather than bare (Lagrangian) masses $m_{0i}$.

Beyond its phenomenological motivations, large-$N_c$ scalar QCD$_2$ also stands out as a theoretically rich framework with distinctive features. In contrast to the fermionic 't Hooft model, where current correlators show softer behavior than short-distance estimates would suggest \cite{Callan:1976PhysRev}, the scalar theory develops a non-vanishing leading term in the short-distance expansion \cite{Shei:1977ci}, making the comparison with QCD$_4$ somewhat more natural. Another key difference is the absence of massless mesons in scalar QCD$_2$. In this theory, the mass parameter range is restricted to $\alpha_i\geq0$, since for $\alpha_i<0$ the Hamiltonian $\mathcal{H}$ \eqref{tHooft-eq-bosons} becomes non-Hermitian (see Section \ref{Background}). In contrast, in the fermionic ’t Hooft model the allowed range is $\alpha_i\geq-1$, and at the boundary value $\alpha_1=\alpha_2=-1$ (chiral limit $m_i=0$) one finds a single massless meson (the “pion”) \cite{THOOFT1974461}. The absence of massless states in the scalar case was first observed numerically in \cite{Demeterfi:1993rs} and later proven analytically in \cite{Aoki:1995dh}. It is also established that, for equal quark-antiquark masses, the resulting meson is lightest in the scalar case and heaviest in the fermionic case \cite{Aoki:1995dh}.

Moreover, the Bethe-Salpeter equation governing the meson spectrum in scalar QCD$_2$ \eqref{tHooft-eq-bosons} also arises in other contexts. It provides a good approximation to the lowest glueball state in the two-particle sector of adjoint QCD$_2$ with scalar fields \cite{Demeterfi:1993rs}. In addition, we point out that the same equation arises in the two-particle approximation of interchain mesons in the doubled Ising model coupled via spin-spin interaction term \cite{Gao:2025mcg}\cite{Litvinov:2025geb}. Thus, our analysis of scalar QCD$_2$ simultaneously extends to this setting, where the equation holds only approximately. These connections highlight the ubiquity of scalar QCD$_2$ with large $N_c$ as a framework for studying confinement dynamics and bound states in $1+1$ dimensions. 

The rest of the paper is organized as follows. In Section \ref{BS-and-Integr}, we first outline the key properties of the Bethe-Salpeter equation, establish its correspondence with the interchain meson equation in the two-particle approximation of the double Ising model, and then turn to a detailed analysis using the FLZ method. The framework is built around two central objects—the $Q$-function and the spectral determinant—which encode spectral data in an algebraic form. Within this approach, the bound-state equation can be cast as a TQ-equation, which in the single-flavor case ($\alpha_1=\alpha_2=\alpha$) coincides with the TQ-equation for IFT, and whose solutions are identical to those of IFT \cite{Litvinov:2025geb}. This reformulation provides a systematic procedure for extracting spectral information via nontrivial relations connecting spectral determinants with the $Q$-functions, which here are modified versions of the formulas derived for IFT \cite{Litvinov:2025geb}. Section \ref{Analytical-results} presents the analytical results and their comparison with numerical data, including explicit expressions for the spectral sums $\mathcal{G}^{(s)}_{\pm}$ and a systematic large-$n$ WKB expansion. In Section \ref{Limiting-cases-section}, these results are examined in two physically interesting limits: near the critical quark mass $m\to g/\sqrt{\pi}$ ($\alpha\to0$, corresponding to the massless limit in the double IFT) and in the heavy-quark regime $m\gg g$ ($\alpha\to\infty$). This analysis provides a nontrivial consistency check by allowing comparison with known results in these limits. In Section \ref{BS-complex-Section}, we examine the meson spectrum $\lambda_n(\alpha)$ under analytic continuation of the mass parameter $\alpha$ into the complex plane. Using the analytic expressions for the spectral sums $\mathcal{G}^{(s)}_\pm$, we identify two infinite sequences of special points, $\alpha^*_k$ and $\widetilde{\alpha}_k$, where one of the mesons becomes massless. The series $\alpha^*_k$ further generates square-root branch cuts in the complex $\alpha$-plane. Following \cite{Fonseca:2006au,Fateev:2009jf}, these points are conjectured to correspond to infrared ``critical points” governed by nontrivial CFTs. Section \ref{Discussion} concludes with a summary and other possible directions, while technical details and proofs are provided in the appendix.
\section{Bethe-Salpeter equation and Integrability}\label{BS-and-Integr}
This section reviews the properties of the Bethe-Salpeter equation \eqref{tHooft-eq-bosons} and confirms its exact equivalence to the equation describing interchain mesons in the two-particle approximation of the double Ising model coupled via a spin-spin interaction term. We then introduce the FLZ method for large-$N_c$ scalar QCD$_2$: while the framework is formulated for the general case $\alpha_1\ne\alpha_2$, the spectral data extraction formulas are applied specifically to the single-flavor case $\alpha_1=\alpha_2=\alpha$.
\subsection{Background}\label{Background}
We begin with some general remarks on the structure of the Bethe–Salpeter equation \eqref{tHooft-eq-bosons}. By applying the transformation $\phi(x) = \sqrt{x(1-x)}\Phi(x)$ \cite{Shei:1977ci}, the equation can be recast in a symmetric form, in which its Hermitian character becomes manifest:
\begin{equation}\label{tHooft-eq-symmetric}
    2\pi^2\lambda\;\phi(x)=\left(\frac{\alpha_1}{x}+\frac{\alpha_2}{1-x}\right)\phi(x)-\fint_0^1\limits dy\frac{\phi(y)}{(x-y)^2}\frac{(x+y)(2-x-y)}{4\sqrt{y(1-y)}\sqrt{x(1-x)}},
\end{equation}
together with the standard norm
\begin{equation}\label{norm}
    \|\phi\|^2=\int_0^1\limits dx\; \phi^*(x)\phi(x)=1.
\end{equation}
In this setting, the Hamiltonian $\mathcal{H}$ is Hermitian with respect to the metric \eqref{norm}, provided the wave functions obey the appropriate boundary conditions \cite{Shei:1977ci},
\begin{equation}\label{phi-boundary-asympt}
    \phi(x)\sim
    \begin{cases}
        x^{\beta_1},\quad &x\to0;\\
        (1-x)^{\beta_2},\quad &x\to1,
    \end{cases}
\end{equation}
where the exponents $\beta_i$ are determined by the roots of a transcendental equation\footnote{This boundary behavior of the wave functions \eqref{phi-boundary-asympt} coincides with that found in the two-particle approximation of the Ising Field Theory \cite{Litvinov:2025geb}.}
\begin{equation}
    \pi\beta_i\tan\pi\beta_i-\alpha_i=0,\quad 0\leq\beta_i<1.
\end{equation}
 
More specifically, one has \cite{Aoki:1995dh}
\begin{multline}
    (\phi,\mathcal{H}\psi)=(\mathcal{H}\phi,\psi)=\int_0^1\limits dx\left(\frac{\alpha_1}{x}+\frac{\alpha_2}{1-x}+\frac{\pi}{4\sqrt{x(1-x)}}\right)\phi^*(x)\psi(x)+
    \\+\frac{1}{2}\int_0^1\limits dx\int_0^1\limits dy \frac{(\phi^*(x)-\phi^*(y))(\psi(x)-\psi(y))}{(x-y)^2}\frac{(x+y)(2-x-y)}{4\sqrt{x(1-x)y(1-y)}}.
\end{multline}
From this relation, it follows immediately that for positive mass parameters $\alpha_i$ (corresponding to $m_i \ge g/\sqrt{\pi}$) the meson mass is strictly positive. A direct comparison with the analogous expression in the 't Hooft model \cite{THOOFT1974461} shows that, for identical quark and antiquark masses, mesons formed from scalar quarks are lighter than those formed from fermions. This will be clearly evident from our large-$n$ WKB expansion (see Section \ref{WKB-section}).

The Hamiltonian spectrum is discrete, but infinite. Consequently, the meson eigenstates $\phi^{(n)}(x)$ with energies $2\pi^2\lambda_n$ form a complete and orthonormal set
\begin{equation}\label{complete_and_orth}
    \sum_{k=0}^{\infty}(\phi^{(k)}(x))^*\phi^{(k)}(x')=\delta(x-x'),\quad \int_0^1\limits dx\;(\phi^{(n)}(x))^*\phi^{(m)}(x)=\delta_{nm}.
\end{equation}
Furthermore, the eigenfunctions can be chosen to be real, $\phi^*(x)=\phi(x)$, in which case they are simultaneously eigenstates of the charge-conjugation operator $\hat{\mathcal{C}}$.

The Bethe-Salpeter equation \eqref{tHooft-eq-bosons} is invariant under the combined transformation $\alpha_1\leftrightarrow\alpha_2$ and $x\leftrightarrow 1-x$. As a result, the corresponding wave functions can be classified as either symmetric or antisymmetric with respect to this operation:
\begin{equation} \label{sym}
    \phi^{(n)}(x|\alpha_1,\alpha_2) = (-1)^n \phi^{(n)}(1-x|\alpha_2,\alpha_1).
\end{equation}

Let us first recast \eqref{tHooft-eq-bosons} in a more convenient form 
\begin{equation}\label{tHooft-eq-symmetric-1}
    2\pi^2\lambda\;\phi(x)=\left(\frac{\alpha_1}{x}+\frac{\alpha_2}{1-x}\right)\phi(x)+\frac{1}{4}\int_0^1\limits dy\frac{\phi(y)}{\sqrt{y(1-y)}\sqrt{x(1-x)}}+\fint_0^1\limits dy\frac{\phi(y)}{(x-y)^2}\frac{2xy-x-y}{2\sqrt{y(1-y)}\sqrt{x(1-x)}},
\end{equation}
which closely resembles the integral equation governing the two-particle approximation in the IFT \cite{Litvinov:2025geb} (eq. (2.8)). The difference lies in the structure of the projector term, and in IFT only a single flavor is present ($\alpha_1=\alpha_2=\alpha$), since the fermions are Majorana and share the same mass.

It is then natural to rewrite the equation in terms of the variable $\theta=\frac{1}{2}\log\left(\frac{x}{1-x}\right)$, which, in the equal-mass case, has the natural interpretation of the rapidity in the center-of-mass frame:
\begin{equation}\label{tHooft-eq-like-Ising}
    \left(\frac{\pi^2\lambda}{2\cosh^2{\theta}}-\frac{\alpha_1+\alpha_2}{2}+\frac{\alpha_1-\alpha_2}{2}\tanh{\theta}\right)\phi(\theta)=-\fint_{-\infty}^{\infty}\limits\frac{d\theta'}{2\pi}\left(\frac{2\cosh{(\theta-\theta')}}{\sinh^2(\theta-\theta')}-\frac{1}{2\cosh{\theta}\cosh{\theta'}}\right)\phi(\theta').
\end{equation}
The similarity between \eqref{tHooft-eq-like-Ising} and the Bethe-Salpeter equation in the two-particle approximation of IFT is not accidental. In fact, this equation coincides exactly with the one describing the interchain meson in the double Ising model coupled via the spin-spin interaction term $\sigma^{(1)}(x)\sigma^{(2)}(x)$, within the two-particle approximation. The corresponding action reads
\begin{equation}\label{2IFT_action}
    \begin{gathered}
    \mathcal{A}=\mathcal{A}_{\text{FF}}^{(1)}+\mathcal{A}_{\text{FF}}^{(2)}+\mu\int\sigma^{(1)}(x)\sigma^{(2)}(x)\;d^2x,
    \\
    \mathcal{A}_{\text{FF}}^{(i)}=\frac{1}{2\pi}\int \left[\psi^{(i)}\bar{\partial}\psi^{(i)}+\bar{\psi}^{(i)}\partial\bar{\psi}^{(i)}+m_i\bar{\psi}^{(i)}\psi^{(i)}\right]\;d^2x,
    \end{gathered}
\end{equation}
where $\mathcal{A}^{(i)}_{\text{FF}}$ denotes the free fermionic action with mass $m_i$, and $\mu$ controls the strength of the interchain coupling. In the context of the coupled Ising model, equation \eqref{tHooft-eq-like-Ising} was first derived and studied in \cite{Gao:2025mcg} (in one flavor form $\alpha_1=\alpha_2$) and subsequently presented in a two-flavor form in \cite{Litvinov:2025geb}. Thus, the analysis of scalar QCD$_2$ presented in this work also provides, as a by-product, a parallel analysis of the two-particle approximation in the coupled-Ising model, where the same structure governs the spectrum of interchain mesons. 

Having established this structural correspondence, we now turn to a detailed study of the Bethe-Salpeter equation itself. To this end, it is convenient to reformulate the problem in $\nu$-space through a Fourier transform with variable $\theta$, which provides the natural starting point for applying the FLZ method \cite{Fateev:2009jf,Litvinov:2024riz,Artemev:2025cev,Litvinov:2025geb}. Specifically, we define
\begin{equation}\label{Fourier-def}
    \begin{aligned}
        \Psi(\nu)&\overset{\text{def}}{=}\mathcal{F}[\phi(\theta)]=\int_{-\infty}^{\infty}\limits d\theta\;\phi(\theta)e^{-i\nu\theta}=\int_0^1\limits\frac{dx}{2x(1-x)}\left(\frac{x}{1-x}\right)^{-\frac{i\nu}{2}}\phi(x),
        \\
        \phi(\theta)&\overset{\text{def}}{=}\mathcal{F}^{-1}[\Psi(\nu)]=\int_{-\infty}^{\infty}\limits\frac{d\nu}{2\pi}\;\Psi(\nu)e^{i\nu\theta}=\int_{-\infty}^{\infty}\limits \frac{d\nu}{2\pi}\;\left(\frac{x}{1-x}\right)^{\frac{i\nu}{2}}\Psi(\nu),
    \end{aligned}
\end{equation}
and multiplying \eqref{tHooft-eq-symmetric} by $x(1-x)$ before transforming, one arrives at the following representation:
\begin{multline}\label{BS-eq}
    \left(\frac{2\alpha}{\pi}+\nu\tanh{\frac{\pi\nu}{2}}\right)\Psi(\nu)-\frac{2i\beta}{\pi}\fint_{-\infty}^{\infty}\limits d\nu'\frac{1}{2\sinh{\frac{\pi(\nu-\nu')}{2}}}\Psi(\nu')+\frac{1}{8}\frac{1}{\cosh{\frac{\pi\nu}{2}}}\int_{-\infty}^{\infty}\limits d\nu'\frac{1}{\cosh{\frac{\pi\nu'}{2}}}\Psi(\nu')=
    \\=\lambda\int_{-\infty}^{\infty}\limits d\nu'\frac{\pi(\nu-\nu')}{2\sinh{\frac{\pi(\nu-\nu')}{2}}}\Psi(\nu'),
\end{multline}
where we have introduced $\alpha=\frac{\alpha_1+\alpha_2}{2}$ and $\beta=\frac{\alpha_2-\alpha_1}{2}$.

A closer analysis of \eqref{BS-eq} shows that $\Psi(\nu)$ is a meromorphic function of the complex variable $\nu$, with simple poles located at
\begin{equation}\label{Psi-all-poles}
    \pm i\nu_k^*(\alpha\pm\beta)\pm2iN,\quad N=0,1,2,\ldots,
\end{equation}
where $i\nu^*_k(\alpha)$ denotes the $k$-th root of the transcendental equation
\begin{equation}\label{main-transcendental-equation}
    \frac{2\alpha}{\pi}+\nu\tanh\left(\frac{\pi\nu}{2}\right)=0.
\end{equation}
The first such roots, $\pm i\nu^*_1(\alpha\pm\beta)$, lie within the strip $\textrm{Im }\nu\in[-1,1]$ for positive real $\alpha$ and any $\beta$ satisfying $\alpha\pm\beta>0$. As in the two-particle approximation of the IFT \cite{Litvinov:2025geb}, the corresponding pole of $\Psi(\nu)$ dictates the asymptotic form of the wave function $\phi(x)$ \eqref{phi-boundary-asympt}.

Because $\Psi_n(\nu)$ has no poles at $\nu=\pm i$, one obtains the quantization condition
\begin{equation}\label{Psi_quantization_cond}
    \Psi_n(\pm i)=\frac{1}{8}\int_{-\infty}^{\infty}\limits d\nu'\frac{1}{\cosh{\frac{\pi\nu'}{2}}}\Psi_n(\nu'),
\end{equation}
which vanishes identically for odd $\Psi_n(\nu)$. We stress that no analogous condition appears in the 't Hooft model, where the absence of this feature is directly tied to the different structure of \eqref{BS-eq}—in particular, to the extra integral term present on its left-hand side.
\subsection{\texorpdfstring{$Q$}{}-function and TQ equation}
To proceed, we introduce the $Q$-function, a central element of the FLZ method, which forms part of the key equation \eqref{2.17-new}, allowing us to extract spectral data from the theory
\begin{equation}\label{Qdef}
    Q(\nu)\overset{\text{def}}{=}\left(\frac{2\alpha}{\pi}\cosh{\frac{\pi\nu}{2}}+\nu\sinh{\frac{\pi\nu}{2}}\right)\Psi(\nu)=\cosh{\frac{\pi\nu}{2}}\left(\frac{2\alpha}{\pi}+\nu\tanh{\frac{\pi\nu}{2}}\right)\Psi(\nu)
\end{equation}
and consider its meromorphic continuation to the maximal domain of analyticity. In terms of $Q(\nu)$, the Bethe-Salpeter equation \eqref{BS-eq} takes the form
\begin{multline}\label{BS-eq-2}
    Q(\nu)-\frac{2i\beta}{\pi}\cosh{\frac{\pi\nu}{2}}\fint_{-\infty}^{\infty}\limits \frac{1}{2\sinh{\frac{\pi(\nu-\nu')}{2}}}\frac{Q(\nu')}{\cosh{\frac{\pi\nu'}{2}}(\frac{2\alpha}{\pi}+\nu'\tanh{\frac{\pi\nu'}{2}})}+\int_{-\infty}^{\infty}\limits d\nu'\frac{1}{8\cosh^2{\frac{\pi\nu'}{2}}}\frac{Q(\nu')}{\frac{2\alpha}{\pi}+\nu'\tanh{\frac{\pi\nu'}{2}}}=
    \\=\lambda\cosh{\frac{\pi\nu}{2}}\int_{-\infty}^{\infty}\limits d\nu'\frac{\pi(\nu-\nu')}{2\sinh{\frac{\pi(\nu-\nu')}{2}}}\frac{Q(\nu')}{\cosh{\frac{\pi\nu'}{2}}\left(\frac{2\alpha}{\pi}+\nu'\tanh{\frac{\pi\nu'}{2}}\right)}.
\end{multline}

Equation \eqref{BS-eq} (or \eqref{BS-eq-2}) and finiteness of the norm of the original function $\|\phi\|<\infty$ imply the following analytic properties of the function $Q(\nu)$ valid in the strip $\textrm{Im } \nu\in[-2, 2]$:
\begin{enumerate}
    \item $Q(\nu)$ grows slower than any exponential as $|\textrm{Re}\,\nu| \to \infty$, implying that it stays bounded in this limit as
    \begin{equation}\label{Q-is-bounded}  
        \forall\epsilon>0\quad Q(\nu)=\mathcal{O}(e^{\epsilon|\nu|}), \quad|\textrm{Re}\,\nu|\to\infty.
    \end{equation}
    The proof for $\beta\ne0$ proceeds in the same way as in the case $\beta = 0$, discussed in Section 2.2 of \cite{Litvinov:2025geb} in the context of the Ising Field Theory;
    \item $\beta\ne0$: $Q(\nu)$ has (see Appendix A of \cite{Artemev:2025cev} for an explanation of the analogous statement in the ’t Hooft model)
    \begin{equation}\label{essential-zeros&poles}
    \begin{aligned}
        \text{necessarily zeros}&: \quad \pm i\nu^*_1(\alpha),\\
        \text{possible poles}&: \quad i\nu^*_1(\alpha+\beta),\quad -i\nu^*_1(\alpha-\beta),
    \end{aligned}
    \end{equation}
    where $i\nu^*_1(\alpha)$ is the first root (closest to the origin) of \eqref{main-transcendental-equation}. Other zeros are also allowed. On the other hand, poles are only allowed at the listed points, but there might be no poles at all.
    
    $\beta=0$: $Q(\nu)$ is analytic in the strip $\textrm{Im }\nu \in [-2,2]$, which follows from the limit $\beta\to0$: mandatory zeros will always cancel the allowed poles; 
    \item the quantization condition of the function $\Psi(\nu)$ \eqref{Psi_quantization_cond} extends to $Q(\nu)$:
        \begin{equation}\label{Q_quantization_cond}
            Q(\pm i)=-\frac{1}{8}\int_{-\infty}^{\infty}\limits d\nu'\frac{1}{\cosh{\frac{\pi\nu'}{2}}}\frac{Q(\nu')}{\frac{2\alpha}{\pi}\cosh{\frac{\pi\nu'}{2}}+\nu'\sinh{\frac{\pi\nu'}{2}}}.
        \end{equation}
\end{enumerate}

\begin{figure}
    \centering
    \includegraphics[width=1.\linewidth]{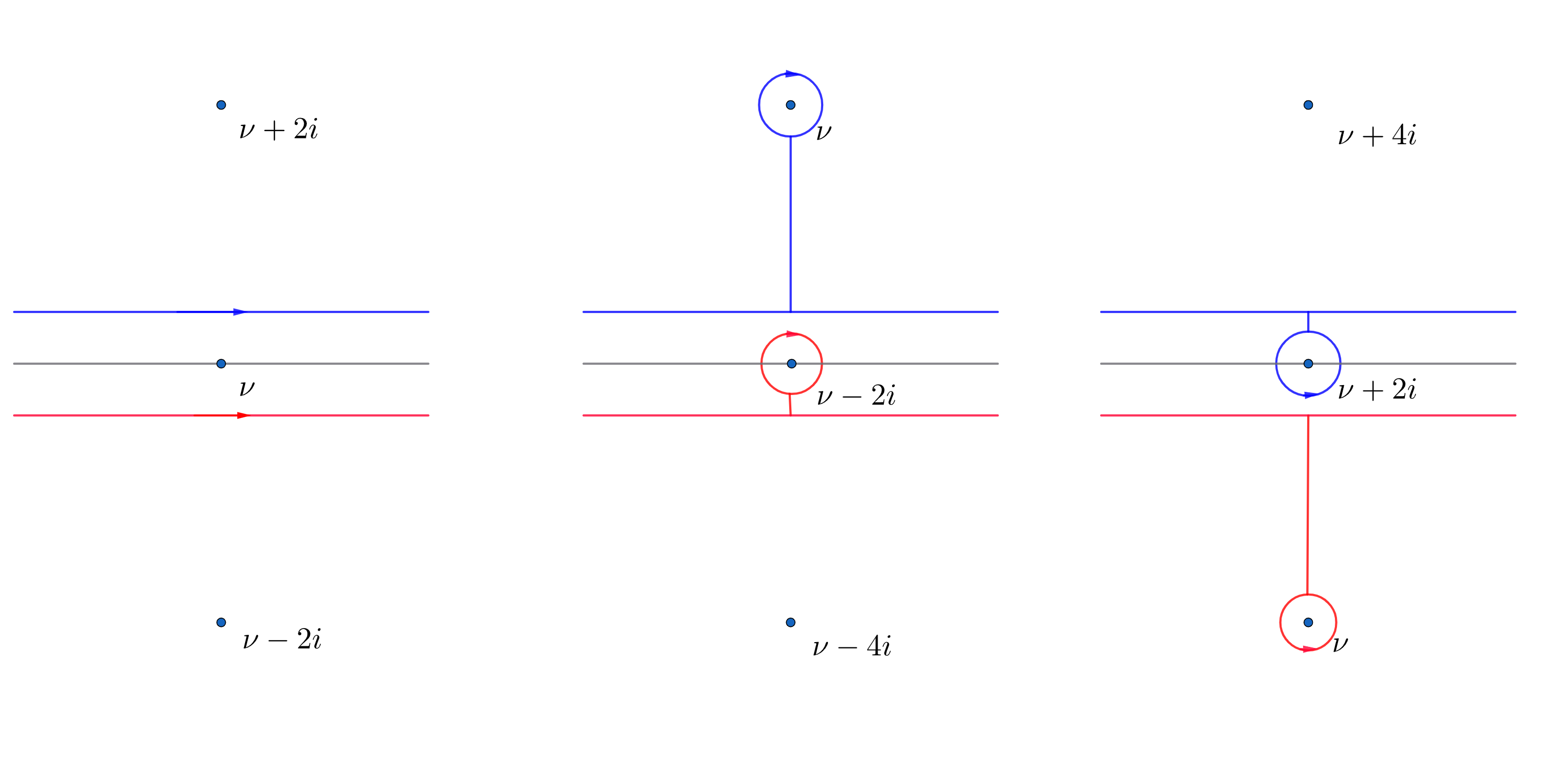}
    \caption{TQ derivation. Analytic continuation $\nu\to\nu\pm2i$ from real values brings additional terms, give by halfresidues of the poles crossing the principal value integration contour.}
    \label{fig:TQ}
\end{figure}

A difference equation satisfied by $Q(\nu)$ can be derived by considering the following linear combination:
\begin{equation}\label{Q-linear-comb}
    Q(\nu+2i)+Q(\nu-2i)-2Q(\nu).
\end{equation}
Performing the analytic continuation to $\nu \pm 2i$ in \eqref{BS-eq-2} (see Fig. \ref{fig:TQ}) 
gives rise to additional contributions originating from half-residues at the poles of $\frac{1}{\sinh\frac{\pi(\nu-\nu')}{2}}$ (located at $\nu'=\nu,\quad \nu\pm 2i$) and $\frac{\nu-\nu'}{\sinh\frac{\pi(\nu-\nu')}{2}}$ (at $\nu'=\nu\pm 2i$), which cross the integration contour. Moreover, upon substitution into \eqref{Q-linear-comb}, the integral contributions mutually cancel. The resulting expression takes the form
\begin{equation}\label{TQ-equation}
    \left(1+\frac{\beta x}{\nu+2i+\alpha x}\right)Q(\nu+2i)+\left(1-\frac{\beta x}{\nu-2i+\alpha x}\right)Q(\nu-2i)-2Q(\nu)=-\frac{2z}{\nu+\alpha x}Q(\nu),
\end{equation}
where
\begin{equation}\label{x-z-notation}
    x=\frac{2}{\pi}\coth{\frac{\pi\nu}{2}},\quad z=2\pi\lambda\coth{\frac{\pi\nu}{2}}.
\end{equation}

Difference equations of the type \eqref{TQ-equation}, commonly known as TQ equations, were originally introduced by Baxter in his seminal study of the eight-vertex model \cite{Baxter:1972hz}. Within the framework of exactly solvable lattice models, such equations provide a fundamental tool for determining the eigenvalues of transfer matrices, with the analytic properties of the $Q$-function playing a decisive role. In the simplest instances, the $Q$-function takes the form of a polynomial, whose zeros are fixed in such a way that the corresponding $T$-function is likewise polynomial (here $T(\nu)=2-\frac{2z}{\nu+\alpha x}$). A similar strategy is employed in the present work. The resulting TQ equation is remarkably similar to that arising in the 't Hooft model \cite{Fateev:2009jf,Litvinov:2024riz,Artemev:2025cev}, with the only modification being the replacement $\coth\frac{\pi\nu}{2}\to\tanh\frac{\pi\nu}{2}$, and is also a direct generalization of the TQ equation in the context of the two-particle approximation of the Ising Field Theory \cite{Litvinov:2025geb}.

Any normalizable solution of the Bethe-Salpeter equation \eqref{BS-eq} must satisfy the TQ equation \eqref{TQ-equation}. Conversely, suppose one considers a solution of the TQ equation that possesses the analytic properties specified in \eqref{Q-is-bounded}–\eqref{essential-zeros&poles}, but without imposing the quantisation condition \eqref{Q_quantization_cond}. Introducing the operator $\hat{O}$
\begin{equation}\label{O-def}
    \hat{O}\cdot f(\nu)\overset{\text{def}}{=}\int_{-\infty}^{\infty}\limits d\nu'\frac{\pi(\nu-\nu')}{2\sinh{\frac{\pi(\nu-\nu')}{2}}}\frac{f(\nu')}{\cosh{\frac{\pi\nu'}{2}}}
\end{equation}
and applying it to both sides of \eqref{TQ-equation}, the first two terms on the left-hand side, after deforming the principal value contour, reduce to
\begin{multline}
    \fint_{-\infty}^{\infty}\limits d\nu'\frac{\pi(\nu-\nu')}{2\sinh{\frac{\pi(\nu-\nu')}{2}}}\frac{Q(\nu'\pm 2i)}{\cosh{\frac{\pi\nu'}{2}}}\frac{\nu'\pm 2i+(\alpha\pm\beta)x'}{\nu'\pm2i+\alpha x'}=\fint_{-\infty}^{\infty}\limits d\nu'\frac{\pi(\nu-\nu'\pm 2i)}{2\sinh{\frac{\pi(\nu-\nu')}{2}}}\frac{Q(\nu')}{\cosh{\frac{\pi\nu'}{2}}}\frac{\nu'+(\alpha\pm\beta)x'}{\nu'+\alpha x'}\mp
    \\
    \mp2\pi i\left(\underset{\nu'=\mp i}{\text{Res}}+\frac{1}{2}\underset{\nu'=\nu\mp2i}{\text{Res}}\right)\frac{\pi(\nu-\nu')}{2\sinh{\frac{\pi(\nu-\nu')}{2}}}\frac{Q(\nu'\pm2i)}{\cosh{\frac{\pi\nu'}{2}}}\frac{\nu'\pm 2i+(\alpha\pm\beta)x'}{\nu'\pm 2i+\alpha x'}.
\end{multline}
The pole at $\nu'\pm 2i+\alpha x'=0$ is compensated by a corresponding zero of the $Q$-function. Conversely, any potential poles of $Q(\nu)$ are canceled by the zeros of the factor $(\nu'\pm 2i+(\alpha\pm\beta)x')$. The resulting residue contributions take the form
\begin{equation}
    \mp\frac{i\pi}{\cosh\frac{\pi\nu}{2}}\left(2(\nu\pm i)Q(\pm i)\mp2iQ(\nu)\frac{(\nu+(\alpha\pm\beta)x)}{\nu+\alpha x}\right).
\end{equation}

Upon collecting all contributions, the action of $\hat{O}$ transforms equation \eqref{TQ-equation} into an inhomogeneous Fredholm integral equation
\begin{multline}\label{BS-eq-inhomogeneous}
    \frac{Q(\nu)}{\cosh{\frac{\pi\nu}{2}}}-\frac{2i\beta}{\pi}\fint_{-\infty}^{\infty}\limits d\nu'\frac{1}{2\sinh{\frac{\pi(\nu-\nu')}{2}}}\frac{Q(\nu')}{\cosh{\frac{\pi\nu'}{2}}(\frac{2\alpha}{\pi}+\nu'\tanh{\frac{\pi\nu'}{2}})}-
    \\-\lambda\int_{-\infty}^{\infty}\limits d\nu'\frac{\pi(\nu-\nu')}{2\sinh{\frac{\pi(\nu-\nu')}{2}}}\frac{Q(\nu')}{\cosh{\frac{\pi\nu'}{2}}\left(\frac{2\alpha}{\pi}+\nu\tanh{\frac{\pi\nu}{2}}\right)}=F(\nu),
\end{multline}
with the inhomogeneous part 
\begin{equation}
    F(\nu)\overset{\text{def}}{=}\frac{Q(i)+Q(-i)}{2}\frac{1}{\cosh{\frac{\pi\nu}{2}}}+\frac{Q(i)-Q(-i)}{2i}\frac{\nu}{\cosh{\frac{\pi\nu}{2}}}.
\end{equation}

Once the values $Q(\pm i)$ are specified, equation \eqref{BS-eq-inhomogeneous} admits a unique solution. Consequently, the space of functions $Q(\nu|\lambda)$ is two-dimensional, with a natural choice of basis given by symmetric and antisymmetric solutions,
\begin{equation}\label{Qpm-symmetry}
    Q_\pm(-\nu)=\pm Q_{\pm}(\nu).
\end{equation}
Each of these solutions corresponds to a particular inhomogeneous term in \eqref{BS-eq-inhomogeneous}
\begin{equation}
    F_+(\nu)=\frac{1}{\cosh\frac{\pi\nu}{2}}\quad \text{and} \quad F_-(\nu) = \frac{\nu}{\cosh\frac{\pi\nu}{2}},
\end{equation}
which in turn is associated with one of the normalization conditions
\begin{equation}\label{Qpm-normalization-conditions}
    Q_+(i)=1 \quad \text{and} \quad Q_-(i)=i,    
\end{equation}
respectively.

In the preceding analysis we considered the general two-flavor case with $\alpha_1 \neq \alpha_2$. Since the two-flavor setting introduces a substantial additional complexity, in what follows we restrict our attention to the simplified case $\alpha_1=\alpha_2=\alpha$ (i.e., $\beta=0$), leaving the fully general situation for future investigation.

The FLZ method \cite{Fateev:2009jf} relies on constructing solutions to the TQ-equation \eqref{TQ-equation} in two asymptotic regimes---at small and large\footnote{It should be emphasized that the region $\lambda\to-\infty$ does not correspond to the physical domain of eigenvalues of the Bethe-Salpeter equation \eqref{BS-eq}. Ultimately, our analysis will involve an analytic continuation to positive values of $\lambda$.} energies—in the form of formal power series
\begin{equation}
    \begin{aligned}
        \lambda\to 0:\quad &Q_{\pm}(\nu|\lambda)=\sum_{k=0}^{\infty}q_{\pm}^{(k)}(\nu)\lambda^k,\qquad\\ 
        \lambda\to-\infty:\quad &Q_{\pm}(\nu|\lambda)=\sum_{k=0}^{\infty}\left((-\lambda)^{-\frac{i\nu}{2}}Q_{\pm}^{(k)}(\nu)\pm(-\lambda)^{\frac{i\nu}{2}}Q_{\pm}^{(k)}(-\nu)\right)\lambda^{-k},
    \end{aligned}
\end{equation}
where each coefficient $q_{\pm}^{(k)}(\nu)$, $Q_{\pm}^{(k)}(\nu)$ is analytic in the strip $\mathrm{Im},\nu \in [-2,2]$ and satisfies the boundedness condition \eqref{Q-is-bounded}. The symmetry constraint \eqref{Qpm-symmetry}, together with the normalization conditions \eqref{Qpm-normalization-conditions}, guarantees the uniqueness of the solution with the required properties. This construction is expected to be fully equivalent to the iterative solution of the inhomogeneous integral equation \eqref{BS-eq-inhomogeneous}, although the direct evaluation of the corresponding integrals is technically cumbersome. For the case $\beta=0$, explicit solutions of the TQ-equation were obtained in \cite{Litvinov:2025geb} (Section 3) by the method originally proposed in \cite{Fateev:2009jf} and subsequently generalized in \cite{Litvinov:2024riz,Artemev:2025cev}. Here we do not reproduce these solutions explicitly; instead, we define the auxiliary integrals $\mathtt{i}_{2k-1}(\alpha)$, $\mathtt{i}_{2k}(\alpha)$ and $\mathtt{u}_{2k-1}(\alpha)$\footnote{Note that
\begin{equation}
    \mathtt{i}_{2k}(\alpha)=2\mathtt{u}_{2k-1}(\alpha)-2\alpha\mathtt{u}_{2k+1}(\alpha)+2\fint_{-\infty}^{\infty}\limits dt \frac{\cosh{t}}{t\sinh^{2k+1}{t}}.
\end{equation}} that enter their construction and appear directly in the analytic expressions for the spectral data of the theory
\begin{equation}\label{i-def}
    \mathtt{i}_{2k-1}(\alpha)\overset{\text{def}}{=}\int_{-\infty}^{\infty}\limits\frac{\sinh 2t+2t}{t\cosh^{2k-1}t(\alpha\cosh t+t\sinh t)}dt, \quad \mathtt{i}_{2k}(\alpha)\overset{\text{def}}{=}\fint_{-\infty}^{\infty}\limits\frac{\cosh{t}(\sinh2t+2t)}{t\sinh^{2k}{t}(\alpha\cosh t+t\sinh t)}dt,
\end{equation}
\begin{equation}\label{u-def}
    \mathtt{u}_{2k-1}(\alpha)\overset{\text{def}}{=}\fint_{-\infty}^{\infty}\limits\frac{\cosh^2{t}}{t\sinh^{2k-1}{t}\cdot(\alpha\cosh{t}+t\sinh{t})}dt.
\end{equation}
The application of the $Q_\pm(\nu)$ solutions to extract spectral data non-perturbative with respect to the mass parameter $\alpha$ will be discussed in Section \ref{D-Q-relations}, with explicit analytical results presented in Section \ref{Analytical-results}.
\subsection{Spectral sums and spectral determinants}
One of the main results of this study is the explicit derivation of analytic formulas for the spectral sums associated with the integral operator in equation \eqref{BS-eq} (see Section \ref{Analytical-results}) . These spectral sums, or spectral zeta functions, for the operator $\hat{\mathcal{O}}$ are introduced in the conventional way as
\begin{equation}
    \zeta_\mathcal{O}(s)\overset{\text{def}}{=}\text{tr }\hat{\mathcal{O}}^{-s}=\sum_{i}\frac{1}{\lambda_i^s},
\end{equation}
initially defined for those values of $s$ where the series is convergent and subsequently extended to other domains by analytic continuation. In the present work, our attention is restricted to integer values of $s$, with particular emphasis on the even and odd spectral sums, which are defined as the traces of the spectral operator over the even and odd sectors, respectively. Specifically,
\begin{equation}\label{spectral-sums-def}
    \mathcal{G}_{+}^{(s)}\overset{\text{def}}{=}\text{tr}_+\;\hat{\mathcal{O}}^{-s}=\sum\limits_{n=0}^\infty\left(\frac{1}{\lambda_{2n}^s}-\frac{\delta_{s,1}}{n+1}\right),\quad \mathcal{G}_{-}^{(s)}\overset{\text{def}}{=}\text{tr}_-\;\hat{\mathcal{O}}^{-s}=\sum\limits_{n=0}^\infty\left(\frac{1}{\lambda_{2n+1}^s}-\frac{\delta_{s,1}}{n+1}\right).
\end{equation}
The subtraction at $s=1$ is necessary to guarantee convergence, as the spectrum grows asymptotically linearly, $\lambda_n \sim n/2$ as $n \to \infty$ \cite{Shei:1977ci}. This behavior will be demonstrated and further systematically generalized in Section \ref{Analytical-results}.

We now introduce another key object within the FLZ method—the spectral determinants\footnote{It should be emphasized that the spectral determinants defined in \eqref{Determinants-def} differ from the standard definition of a spectral determinant for an operator $\mathcal{O}$, which is usually expressed as $\det \mathcal{O} = e^{-\zeta'_\mathcal{O}(0)}$.} (also referred to as Fredholm determinants)—defined as functions whose zeros coincide with the even and odd eigenvalues of the Bethe-Salpeter equation \eqref{BS-eq}
\begin{equation}\label{Determinants-def}
    \mathcal{D}_+(\lambda)\overset{\text{def}}{=}\left(\frac{2\pi}{e}\right)^\lambda\prod_{n=0}^\infty\left(1-\frac{\lambda}{\lambda_{2n}}\right)e^{\frac{\lambda}{n+1}},\quad \mathcal{D}_-(\lambda)\overset{\text{def}}{=}\left(\frac{2\pi}{e}\right)^\lambda\prod_{n=0}^\infty\left(1-\frac{\lambda}{\lambda_{2n+1}}\right)e^{\frac{\lambda}{n+1}}.
\end{equation}
The associated infinite product is convergent at least in the regime of small $\lambda$; in particular, it admits the following representation\footnote{For large $\lambda$, the spectral determinants take the alternative form given in \eqref{F_definition}.}
\begin{equation}\label{D-trough-G-def}
    \mathcal{D}_\pm(\lambda)=\left(\frac{2\pi}{e}\right)^\lambda\exp\left[-\sum_{s=1}^\infty s^{-1}\mathcal{G}^{(s)}_{\pm}\lambda^s\right].
\end{equation}
In Section \ref{D-Q-relations}, we will establish several important relations connecting the solutions of the TQ equation, $Q_\pm$, with the spectral determinants $\mathcal{D}_\pm$. These relations provide a means to extract the relevant spectral data for large-$N_c$ scalar QCD in $1+1$ dimensions.

Now, following \cite{Litvinov:2025geb}, we will specify the operator relevant to our spectral problem and, for brevity, adopt the following notation:
\begin{equation}
    t=\frac{\pi\nu}{2},\quad \phi(t)=\sqrt{f(t)}\Psi(t),\quad f(t)=\alpha+t\cdot\tanh t,
\end{equation}
along with the integral operators $\hat{A}$ and $\hat{K}$ defined as:
\begin{equation}\label{A-K-def}
    (\hat{A}g)(t)\overset{\text{def}}{=}-\frac{1}{8}\int_{-\infty}^{\infty}\limits\frac{1}{\cosh t\cosh t'}\frac{g(t')}{\sqrt{f(t)f(t')}}dt',\quad (\hat{K}g)(t)\overset{\text{def}}{=}\int_{-\infty}^{\infty}\limits\frac{t-t'}{\sinh{(t-t')}}\frac{g(t')}{\sqrt{f(t)f(t')}}dt'.
\end{equation}
In this notation, the bound state equation \eqref{BS-eq} and the corresponding spectral problem can be expressed in compact form
\begin{equation}\label{spectral_problem}
    \phi(t)=(\hat{A}\phi)(t)+\lambda(\hat{K}\phi)(t) \quad \Rightarrow \quad \hat{\mathcal{K}}\phi\overset{\text{def}}{=}(\hat{1}-\hat{A})^{-1}\hat{K}\phi=\frac{1}{\lambda}\phi.
\end{equation}
We will occasionally refer to $\hat{\mathcal{K}}$ and $\hat{K}$ as the ``physical" and ``non-physical" operators, respectively, since $\hat{\mathcal{K}}$ encodes the meson spectrum of the original problem (more precisely, its inverse spectrum), while $\hat{K}$ corresponds to the spectrum without the operator $\hat{A}$.

Let us consider the effect of $\hat{A}$ on the spectral sums, noting that it is proportional to the projector $\hat{P}$ onto a particular even state $\ket{p}$
\begin{equation}\label{A-proector}
\begin{gathered}
    \ket{p}=\frac{1}{\sqrt{8\mathtt{v}(\alpha)}}\frac{1}{\cosh{t}\sqrt{f(t)}},\\
    \hat{A}=-\mathtt{v}(\alpha)\hat{P},\quad \hat{P}^2=\hat{P},\quad \braket{p|p}=1,
\end{gathered}
\end{equation}
where
\begin{equation}\label{Proector-def}
    \mathtt{v}(\alpha)=\frac{1}{8}\int_{-\infty}^{\infty}\limits\frac{dt}{f(t)} \frac{1}{\cosh^2t},\quad (\hat{P}g)(t)=\frac{1}{8\mathtt{v}(\alpha)}\int_{-\infty}^{\infty}\limits dt'\frac{1}{\cosh t \cosh t'} \frac{g(t')}{\sqrt{f(t)f(t')}}.
\end{equation}
Employing \eqref{A-proector}, the spectral operator $\hat{\mathcal{K}}$ can be expressed in the simplified form 
\begin{equation}
    (\hat{1}-\hat{A})^{-1}=\hat{1}+\sum\limits_{n=1}^\infty\hat{A}^n=\hat{1}+\sum\limits_{n=1}^\infty(-1)^n\mathtt{v}^n(\alpha)\hat{P}^n=\hat{1}-\frac{\mathtt{v}(\alpha)}{1+\mathtt{v}(\alpha)}\hat{P}\quad\Rightarrow\quad \hat{\mathcal{K}}=\left(\hat{1}+\mathcal{V}(\alpha)\hat{P}\right)\hat{K},
\end{equation}
where
\begin{equation}\label{xi-def}
    \mathcal{V}(\alpha)\overset{\text{def}}{=}-\frac{\mathtt{v}(\alpha)}{1+\mathtt{v}(\alpha)}.
\end{equation}

We will denote the spectral sums of the operator $\hat{K}$ by $G_{\pm}^{(s)}$, and those of $\hat{\mathcal{K}}$ by $\mathcal{G}_{\pm}^{(s)}$. Because $\hat{P}$ projects onto an even state $\ket{p}$, the odd spectral sums for both operators coincide, $G_{-}^{(s)} = \mathcal{G}_{-}^{(s)}$. In contrast, the even spectral sums differ by contributions involving matrix elements of the form $\bra{p}\hat{K}^n\ket{p}$. These relations are obtained by inserting a complete set of states into the trace expression for $\hat{\mathcal{K}}$. For instance, the first even spectral sum is given by $\mathcal{G}_{+}^{(1)}$
\begin{equation}
    \mathcal{G}_{+}^{(1)}=G_{+}^{(1)}+\mathcal{V}(\alpha)\bra{p}\hat{K}\ket{p},\quad \bra{p}\hat{K}\ket{p}=\frac{1}{8\mathtt{v}(\alpha)}\int_{-\infty}^{\infty}\limits\frac{dt}{f(t)}\frac{1}{\cosh{t}}\int_{-\infty}^{\infty}\limits \frac{dt'}{f(t')}\frac{t-t'}{\sinh{(t-t')}}\frac{1}{\cosh{t'}}.
\end{equation}
The general expression for $\mathcal{G}_{+}^{(s)}$ takes the following form
\begin{equation}\label{Gp-with-matr-el}
    \mathcal{G}_{+}^{(s)}=G_{+}^{(s)}+s\cdot\sum\limits_{m=1}^s\frac{\mathcal{V}^m(\alpha)}{m}\sum_{\substack{i_1j_1+\ldots+i_mj_m=s
    \\j_1+\ldots+j_m=m
    \\i_1\neq i_2\neq\ldots\neq i_m}}\frac{m!}{j_1!\ldots j_m!}\langle p|\hat{K}^{i_1}\ket{p}^{j_1}\cdot\ldots\cdot\bra{p}\hat{K}^{i_m}\ket{p}^{j_m},
\end{equation}
where additional terms involving matrix elements can be interpreted combinatorially as all possible partitions of the integer $s$, each weighted according to certain symmetry factors.

We now establish the relation between $D_\pm$ and $\mathcal{D}_\pm$, the spectral determinants of the operators $\hat{K}$ and $\hat{\mathcal{K}}$, respectively. It is immediately evident that the odd determinants coincide, $\mathcal{D}_-(\lambda)=D_-(\lambda)$, since the operator $\hat{A}$ in \eqref{A-K-def} does not contribute to the Bethe-Salpeter equation \eqref{BS-eq} for odd wave functions. The relation for the even determinants, however, is more subtle. Specifically, the sum appearing on the right-hand side of \eqref{Gp-with-matr-el} can be interpreted as the coefficient in the power series expansion of the function $g(\lambda)=\sum_{k=1}^{\infty}\limits g^{(k)}\lambda^k$, where the coefficients are given by $g^{(k)}=\mathcal{V}(\alpha)\bra{p}\hat{K}^{k}\ket{p}$:
\begin{equation}
    \frac{1}{m}(g(\lambda))^m=\frac{1}{m}\sum_{k_1,\dots,k_m=1}^{\infty}g^{(k_1)}\ldots g^{(k_m)}\lambda^{k_1+\ldots+k_m}=\frac{1}{m}\sum_{n=m}^{\infty}\sum_{\substack{i_1j_1+\ldots+i_mj_m=n\\j_1+\ldots+j_m=m\\i_1\ne\ldots\ne i_m}}^{\infty}\frac{m!}{j_1!\ldots j_m!}(g^{(i_1)})^{j_1}\ldots(g^{(i_m)})^{j_m}\lambda^{n}.
\end{equation}
The function $g(\lambda)$ can be determined explicitly (see Appendix \ref{matrix-el-of-K}, formula \eqref{g-func-explicit})
\begin{equation}\label{g-for-appendix}
    g(\lambda)=-\mathcal{V}(\alpha)+\frac{i}{1+\mathtt{v}(\alpha)}\frac{Q'_+(i)}{2\pi^2\lambda}=\frac{1}{1+\mathtt{v}(\alpha)}\left(\mathtt{v}(\alpha)+\frac{iQ'_+(i)}{2\pi^2\lambda}\right).
\end{equation}
Using the established relation between spectral sums and determinants \eqref{D-trough-G-def}, we then arrive at
\begin{equation}\label{newDet-oldDet}
    \mathcal{D}_-(\lambda)=D_-(\lambda),\quad \mathcal{D}_+(\lambda)=\frac{1}{1+\mathtt{v}(\alpha)}\left(1-\frac{iQ'_+(i)}{2\pi^2\lambda}\right)D_+(\lambda).
\end{equation}

We emphasize that the relation between $D_+$ and $\mathcal{D}_+$ was derived as a small-$\lambda$ expansion; nevertheless, it remains valid for all positive values of $\lambda$. By definition, $\mathcal{D}_+(\lambda)$ is a function whose zeros coincide with the spectrum $\lambda_{2n}$ of the Bethe-Salpeter equation \eqref{BS-eq}. In contrast, $D_+(\lambda)$ corresponds to the spectral problem without the operator $\hat{A}$, and hence its zeros generally do not match the physical meson masses. Consequently, the correct meson masses are determined by the additional factor appearing in \eqref{newDet-oldDet}
\begin{equation}\label{Quant_cond2}
    1-\frac{iQ'_+(i)}{2\pi^2\lambda}=0.
\end{equation}
Moreover, equation \eqref{Quant_cond2} coincides with the quantization condition \eqref{Q_quantization_cond} for the $Q$-function, providing a non-perturbative verification (up to a factor independent of the zeros in $\lambda$) of the relation between the spectral determinants; further details are given in Appendix \ref{matrix-el-of-K}.

Let us comment on the similarities and differences between the spectral problems in scalar QCD$_2$ and in the Ising Field Theory. In the two-particle approximation of the Ising model \cite{Fonseca:2006au,Litvinov:2025geb}, the operator $\hat{A}$ projected onto an odd state. Consequently, matrix-element terms appeared in the odd spectral sums $\mathcal{G}^{(s)}_-$, and in the analogue of relation \eqref{newDet-oldDet} the roles of the spectral determinants were interchanged: the even determinants coincided for the physical operator $\hat{\mathcal{K}}$ and non-physical $\hat{K}$, whereas the odd ones differed by a factor of the type given in \eqref{Quant_cond2}. By contrast, if one disregards the operator $\hat{A}$ and considers the spectral problem for the non-physical operator $\hat{K}$, the structure is identical in both theories. As a result, the bare spectral sums $G^{(s)}_\pm$ (prior to inclusion of matrix-element contributions) coincide in scalar QCD$_2$ and the Ising model. The first five spectral sums in the Ising case were computed in \cite{Litvinov:2025geb}, and we will reproduce them in Section \ref{Analytical-results}. The physical spectral sums, however, acquire additional matrix-element contributions that cannot be neglected, with the explicit form of these elements depending on the particular theory.
\subsection{Relation between \texorpdfstring{$Q$}{}-functions and spectral determinants}\label{D-Q-relations}
In this subsection, we examine how the solutions of the TQ equation \eqref{TQ-equation}, $Q_\pm(\nu)$, derived in Section 3 of \cite{Litvinov:2025geb} for the case $\beta=0$, can be utilized to investigate the meson mass spectrum in large-$N_c$ scalar QCD$_2$. Specifically, we present three non-perturbative relations—valid at least in both asymptotic regimes ($\lambda \to 0$ and $\lambda \to -\infty$)—that connect the spectral determinants $\mathcal{D}_\pm$ with the functions $Q_\pm(\nu)$.

It is sufficient to establish these relations for the determinants $D_\pm$, defined via the spectrum of the non-physical operator $\hat{K}$ \eqref{A-K-def}, and then apply the kinematic relation \eqref{newDet-oldDet} to connect them with the physical determinants $\mathcal{D}_\pm$. Since the equation without the operator $\hat{A}$ is identical to that appearing in the two-particle approximation of the Ising Field Theory, the relations between $D_\pm$ and $Q_\pm$ coincide with those derived in \cite{Litvinov:2025geb} (Section 4). For this reason, we reproduce the formulas here without detailed derivations, highlighting only the essential facts relevant to our analysis.
\paragraph{Integral relations.}
The spectral determinants $D_\pm$ and the $Q$-functions are connected by the following integral relations:
\begin{equation}\label{DD-integral}
    \begin{aligned}
        \partial_\lambda\log(D_+D_-)&=2-\fint_{-\infty}^\infty\limits d\nu\left[\frac{1}{f(\nu)}\left(Q_+(\nu)\partial_\nu Q_-(\nu)-Q_-(\nu)\partial_\nu Q_+(\nu)\right)-\frac{1}{\nu\tanh\frac{\pi\nu}{2}}\right],
        \\
        \partial_\lambda\log\left(\frac{D_+}{D_-}\right)&=-\int_{-\infty}^{\infty}\limits\frac{\pi}{f(\nu)}\frac{Q_+(\nu)Q_-(\nu)}{\sinh{\pi\nu}}d\nu.
    \end{aligned}
\end{equation}
The proof relies on the fact that the operator $\hat{K}$ belongs to the class of so-called ``completely integrable" operators \cite{Its:1980,Its:1990MPhysB}, whose resolvent can be expressed explicitly in terms of the solutions of the inhomogeneous Fredholm integral equation \eqref{BS-eq-inhomogeneous} \cite{Its:1980,Its:1990MPhysB,zbMATH01284258}. A perturbative derivation of this property, in the context of the ’t Hooft model and based on the Liouville-Neumann series, is presented in Appendix A of \cite{Fateev:2009jf}. The argument for scalar QCD$_2$ proceeds in a fully analogous manner.

Employing the explicit $Q_\pm(\nu)$ solutions of the TQ equation \eqref{TQ-equation}, obtained in Section 3 of \cite{Litvinov:2025geb}, within the integral identities \eqref{DD-integral} enables one to derive the first spectral sums $G^{(1)}_\pm$ in closed analytic form and to approximate higher-order sums $G^{(s)}_\pm$ numerically. Nevertheless, this procedure is not practically efficient. More importantly, in line with the discussion in Appendix A of \cite{Litvinov:2024riz}, these identities play the role of a nontrivial cross-check, complementing the stronger relations developed in the following paragraphs.
\paragraph{Rational relations.}
The following rational relations establish the link between $D_\pm$ and the $Q_\pm$:
\begin{equation}\label{DD-QQ-relation}
    \frac{Q_-(i)}{Q_-(2i)} = \frac{1}{2} \frac{D_-(\lambda)}{D_+(\lambda)} \quad \text{and} \quad \frac{Q_+(i)}{Q_+(0)}= \frac{D_+(\lambda)}{D_-(\lambda)}.
\end{equation}
The derivation of this fact is based on the fact that Quantum Wronskian $W(\nu)=Q_-(\nu+i)Q_+(\nu-i)-Q_+(\nu+i)Q_-(\nu-i)$ is a constant. An analogous relation also holds in the case of the single-flavor 't Hooft model \cite{Fateev:2009jf,Litvinov:2024riz}. In the two-flavor case, however, the Wronskian ceases to be constant \cite{Artemev:2025cev}; nevertheless, its analysis still yields a closely related, though modified, version of the relation rather than the original formula itself.

Using \eqref{newDet-oldDet}, one can express the ratio of the spectral determinants $\mathcal{D}_{\pm}(\lambda)$ 
\begin{equation}\label{DD-QQ-relation-new}
    \frac{Q_-(i)}{Q_-(2i)}=\frac{1}{2(1+\mathtt{v}(\alpha))}\left(1-\frac{iQ'_+(i)}{2\pi^2\lambda}\right)\frac{\mathcal{D}_-(\lambda)}{\mathcal{D}_+(\lambda)} \quad \text{and}\quad\frac{Q_+(i)}{Q_+(0)}=(1+\mathtt{v}(\alpha))\left(1-\frac{iQ'_+(i)}{2\pi^2\lambda}\right)^{-1}\frac{\mathcal{D}_+(\lambda)}{\mathcal{D}_-(\lambda)}.
\end{equation}
The key advantage of \eqref{DD-QQ-relation-new} is that it provides access to the ratio of normalization factors of the spectral determinants \eqref{dm/dp}, as well as explicit formulas for the differences of the spectral sums $\mathcal{G}^{(s)}_+-\mathcal{G}^{(s)}_-$ and the differences of $F^{(s)}_+-F^{(s)}_-$ \eqref{F_definition}, the latter being the coefficients in the large-$\lambda$ expansion of $\log{\mathcal{D}_\pm}$.
\paragraph{Log-derivative relations.}
In the case of the Ising Field Theory \cite{Litvinov:2025geb}, we were able to establish the following relations of this kind for the spectral determinants $D_\pm$ of the operator $\hat{K}$ \eqref{A-K-def}
\begin{equation}\label{2.17-old}
\begin{aligned} 
    &\partial_\lambda\log{D_-(\lambda)}-\frac{\alpha\mathtt{i}_2(\alpha)}{4}=2i\left(1-\frac{\alpha}{\pi^2}\lambda^{-1}\right)\partial_\nu\log{Q_+(\nu)}\Big|_{2i},\\
    &\partial_\lambda\log{D_+(\lambda)}-\frac{\alpha\mathtt{i}_2(\alpha)}{4}=2i\left(1+\frac{1}{1-\frac{\pi^2\lambda}{\alpha}}\right)\partial_\nu\log{Q_-(\nu)}\Big|_{2i},
\end{aligned}
\end{equation}
where $\mathtt{i}_2(\alpha)$ is defined in \eqref{i-def}. Using the relations between the spectral determinants $D_\pm$ and $\mathcal{D}_\pm$ \eqref{newDet-oldDet}, these formulas can be rewritten for the spectral determinants of the operator $\hat{\mathcal{K}}$ \eqref{spectral_problem}
\begin{equation}\label{2.17-new}
\begin{aligned} 
    &\partial_\lambda\log{\mathcal{D}_-(\lambda)}-\frac{\alpha\mathtt{i}_2(\alpha)}{4}=2i\left(1-\frac{\alpha}{\pi^2}\lambda^{-1}\right)\partial_\nu\log{Q_+(\nu)}\Big|_{2i},\\
    &\partial_\lambda\log{\mathcal{D}_+(\lambda)}-\frac{\alpha\mathtt{i}_2(\alpha)}{4}=2i\left(1+\frac{1}{1-\frac{\pi^2\lambda}{\alpha}}\right)\partial_\nu\log{Q_-(\nu)}\Big|_{2i}+\partial_\lambda\log{\left(1-\frac{iQ'_+(i)}{2\pi^2\lambda}\right)}.
\end{aligned}
\end{equation}
A relation of this type was first derived for the 't Hooft model in the case $\alpha=0$ in \cite{Fateev:2009jf}, and was subsequently generalized to arbitrary $\alpha$ \cite{Litvinov:2024riz} and to the general two-flavor case \cite{Artemev:2025cev}. The 't Hooft equation is structured so that there is no notion of non-physical spectral determinants; all spectral determinants are directly associated with the physical spectrum. In contrast, this additional complication appears in the Ising Field Theory and was resolved in \cite{Litvinov:2025geb}.

Unlike the relations discussed in the preceding paragraphs, formula \eqref{2.17-new} currently lacks a rigorous derivation. Nevertheless, it has been confirmed to high accuracy through extensive numerical and analytical checks (see Sections \ref{Analytical-results} and \ref{Limiting-cases-section}). This formula plays a central role by linking the spectral determinants with the $Q$-functions and provides the foundation for extracting all spectral results presented in this work.
\section{Analytical and numerical results}\label{Analytical-results}
In this section, we present the main analytic results for the spectrum of large-$N_c$ scalar QCD in $1+1$ dimensions and validate them using two complementary numerical methods, each compensating for the limitations of the other.
\subsection{Spectral sums}
The coefficients appearing in the small-$\lambda$ expansion of $\log \mathcal{D}_\pm(\lambda)$ can be identified with the spectral sums $\mathcal{G}^{(s)}_\pm$, which follows directly from \eqref{D-trough-G-def}. Although these sums themselves lack a direct physical interpretation, they serve as convenient benchmark ``observables” for testing the validity of our analytic results. Furthermore, when combined with the WKB expansion presented in the next subsection, they offer a practical method for improving numerical estimates of the spectrum—especially for the lowest energy levels, where the WKB approach alone is least accurate.

Our proposed formulas yield non-perturbative predictions for the spectral sums $G^{(s)}_\pm$ of the auxiliary (``non-physical'') operator $\hat{K}$ \eqref{A-K-def}, as well as for the corresponding sums $\mathcal{G}^{(s)}_\pm$ of the ``physical'' operator $\hat{\mathcal{K}}$ \eqref{spectral_problem}. Utilizing the logarithmic derivative identities \eqref{2.17-new}, we are able to precisely determine these sums by equating the coefficients at $\lambda^s$ in the left and right sides of the equation. The first three explicit spectral sums are presented below
\begin{equation}\label{Gpm1}
    \begin{aligned}
        &\mathcal{G}^{(1)}_+=\log(2\pi)-3-\frac{\alpha\mathtt{i}_2(\alpha)}{4}+\mathcal{V}(\alpha)\bra{p}\hat{K}\ket{p},\\
        &\mathcal{G}^{(1)}_-=\log(2\pi)-1-\frac{2\alpha\zeta(3)}{\pi^2}-\frac{\alpha\mathtt{i}_2(\alpha)}{4}-\alpha^2\mathtt{u}_3(\alpha),
    \end{aligned}
\end{equation}
\begin{equation}\label{Gpm2}
    \begin{aligned}
        \mathcal{G}^{(2)}_+=&\;2+\frac{\pi^2}{\alpha}-\frac{4\alpha}{3}+\frac{8\alpha^2\zeta(3)}{3\pi^2}-\frac{4\alpha^2\zeta(5)}{\pi^4}+2\alpha^3\left(\mathtt{u}_3(\alpha)+\mathtt{u}_5(\alpha)\right)+2\mathcal{V}(\alpha)\bra{p}\hat{K}^2\ket{p}+\mathcal{V}^2(\alpha)\bra{p}\hat{K}\ket{p}^2,\\
        \mathcal{G}^{(2)}_-=&-\frac{\pi^2}{\alpha}+\frac{8\alpha}{3}-\frac{8\alpha}{3\pi^2}(3+2\alpha)\zeta(3)+\frac{4\alpha^2\left(\zeta^2(3)+2\zeta(5)\right)}{\pi^4}-4\alpha^2\mathtt{u}_3(\alpha)\left(1+\alpha-\frac{\alpha\zeta(3)}{\pi^2}\right)+
        \\&+\alpha^4\mathtt{u}^2_3(\alpha)-4\alpha^3\mathtt{u}_5(\alpha),
    \end{aligned}
\end{equation}
\begin{equation}\label{Gpm3}
    \begin{aligned}
        \mathcal{G}^{(3)}_+=&\;\frac{10\pi^2}{9}-\frac{3\pi^2}{\alpha}-\frac{4}{15}\left(10+6\alpha^2-5\alpha(2-\zeta(3))\right)+\frac{184\alpha^3\zeta(3)}{45\pi^2}+\frac{2\alpha}{\pi^2}\left(1-\frac{16\alpha^2}{3\pi^2}\right)\zeta(5)+\frac{8\alpha^3\zeta(7)}{\pi^6}-
        \\
        &-\alpha^2\left(\left(\pi^2-4\alpha^2\right)\mathtt{u}_3(\alpha)+\left(\pi^2-8\alpha^2\right)\mathtt{u}_5(\alpha)-4\alpha^2\mathtt{u}_7(\alpha)\right)+3\mathcal{V}(\alpha)\bra{p}\hat{K}^3\ket{p}+
        \\&+3\mathcal{V}^2(\alpha)\bra{p}\hat{K}^2\ket{p}\bra{p}\hat{K}\ket{p}+\mathcal{V}^3(\alpha)\bra{p}\hat{K}\ket{p}^3,\\
        \mathcal{G}^{(3)}_-=&-\frac{4\pi^2}{3}+\frac{4\alpha(5+3\alpha)}{5}+2\alpha^2(3+2\alpha)\left(\frac{2\zeta(3)}{\pi^2}+\alpha\mathtt{u}_3(\alpha)\right)^2-\alpha^3\left(\frac{2\zeta(3)}{\pi^2}+\alpha\mathtt{u}_3(\alpha)\right)^3-
        \\&
        -\frac{12\zeta(7)\alpha^3}{\pi^6}-\frac{\alpha\left(\pi^2-4\alpha(3+2\alpha)\right)}{3}\left(\frac{6\zeta(5)}{\pi^4}-\alpha\left(\mathtt{u}_3(\alpha)+3\mathtt{u}_5(\alpha)\right)\right)+
        \\&+2\left(\frac{2\zeta(3)}{\pi^2}+\alpha\mathtt{u}_3(\alpha)\right)\left(\pi^2+\frac{(\pi^2-9)\alpha}{3}-\frac{(90+23\alpha)\alpha^2}{15}-\frac{6\alpha^3\zeta(5)}{\pi^4}+\alpha^4\left(\mathtt{u}_3(\alpha)+3\mathtt{u}_5(\alpha)\right)\right)-
        \\&-\frac{2\alpha^4}{15}\left(2\mathtt{u}_3(\alpha)+30\mathtt{u}_5(\alpha)+45\mathtt{u}_7(\alpha)\right).
    \end{aligned}
\end{equation}
For the sake of clarity, we do not present the full expressions for the matrix elements $\bra{p}\hat{K}^n\ket{p}$ here, as their inclusion would unnecessarily complicate the formulas. The explicit forms of the first three matrix elements are provided in Appendix \ref{matrix-el-of-K}. A more extensive compilation, containing the initial spectral sums together with the matrix elements up to $s=5$, is available in the supplementary file \texttt{Spectral-sums-scalar-QCD.nb}. Numerical verification based on a standard discretization procedure of the integral equation \eqref{BS-eq} confirmed the expressions with high accuracy. Analytic expressions can, in principle, be derived for any value of $s$; however, the complexity of the calculations grows rapidly as $s$ increases. A comparison of analytical expressions for spectral sums with numerical results is presented in Table \ref{tab:G}.

It should be emphasized that, in principle, the spectral sums $G^{(s)}_\pm$ can be obtained analytically from the integral relations \eqref{DD-integral} by substituting the asymptotic expansions of the $Q$-functions in $\lambda$. For instance, determining the first spectral sums requires only the leading terms,
\begin{equation}
    Q_+(\nu)=1+\mathcal{O}(\lambda),\quad Q_-(\nu)=\nu+\mathcal{O}(\lambda).
\end{equation}
In practice, however, the direct evaluation of the integrals in \eqref{DD-integral} becomes increasingly inefficient for higher-order sums $G^{(s)}_\pm$. As $s$ grows, the explicit calculations rapidly turn cumbersome. In addition, the rational identities \eqref{DD-QQ-relation-new} allow one to obtain explicit expressions for the differences of spectral sums. 
\subsection{WKB expansion}\label{WKB-section}
To the best of our knowledge, no analytic corrections to the leading result \eqref{WKB-old} have previously been presented for large-$N_c$ scalar QCD$_2$. This highlights that, compared to the extensively studied 't Hooft model, the scalar theory has received far less attention. In the case of the ’t Hooft model, the leading and subleading corrections to the semiclassical spectrum were worked out long ago \cite{THOOFT1974461,Brower:1979PhysRevD}, and only with the more recent works \cite{Fateev:2009jf,Litvinov:2024riz,Artemev:2025cev} was a systematic framework developed to obtain nonperturbative corrections in the parameter $\alpha$ to the linear spectrum.

For sufficiently large $n$ (highly excited states), approximate analytic expressions for both meson wave functions and eigenvalues can be derived using a semiclassical (WKB) approach \cite{Shei:1977ci}
\begin{equation}\label{WKB-old}
\phi_{n}(x)\approx\sqrt{2}\sin{\left(\pi n x\right)},\quad \lambda_n\approx \tfrac{1}{2} n .
\end{equation}
By analogy with the ’t Hooft model \cite{Brower:1979PhysRevD}, the validity of this approximation for the wavefunction should be restricted to the region $\tfrac{1}{n}\lesssim x\lesssim 1-\tfrac{1}{n}$; however, this refinement has not yet been studied in the scalar case. For the fermionic theory, an additional phase shift in the sine function was also derived in \cite{Brower:1979PhysRevD}. In what follows, we apply the FLZ method to extend these approximate results and derive systematic corrections to \eqref{WKB-old} that are nonperturbative in the mass parameter $\alpha$.

For large negative $\lambda$, the spectral determinants $\mathcal{D}_\pm(\lambda)$ possess the following asymptotic expansion, which follows directly from their definition \eqref{Determinants-def}
\begin{equation}\label{F_definition}
    \mathcal{D}_{\pm}(\lambda)=d_{\pm}\left(2\pi e^{-2+\gamma_E}\right)^{\lambda}(-\lambda)^{\lambda+\frac{1}{8}\pm\frac{1}{4}}
    \exp\Bigl(F^{(0)}_{\pm}(L)+F^{(1)}_{\pm}(L)\lambda^{-1}+F^{(2)}_{\pm}(L)\lambda^{-2}+\dots\Bigr),
\end{equation}
with $F^{(k)}_{\pm}(L)$ being the polynomials in $L=\log(-2\pi\lambda)+\gamma_E$. The prefactor $(2\pi e^{-2+\gamma_E})^\lambda$ in \eqref{F_definition} uniquely determines the appearance of the additional term $-\frac{\alpha \mathtt{i}_2(\alpha)}{4}$ on the left-hand side of the logarithmic derivative identities \eqref{2.17-new}. Its value is fixed by matching the $\mathcal{O}(\lambda^0)$ contributions in the limit $\lambda \to -\infty$ on both sides of the relation. Utilizing the asymptotic expansion of the $Q$-functions which was derived in \cite{Litvinov:2025geb} and integrating the logarithmic derivative identities with respect to $\lambda$, one can systematically extract the coefficients $F^{(k)}_\pm(L)$ in powers of $1/\lambda$; for instance:
\begin{equation}
    \begin{aligned}
        F^{(0)}_{\pm}(L)=&\;-\frac{\alpha L^2}{2\pi^2}+\frac{\alpha\mathtt{i}_2(\alpha)L}{4\pi^2},
        \\
        F^{(1)}_{\pm}(L)=&\;\pm\frac{\left(\pi^2\pm4\alpha^2\right)L}{8\pi^4}-\frac{8\alpha^2\pm2\pi^2(2-4\log2+8\alpha \pm(1+\alpha))+4\alpha^3\mathtt{i}_2(\alpha)\pm\pi^2\alpha\mathtt{i}_1(\alpha)}{32\pi^4},
    \end{aligned}
\end{equation}
where the integrals $\mathtt{i}_1(\alpha)$ and $\mathtt{i}_2(\alpha)$ are defined in \eqref{i-def}.

The identity \eqref{DD-QQ-relation-new} allows us to derive explicit expressions for the differences $F^{(k)}_+(L)-F^{(k)}_-(L)$, which provide a useful and nontrivial consistency check. Moreover, it gives access to the ratio\footnote{For the spectral determinants $D_\pm(\lambda)$ of the operator $\hat{K}$ \eqref{A-K-def}, this ratio equals $\frac{\sqrt{2\alpha}}{\pi}$.}
\begin{equation}\label{dm/dp}
    \frac{d_-}{d_+}=(1+\mathtt{v}(\alpha))\frac{\sqrt{2\alpha}}{\pi},
\end{equation}
even though the individual quantities $d_+$ and $d_-$ cannot be determined separately by the method employed in this work. The coefficient in \eqref{dm/dp} admits a representation as a rapidly convergent infinite product, obtained directly from the definition \eqref{Determinants-def} together with the leading asymptotic behavior of the spectrum (which will be derived below)
\begin{equation}\label{dm/dp-prod}
    \frac{d_-}{d_+}=\frac{\Gamma({\frac{5}{8}})}{\Gamma({\frac{1}{8}})}\left(\prod_{m=0}^{\infty}\limits\,\frac{m+\frac{5}{8}}{\lambda_{2m+1}}\right)\cdot\left(\prod_{m=0}^{\infty}\limits\,\frac{m+\frac{1}{8}}{\lambda_{2m}}\right)^{-1}.
\end{equation}
Table \ref{tab:dm/dp} presents a comparison of the numerical results obtained from \eqref{dm/dp-prod} for different values of $\alpha$ with the analytic expression \eqref{dm/dp}.

The formulas in \eqref{F_definition} hold asymptotically only in the regime of large negative $\lambda$, and thus are not applicable in the vicinity of the positive spectral points $\lambda_n>0$, associated with the meson masses. However, in the limit of large positive $\lambda$, the spectral determinants $\mathfrak{D}_\pm(\lambda)$ can alternatively be expressed in the following form:
\begin{equation}
    \mathfrak{D}_{\pm}(\lambda)=\frac{1}{2}
    \left(\mathcal{D}_{\pm}(-e^{-i\pi}\lambda)+\mathcal{D}_{\pm}(-e^{+i\pi}\lambda)\right).
\end{equation}
Here, the two contributions correspond to analytic continuations of \eqref{F_definition} through the upper and lower half-planes, respectively. This continuation procedure was first introduced for the ’t Hooft model in \cite{Fateev:2009jf}, motivated by the condition that the associated $Q$-function vanishes as $\nu \to \infty$. Its validity was later confirmed, and the approach was successfully employed in subsequent works \cite{Litvinov:2024riz,Artemev:2025cev,Litvinov:2025geb}. In scalar QCD$_2$, this leads to
\begin{equation}\label{D-analytical-continuation}
    \mathfrak{D}_{\pm}(\lambda)=2d_{\pm}
    \left(2\pi e^{-2+\gamma_E}\right)^{\lambda}
    \lambda^{\lambda+\frac{1}{8}\pm\frac{1}{4}}\exp{\left(\sum_{k=0}^{\infty}\limits\Xi^{(k)}_{\pm}(l)\lambda^{-k}\right)}\cos\left(
    \frac{\pi}{2}\left[2\lambda+\frac{1}{4}\pm\frac{1}{2}+\sum_{k=0}^{\infty}\Phi^{(k)}_{\pm}(l)\lambda^{-k}\right]\right),
\end{equation}
where $\Xi^{(k)}_{\pm}(l)$ and $\Phi^{(k)}_{\pm}(l)$ are polynomials in
\begin{equation}
    l=\log{(2\pi\lambda)}+\gamma_E,
\end{equation}
and represent symmetrized and antisymmetrized combinations of $F^{(k)}_{\pm}(L)$, respectively,
\begin{equation}
    \Xi^{(k)}_{\pm}(l)\overset{\text{def}}{=}\frac{1}{2}\left(F^{(k)}_{\pm}(l+i\pi)+F^{(k)}_{\pm}(l-i\pi)\right),\quad \Phi^{(k)}_{\pm}(l)\overset{\text{def}}{=}\frac{i}{\pi}\left(F^{(k)}_{\pm}(l-i\pi)-F^{(k)}_{\pm}(l+i\pi)\right).
\end{equation}

The zeros of $\mathfrak{D}_{\pm}(\lambda)$ are determined by the cosine factor $\cos(\dots)$, which effectively enforces the ``quantization conditions” on $\lambda$
\begin{equation}\label{quantisation-condition-lambda}
    2\lambda+\frac{1}{4}\pm\frac{1}{2}+\Phi^{(0)}_{\pm}(l)+\Phi^{(1)}_{\pm}(l)\lambda^{-1}+\ldots=2m+1,\quad m=0,1,2,\dots
\end{equation}
The initial few phases $\Phi_\pm^{(k)}(l)$ can be expressed explicitly as follows:
\begin{equation}\label{Phi-pm-phases}
    \resizebox{\textwidth}{!}{$
    \begin{aligned}
        \Phi^{(0)}_{\pm}(l)=&\;\frac{\alpha^2}{2\pi^2}\mathtt{i}_2(\alpha)-\frac{2\alpha l}{\pi^2},\quad \Phi^{(1)}_{\pm}(l)=\frac{\alpha^2}{\pi^4}\pm\frac{1}{4\pi^2},
        \\
        \Phi^{(2)}_+(l)=&\;\frac{4\alpha^3+\pi^2(7\alpha+4-6\log2)}{8\pi^6}+\frac{3\alpha}{32\pi^4}\mathtt{i}_1(\alpha)-\frac{3l}{8\pi^4},
        \\
        \Phi^{(2)}_-(l)=&\;\frac{4\alpha^3-\pi^2(5\alpha+3-4\log2)}{8\pi^6}-\frac{\alpha}{16\pi^4}\mathtt{i}_1(\alpha)+\frac{l}{4\pi^4},
        \\
        \Phi^{(3)}_+(l)=&\;\frac{80\alpha^4+2\pi^2(4\alpha(53\alpha+61-108\log2)+53)-37\pi^4}{192\pi^8}-\frac{(23-13\log2)\log2}{8\pi^6}+
        \\&+\frac{\alpha\mathtt{i}_1(\alpha)\left(8(36\alpha+23-26\log2)+13\alpha\mathtt{i}_1(\alpha)\right)}{512\pi^6}-\frac{\left(144\alpha+13\alpha\mathtt{i}_1(\alpha)+92-104\log2\right)l}{64\pi^6}+\frac{13l^2}{32\pi^6},
        \\
        \Phi^{(3)}_-(l)=&\;\frac{80\alpha^4-8\pi^2\alpha(37\alpha+44-72\log2)+25\pi^4-80\pi^2}{192\pi^8}+\frac{(2-\log2)\log2}{\pi^6}-
        \\&-\frac{\alpha\mathtt{i}_1(\alpha)\left(24\alpha+\alpha\mathtt{i}_1(\alpha)+16-8\log4\right)}{64\pi^6}+\frac{\left(12\alpha+\alpha\mathtt{i}_1(\alpha)+8-8\log2\right)l}{8\pi^6}-\frac{l^2}{4\pi^6}.
    \end{aligned}
    $}
\end{equation}
Additional values of $\Phi^{(k)}_\pm(l)$ for $k \leq 5$ are listed in the supplementary file \texttt{Phi.nb}.

Equation \eqref{quantisation-condition-lambda} is non-perturbative in the mass parameter $\alpha$, and its solutions provide a WKB approximation to the spectrum of the theory. At leading order, the spectrum of highly excited states is linear, as shown in \cite{Shei:1977ci} and further discussed in \cite{Tomaras:1978nt,Aoki:1995dh}. For sufficiently large $\lambda$, \eqref{quantisation-condition-lambda} can be inverted to obtain the asymptotic WKB expansion for $\lambda_n$. Since the phases $\Phi^{(k)}_\pm(l)$ begin to differ significantly beyond the second order, the result is presented as two separate series:
\begin{equation}\label{WKB-normal-alpha-even}
    \resizebox{\textwidth}{!}{$
    \begin{aligned}
        \lambda^{\text{even}}_{n}=&\;\frac{1}{2}\mathfrak{n}+\frac{\alpha}{\pi^2}\log\rho+\frac{1}{\mathfrak{n}}\left[\frac{\alpha^2}{\pi^4}(2\log\rho-1)-\frac{1}{4\pi^2}\right]-\frac{1}{\mathfrak{n}^2}\Biggl[\frac{2\alpha^3\log^2{\rho}}{\pi^6}-\frac{\left(24\alpha^3+\pi^2(3+2\alpha)\right)\log{\rho}}{4\pi^6}+\frac{3\alpha^3}{\pi^6}+
        \\&+\frac{4+9\alpha-6\log2}{4\pi^4}+\frac{3\alpha\mathtt{i}_1(\alpha)}{16\pi^4}\Biggl]+\frac{1}{\mathfrak{n}^3}\Biggl[\frac{8\alpha^4\log^3{\rho}}{3\pi^8}-\frac{\left(128\alpha^4+\pi^2(13+8\alpha(3+\alpha))\right)\log^2{\rho}}{8\pi^8}+
        \\&+\frac{\left(384\alpha^4+4\pi^2(40\alpha^2+8\alpha(8-\log8)+23-26\log2)+\pi^2(12\alpha+13)\alpha\mathtt{i}_1(\alpha)\right)\log{\rho}}{16\pi^8}+
        \\&-\frac{464\alpha^4+2\pi^2\left(344\alpha^2+4\alpha(86\alpha+73-126\log2)+39\log^24-276\log2\right)+106\pi^2-31\pi^4}{48\pi^8}-
        \\&-\frac{\alpha\mathtt{i}_1(\alpha)\left(336\alpha+8(23-26\log2)+13\alpha  \mathtt{i}_1(\alpha)\right)}{128\pi^6}\Biggl]+\mathcal{O}\left(\frac{\log^4\mathfrak{n}}{\mathfrak{n}^4}\right),\quad n=0,2,4\dots,
    \end{aligned}
    $}
\end{equation}
\begin{equation}\label{WKB-normal-alpha-odd}
    \resizebox{\textwidth}{!}{$
    \begin{aligned}
        \lambda^{\text{odd}}_{n}=&\;\frac{1}{2}\mathfrak{n}+\frac{\alpha}{\pi^2}\log\rho+\frac{1}{\mathfrak{n}}\left[\frac{\alpha^2}{\pi^4}(2\log\rho-1)+\frac{1}{4\pi^2}\right]-\frac{1}{\mathfrak{n}^2}\Biggl[\frac{2\alpha^3\log^2{\rho}}{\pi^6}-\frac{\left(12\alpha^3-\pi^2(1+\alpha)\right)\log{\rho}}{2\pi^6}+\frac{3\alpha^3}{\pi^6}-
        \\&-\frac{3+7\alpha-4\log2}{4\pi^4}-\frac{\alpha\mathtt{i}_1(\alpha)}{8\pi^4}\Biggl]+\frac{1}{\mathfrak{n}^3}\Biggl[\frac{8\alpha^4\log^3{\rho}}{3\pi^8}-\frac{\left(16\alpha^4-\pi^2(1+\alpha)^2\right)\log^2{\rho}}{\pi^8}+
        \\&+\frac{\left(48\alpha^4-2\pi^2(8\alpha^2+11\alpha+4-(1+\alpha)\log16)+\pi^2\alpha(1+\alpha)\mathtt{i}_1(\alpha)\right)\log{\rho}}{2\pi^8}+
        \\&-\frac{464\alpha^4-8\pi^2\left(64\alpha^2+\alpha(53-84\log2)-24(2-\log2)\log2\right)+80\pi^2+31\pi^4}{48\pi^8}+
        \\&+\frac{\alpha\mathtt{i}_1(\alpha)\left(28\alpha-16(\log2-1)+\alpha\mathtt{i}_1(\alpha)\right)}{16\pi^6}\Biggl]+\mathcal{O}\left(\frac{\log^4\mathfrak{n}}{\mathfrak{n}^4}\right), \quad n=1,3,5\dots,
    \end{aligned}
    $}
\end{equation}
where we have introduced
\begin{equation}\label{WKB-parameter}
    \mathfrak{n}=n+\frac{1}{4}-\frac{\alpha^2\mathtt{i}_2(\alpha)}{2\pi^2},\quad \rho=\pi e^{\gamma_E}\mathfrak{n}.
\end{equation}

Expressions \eqref{WKB-normal-alpha-even}-\eqref{WKB-normal-alpha-odd} represent a systematic generalization of the previously known leading-order term. The FLZ method employed here allows, in principle, the computation of an arbitrary number of terms in this asymptotic series, with the only limitation being available computational resources. For the first five phases $\Phi^{(k)}_\pm(l)$ that we computed, the large-$n$ WKB expansions were obtained including $\mathcal{O}\left(\frac{1}{\mathfrak{n}^5}\right)$ terms (see Mathematica notebook \texttt{WKB.nb} attached to this submission). Table \ref{tab:WKB} presents a comparison between the analytical WKB expressions and the corresponding numerical results.

We note that the expansion parameter $\mathfrak{n}$ \eqref{WKB-parameter} is smaller than its analog in the 't Hooft model \cite{Litvinov:2024riz} for all values of $\alpha$. This can be verified by analyzing the explicit asymptotics of the integrals $\mathtt{i}_2(\alpha)$ (whose definitions differ in the two theories) in both the small- and large-$\alpha$ limits. The result confirms our expectation that mesons in scalar QCD$_2$ are lighter than in the 't Hooft model: a bound state of two quarks of mass $m$ obeying Fermi statistics is heavier than the corresponding bound state of two quarks of mass $m$ with Bose statistics \cite{Aoki:1995dh}.

Clearly, large-$n$ WKB expansions \eqref{WKB-normal-alpha-even}-\eqref{WKB-normal-alpha-odd} are not applicable to all possible values of the parameters $n$ and $\alpha$. At the very least, the expansion requires $\mathfrak{n}>0$. A closer examination of the WKB formulas shows that, for large $\alpha$, the actual expansion parameter is $\alpha/(\pi^2\mathfrak{n})$. This parameter may cease to be small for low values of $n$, limiting the reliability of the approximation in this regime. The issue of describing the low-lying spectrum at large-$\alpha$ will be addressed in the next section.
\subsection{Numerics}\label{Numerics}
\paragraph{Method I: discretization approach.}
We now introduce a straightforward numerical discretization method, previously applied to the 't Hooft model in \cite{Fateev:2009jf,Ambrosino:2023dik}. In the doubled Ising model coupled via the spin-spin interaction term, the same approach was used in \cite{Gao:2025mcg} for equation \eqref{BS-eq-2} with $\beta=0$, describing interchain mesons.

For the eigenvalue problem \eqref{BS-eq}, we truncate the integration domain to a finite interval $[-L,L]$ with $L$ and discretize it on a lattice of $N+1$ points, which reduces the problem to the matrix form
\begin{equation}
    \bm{A}\Psi=\lambda \bm{B}\Psi,
\end{equation}
with matrices $\bm{A}$ and $\bm{B}$ defined as
\begin{equation}\label{A-B-lattice}
    A_{ij}=\left(\frac{2\alpha}{\pi}+\nu_i\tanh{\frac{\pi\nu_i}{2}}\right)\delta_{ij}-\frac{1}{8}\frac{1}{\cosh{\frac{\pi\nu_i}{2}}}\frac{1}{\cosh{\frac{\pi\nu_j}{2}}}d\nu,\quad B_{ij}= \frac{\pi(\nu_i-\nu_j)}{2\sinh{\frac{\pi(\nu_i-\nu_j)}{2}}}d\nu,\quad i,j\in{1,\ldots,N+1},
\end{equation}
where $\nu_i=(-\frac{N}{2}+i-1)d\nu$ and $d\nu=\frac{2L}{N}$. The eigenvalues and eigenfunctions are then obtained by diagonalizing the $N+1\times N+1$ matrix $\bm{A}^{-1}\bm{B}$ (from a computational perspective, it is more efficient to invert matrix $\bm{A}$ to obtain $\lambda^{-1}$ than to invert matrix $\bm{B}$ directly to compute $\lambda$). 

It should be noted that the choice of $L$ must be consistent with the energy level under consideration. For higher levels (or larger values of $\alpha$), the eigenfunctions, initially concentrated near the origin, extend over a broader region. Consequently, increasing $L$ requires a corresponding increase in $N$ to maintain adequate sampling density in the domain where the eigenfunctions have significant support. For $L=50$ and $N=2000$, this procedure yields reasonably accurate results for the first $20$ levels (except for the odd states at $\alpha=0$, see below), with the accuracy improving for lower-lying states. For instance, at $\alpha=1$ the $30$th level has a relative error of about $0.3\%$, corresponding to a discrepancy of roughly half a digit in the second decimal place. The accuracy of this method could, in principle, be improved by extrapolating the number of sites $N$ to infinity, but we will not pursue this here.

\begin{figure}[h!]
    \centering
    \includegraphics[width=0.49\linewidth]{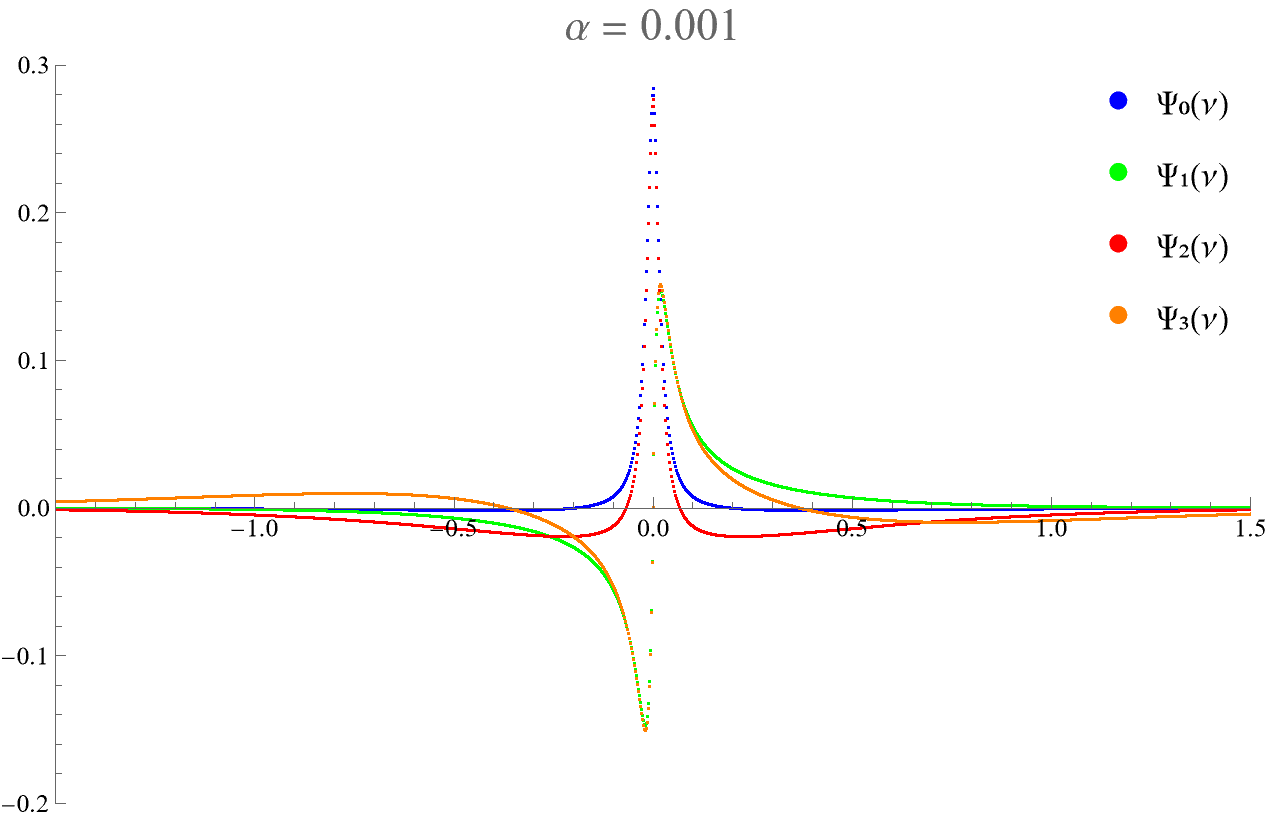}
    \includegraphics[width=0.49\linewidth]{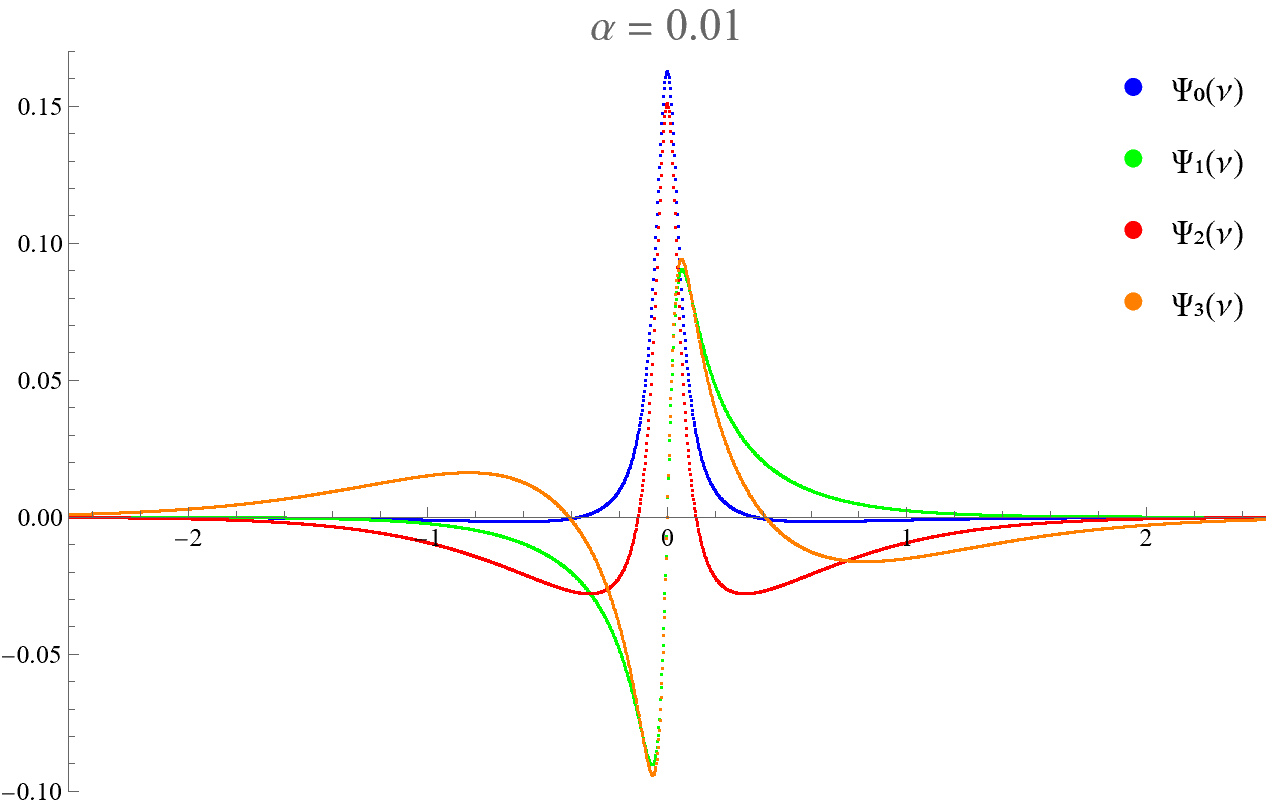}
    \includegraphics[width=0.49\linewidth]{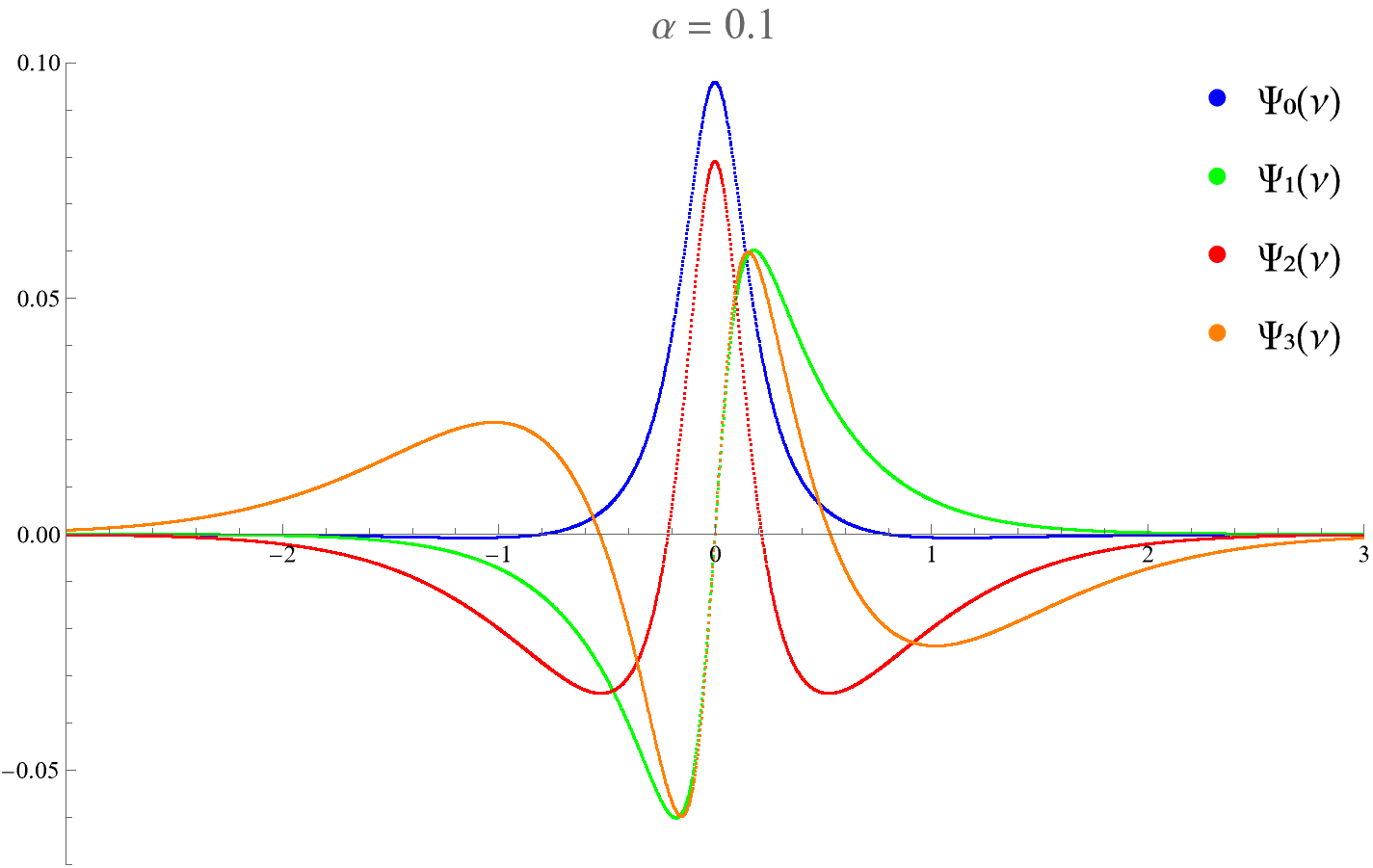}
    \includegraphics[width=0.49\linewidth]{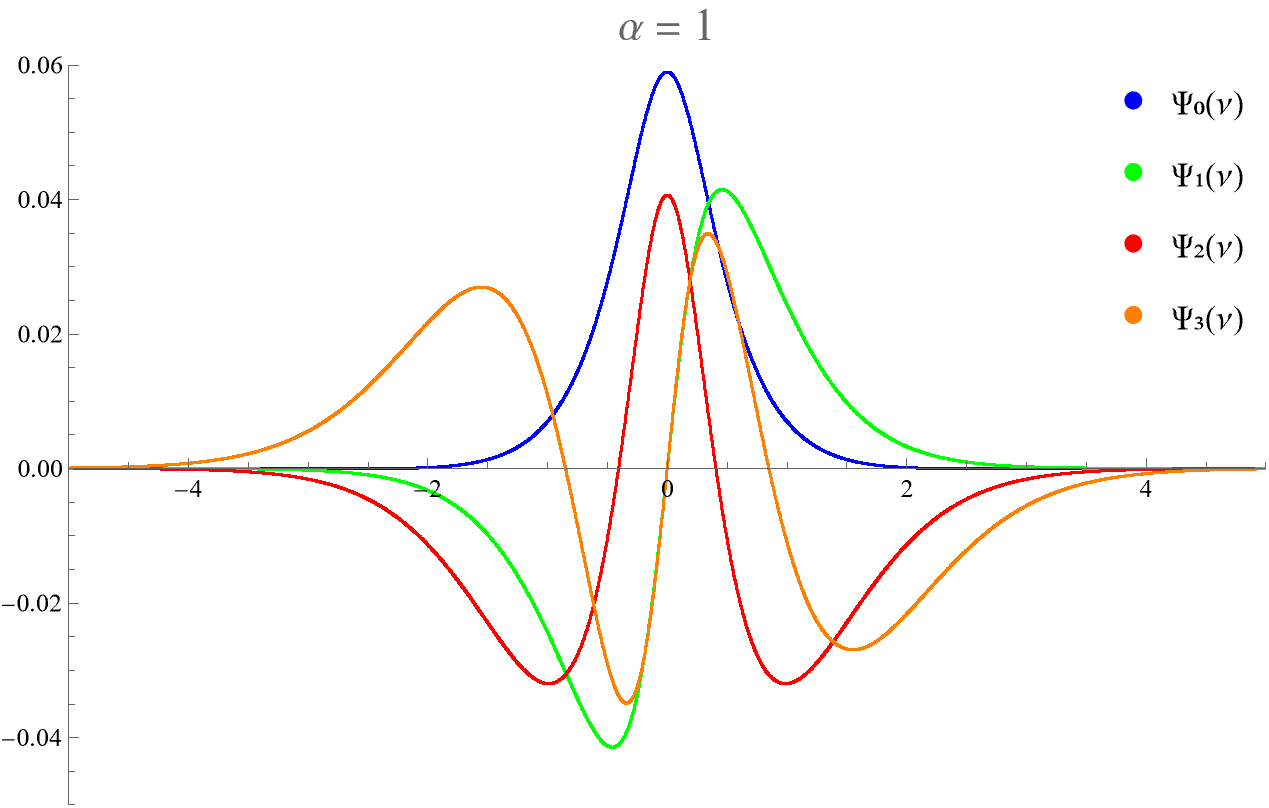}
    \caption{Numerical meson wave functions $\Psi_n$ in $\nu$-space for several values of $n$ and $\alpha$. Results are obtained using Method I with $N=4000$ and $L=5$.}
    \label{fig:wave-func}
\end{figure}

We can use this numerical method not only to determine the spectrum of the theory, but also to extract the meson wave functions $\Psi_n(\nu)$. The lowest eigenfunctions are presented in Fig. \ref{fig:wave-func}. As the quark mass approaches its critical value $m \to g/\sqrt{\pi}$ ($\alpha \to 0$), the wave function of the lowest meson (``pion”) becomes increasingly narrow. However, in contrast to the 't Hooft model \cite{THOOFT1974461}, the pion in scalar QCD$_2$ is not point-like and does not collapse to a delta-function $\delta(\nu)$\footnote{In the 't Hooft model, in the chiral limit $m=0$ ($\alpha=-1$), the pion wave function in $x$-space is $\phi_0(x)=1$ \cite{THOOFT1974461}, which corresponds to $\delta(\nu)$ in $\nu$-space.}. This distinction follows directly from \eqref{BS-eq-2}, since the $\delta$-function is not a valid solution of the equation at $\alpha=\beta=0$. Nevertheless, the sharp profile of $\Psi_0$ makes it possible to extract from \eqref{BS-eq-2} a rough estimate of the ground state energy, $\lambda_0 \approx \frac{1}{8}$ (more accurate value is given in Table \ref{tab:WKB}).

The main drawback of this method is that it produces only a limited number of eigenvalues within a reasonable computation time. Moreover, in the limiting case $\alpha=0$, the discretization approach becomes unreliable in the odd sector of the spectrum. This issue likely originates from the fact that, for $i=\{\frac{N}{2},\frac{N}{2}+1,\frac{N}{2}+2\}$ (i. e. near $\nu=0$), the corresponding diagonal elements of the matrix $\bm{A}$ \eqref{A-B-lattice} become negative, while all other diagonal elements remain positive. For $\alpha>\frac{\pi}{16}d\nu$, no diagonal elements take negative values. A similar difficulty appears in the original 't Hooft model when applying the same discretization-based method in the chiral limit $\alpha=-1$: the analog of matrix $\bm{A}$ becomes degenerate ($\bm{A}$ is diagonal with a zero entry), leading to a rather poor description of the masses in the even sector of the spectrum. 
\paragraph{Method II: Chebyshev polynomials.}
In \cite{Hanson:1976ey}, the Multhopp method was introduced for the ’t Hooft model. It is based on expanding the wavefunction in a Chebyshev polynomial basis, followed by discretization and solving the resulting eigenvalue problem. The method was first applied to scalar QCD$_2$ in \cite{Aoki:1995dh}. In this work, we modify the original approach: instead of discretizing after the Chebyshev expansion, we compute the matrix elements explicitly and then solve the eigenvalue problem.    

We expand the wavefunction $\Phi(x)$ in the basis introduced in \cite{Aoki:1995dh} (here $\Phi(x)$ denotes the original, non-renormalized wavefunction):
\begin{equation}\label{Chebyshev_basis}
    \Phi(x)=\sum_{n}\Phi_n h_n(x) ,\quad h_n(x)=\sin{(n\theta)}=2\sqrt{x(1-x)}U_{n-1}(1-2x), \quad n\geq1,
\end{equation}
where the new variable $\theta \in [0,\pi]$ is related to $x$ by $x=\tfrac{1-\cos\theta}{2}$, and $U_n(x)$ are Chebyshev polynomials of the second kind. This representation reformulates the bound-state equation \eqref{tHooft-eq-bosons} as a matrix eigenvalue problem, which—once truncated to a sufficiently large $N$—can be solved numerically:
\begin{equation}
    2\pi^2\lambda\sum_n M_{mn}\Phi_n =\sum_n H_{mn}\Phi_n, \quad M_{mn}=\braket{h_m|h_n},\quad H_{mn}=\bra{h_m}\mathcal{H}\ket{h_n}.
\end{equation}
The scalar product in this basis takes the standard form
\begin{equation}\label{scalar-product-Chebyshev}
    \braket{h|g}=\int_{0}^{1}\limits dx\; h(x)g(x).
\end{equation}
In this representation, the matrices $\bm{M}$ and $\bm{H}$ can be written in the integral form
\begin{equation}
    M_{mn}=\int_0^1\limits dx\;\sin{(m\theta)}\sin{(n\theta)}=-(1+(-1)^{m+n})\frac{mn}{(m^2-n^2)^2-2(m^2+n^2)+1}   
\end{equation}
and
\begin{equation}\label{H-matrix-Chebyshev}
    H_{mn}=\underbrace{\int_0^1\limits dx\;\sin{(m\theta)}\sin{(n\theta)}\;\alpha\left(\frac{1}{x}+\frac{1}{1-x}\right)}_{H^{(1)}_{mn}}\underbrace{-\int_0^{1}\limits dx\sin{(m\theta)}\fint_0^1\limits dy\frac{\sin{(n\theta')}}{(x-y)^2}\frac{(x+y)(2-x-y)}{4x(1-x)}}_{H^{(2)}_{mn},\quad y=\frac{1-\cos{\theta'}}{2}}.
\end{equation}
Here, the first contribution $H^{(1)}_{mn}$ in \eqref{H-matrix-Chebyshev} has the explicit form (as in 't Hooft model)
\begin{equation}
    H^{(1)}_{mn}=2\alpha\int_0^{\pi}\limits d\theta\;\frac{\sin{(m\theta)}\sin{(n\theta)}}{\sin{\theta}}=
    (1+(-1)^{m+n})\cdot\alpha\sum_{l=\frac{|m-n|}{2}+1}^{\frac{m+n}{2}}\limits\frac{2}{2l-1},
\end{equation}
while the second contribution $H^{(2)}_{mn}$ is given by
\begin{equation}\label{H2-Chebyshev}
\resizebox{\textwidth}{!}{$
\begin{aligned}
    H^{(2)}_{mn}&=-\int^1_0\limits dx \sin{m\theta}\fint_0^1\limits dy\frac{\sin{n\theta'}}{(x-y)^2}\frac{(x+y)(2-x-y)}{4x(1-x)}=\int^1_0\limits dx \sin{m\theta}\cdot2\pi\left[\frac{n\sin{n\theta}}{\sin{\theta}}+\frac{-\cos{\theta}\cos{n\theta}+\delta_{n,1}/8}{\sin^2\theta}\right]=
    \\&=\frac{n\pi^2}{2}\delta_{n,m}-\frac{\pi^2}{2}\left[\delta_{n,m}+(1+(-1)^{m+n})\Theta(m-n-1)\right]+\frac{\pi^2}{8}\delta_{n,1}\frac{1-(-1)^m}{2},
\end{aligned}
$}
\end{equation}
where $\Theta$ is the Heaviside step function.

It is straightforward to see that the matrix $\bm{H}$ is real but not symmetric in the indices $n,m$, i.e. it is non-Hermitian with respect to the chosen scalar product \eqref{scalar-product-Chebyshev}. Consequently, its eigenvalues may, in principle, be complex. In practice, however, the imaginary parts turn out to be negligibly small, and we take the real parts as the relevant numerical values for the spectrum.

The method converges rapidly as $N$, the number of basis functions $h_n$, increases. In practice, it yields a significantly larger portion of the spectrum: for a basis of $N=1000$ functions, the first $\sim600$ eigenvalues can be considered reliable. The main limitation concerns the range of $\alpha$: convergence deteriorates for $\alpha<0$ (which is natural in the 't Hooft model but not in scalar QCD$_2$), since the basis functions do not capture the correct boundary behavior. This issue could, in principle, be addressed by extending the basis with functions that satisfy \eqref{phi-boundary-asympt}, but we do not pursue this modification here. Convergence also worsens for very large $\alpha$. Compared to Method I, this approach yields lower accuracy for the lowest states (except in the odd sector at $\alpha=0$). We attribute this reduction, in contrast to the 't Hooft model where the method is more precise, to the non-Hermitian nature of the matrix $\bm{H}$. Nevertheless, it allows us to compute a significantly larger number of eigenvalues overall.
\paragraph{Numerical Results vs. Analytical Predictions.}
Using a combination of numerical methods, we compute the spectral sums $\mathcal{G}^{(s)}_\pm$ as defined in \eqref{spectral-sums-def} for various values of the parameter $\alpha$. For the first twenty eigenvalues (ten for the even and ten for the odd spectral sums), we use results from Method I, except for the odd sums at $\alpha=0$. The remaining eigenvalues are taken from Method II; for $N=1000$, we trust the first $622$ eigenvalues from this method. Accounting for the first $20$ eigenvalues from Method I, this provides $301$ eigenvalues each for the even and odd sums. In the case $\alpha=0$, the odd spectral sums are computed using $311$ eigenvalues from Method II. All results are summarized in Table \ref{tab:G}. Using the same selection of eigenvalues, we also compare the analytic predictions \eqref{dm/dp} with the numerical evaluation \eqref{dm/dp-prod} for the ratio of spectral determinant constants $d_-/d_+$. The corresponding data are presented in Table \ref{tab:dm/dp}.
\begin{table}[h!]
    \begin{center}
        \resizebox{\textwidth}{!}{
        \begin{tabular}{| c | c | c | c || c | c | c || c | c | c || c | c | c |}
        \hline \rule{0mm}{3.6mm}
        $\alpha$ & \multicolumn{6}{c||}{$0$} & \multicolumn{6}{c|}{$0.5$}  
        \\
        \hline
        $s$ & $\prescript{\rm{an}}{}{}\mathcal{G}_{+}^{(s)}$ & $\delta \mathcal{G}^{(s)}_{+}$ & $\prescript{\rm{num}}{}{}\mathcal{G}_{+}^{(s)}$ & $\prescript{\rm{an}}{}{}\mathcal{G}_{-}^{(s)}$ & $\delta \mathcal{G}^{(s)}_{-}$ & $\prescript{\rm{num}}{}{}\mathcal{G}_{-}^{(s)}$ & $\prescript{\rm{an}}{}{}\mathcal{G}_{+}^{(s)}$ & $\delta \mathcal{G}^{(s)}_{+}$ & $\prescript{\rm{num}}{}{}\mathcal{G}_{+}^{(s)}$ & $\prescript{\rm{an}}{}{}\mathcal{G}_{-}^{(s)}$ & $\delta \mathcal{G}^{(s)}_{-}$ & $\prescript{\rm{num}}{}{}\mathcal{G}_{-}^{(s)}$ \\
        \hline
        $1$ & 8.06529 & $3.5\cdot10^{-4}$ & 8.06248 & 0.837877 & $1.4\cdot10^{-3}$ & 0.836679 & 2.09075 & $8.6\cdot10^{-4}$ & 2.08895 & 0.022743 & $9.1\cdot10^{-3}$ & 0.022536 \\
        $2$ & 69.4004 & $4.6\cdot10^{-5}$ & 69.3972 & 3.28987 & $9.7\cdot10^{-4}$ & 3.28666 & 8.26476 & $3.9\cdot10^{-4}$ & 8.26157 & 1.72435 & $1.9\cdot10^{-3}$ & 1.72116 \\
        $3$ & 561.638 & $9.2\cdot10^{-9}$ & 561.638 & 4.17562 &  $1.8\cdot10^{-6}$ & 4.17563 & 20.1985 & $1.8\cdot10^{-7}$ & 20.1985 & 1.32753 &  $2.8\cdot10^{-6}$ & 1.32753 \\
        $4$ & 4624.75 & $2.1\cdot10^{-12}$ & 4624.75 & 6.18660 &  $4.0\cdot10^{-6}$ & 6.18662 & 53.6805 & $2.6\cdot10^{-9}$ & 53.6805 & 1.25331 & $8.8\cdot10^{-8}$ & 1.25331 \\
        $5$ & 38131.5 & $5.3\cdot10^{-15}$ & 38131.5 & 9.51506 & $5.0\cdot10^{-6}$ & 9.51510 & 144.598 & $6.1\cdot10^{-11}$ & 144.598 & 1.25478 & $3.2\cdot10^{-8}$ & 1.25478 \\
        \hline
        $\alpha$ & \multicolumn{6}{c||}{$1$} & \multicolumn{6}{c|}{$1.5$}  
        \\
        \hline
        $s$ & $\prescript{\rm{an}}{}{}\mathcal{G}_{+}^{(s)}$ & $\delta \mathcal{G}^{(s)}_{+}$ & $\prescript{\rm{num}}{}{}\mathcal{G}_{+}^{(s)}$ & $\prescript{\rm{an}}{}{}\mathcal{G}_{-}^{(s)}$ & $\delta \mathcal{G}^{(s)}_{-}$ & $\prescript{\rm{num}}{}{}\mathcal{G}_{-}^{(s)}$ & $\prescript{\rm{an}}{}{}\mathcal{G}_{+}^{(s)}$ & $\delta\mathcal{G}^{(s)}_{+}$ & $\prescript{\rm{num}}{}{}\mathcal{G}_{+}^{(s)}$ & $\prescript{\rm{an}}{}{}\mathcal{G}_{-}^{(s)}$ & $\delta\mathcal{G}^{(s)}_{-}$ & $\prescript{\rm{num}}{}{}\mathcal{G}_{-}^{(s)}$ \\
        \hline
        $1$ & 1.09147 & $1.5\cdot10^{-3}$ & 1.08982 & $-0.303016$ & $1.5\cdot10^{-4}$ & $-0.303060$ & 0.531069 & $2.7\cdot10^{-3}$ & 0.529617 & $-0.536049$ & $2.8\cdot10^{-4}$ & $-0.535900$ \\
        $2$ & 4.45438 &  $7.2\cdot10^{-4}$ & 4.45117 & 1.31254 & $2.4\cdot10^{-3}$ & 1.30934 & 2.97647 & $1.1\cdot10^{-3}$ & 2.97326 & 1.07362 & $3.0\cdot10^{-3}$ & 1.07042 \\
        $3$ & 7.31282 & $6.7\cdot10^{-7}$ & 7.31281 & 0.809380 &  $6.1\cdot10^{-6}$ & 0.809375 & 3.69852 & $1.4\cdot10^{-6}$ & 3.69851 & 0.559866 & $9.0\cdot10^{-6}$ & 0.559861 \\
        $4$ & 13.5551 & $8.2\cdot10^{-10}$ & 13.5551 & 0.624322 & $1.1\cdot10^{-8}$ & 0.624322 & 5.34349 & $8.3\cdot10^{-10}$ & 5.34349 & 0.370347 & $2.0\cdot10^{-8}$ & 0.370347\\
        $5$ & 25.7222 & $7.1\cdot10^{-11}$ & 25.7222 & 0.515647 & $4.4\cdot10^{-9}$ & 0.515647 & 7.97593 & $9.6\cdot10^{-11}$ & 7.97593 & 0.264213 & $4.2\cdot10^{-9}$ & 0.264213\\
        \hline
        \end{tabular}}
    \end{center}
    \caption{Numerical and analytical values of the first $5$ spectral sums $\mathcal{G}_\pm^{(s)}$ \eqref{spectral-sums-def} for several values of $\alpha$. The ratio $\delta\mathcal{G}^{(s)}_{\pm}= \left|\frac{\prescript{\rm{num}}{}{}\mathcal{G}_{\pm}^{(s)}-\prescript{\rm{an}}{}{}\mathcal{G}_{\pm}^{(s)}}{\prescript{\rm{an}}{}{}\mathcal{G}_{\pm}^{(s)}}\right|$ shows the relative error, where the unrounded values of $\prescript{\rm{an}}{}{}\mathcal{G}_n^{(s)}$ and $\prescript{\rm{num}}{}{}\mathcal{G}_n^{(s)}$ are used in the estimation.}
    \label{tab:G}
\end{table} 
\begin{table}[h!]
    \begin{center}
        \begin{tabular}{| r | c | c | c |}
        \hline \rule{0mm}{3.6mm}
        $\alpha$ & $(d_-/d_+)^{\rm{num}}$ & $(d_-/d_+)^{\rm{an}}$ & $\delta d$\\
        \hline
        $0$ & 0.176785 & 0.176777 & $4.9\cdot10^{-5}$ \\
        $0.5$ & 0.419154 & 0.419579 & $1.0\cdot10^{-3}$ \\
        $1.0$ & 0.533055 & 0.533955 & $1.7\cdot10^{-3}$\\
        $1.5$ & 0.623393 & 0.624873 & $2.4\cdot10^{-3}$\\
        \hline
        \end{tabular}
    \end{center}
    \caption{Numerical and analytical values for ratio $d_-/d_+$ for several values of $\alpha$. The ratio $\delta d=\left|\frac{(d_-/d_+)^{\rm{num}}-(d_-/d_+)^{\rm{an}}}{(d_-/d_+)^{\rm{an}}}\right|$ shows the relative error.
    \label{tab:dm/dp}}
\end{table}

It is important to note that the numerical evaluation of the spectral sums $\mathcal{G}^{(s)}_\pm$ involves several sources of error. The most obvious one arises from the truncation of the series, since in practice we can only sum over a finite number of terms. As no numerical method can provide the full infinite set of eigenvalues, one can, in principle, refine the accuracy by supplementing the calculation with approximate large-$n$ WKB values, which reproduce the true eigenvalues with excellent precision at sufficiently large $n$. The truncation error can be readily estimated as $K^{-s+1}$ for $s>1$ and as $K^{-1}$ for $s=1$, where $K$ denotes the cutoff index (here $K=312$). A second source of error stems from the numerical methods used to obtain the eigenvalues. Each eigenvalue employed in the computation carries its own numerical uncertainty, which is more difficult to quantify.

Comparison of the analytic WKB formulas \eqref{WKB-normal-alpha-even}-\eqref{WKB-normal-alpha-odd} with numerical results (see Table \ref{tab:WKB}) demonstrates high accuracy already for the first excited state at moderately small $\alpha$, with precision further improving for higher excitations. In the regime $\mathfrak{n}\gg 1$, the formulas provide an excellent description of the $n$-th meson mass $\lambda_n$. At $\alpha=0$, however, Method I becomes unreliable in the odd sector, leading to noticeably larger relative errors compared to the even states. A small shift from zero, e.g. $\alpha=0.01$, drastically reduces the discrepancy, indicating that the issue stems from numerical limitations of Method I at $\alpha=0$. Table \ref{tab:WKB} summarizes the results: for the odd sector at $\alpha=0$ we include values from both numerical methods, while in all other cases we rely on Method I. Additional numerical data for other values of $\alpha$ can be found in \cite{Gao:2025mcg}, in the context of the spectrum of interchain mesons in doubled Ising models coupled via the spin-spin interaction term. Our notations are related to theirs by $\frac{\pi}{2\alpha}=\gamma$ and $\frac{\pi^2}{2\alpha}\lambda=\frac{M^2}{4m^2}$.
\begin{table}[h!]
\begin{center}
    \resizebox{\textwidth}{!}{
    \begin{tabular}{| c | l | c | c | c | l||l | c | l|| l | c | l|| l | c | l|| l | c | l|}
    \hline \rule{0mm}{3.6mm}
     $\alpha$ & \multicolumn{5}{c||}{$0$} & \multicolumn{3}{c||}{$0.01$} & \multicolumn{3}{c||}{$0.5$} & \multicolumn{3}{c||}{$1$} & \multicolumn{3}{c|}{$1.5$}\\
    \hline
    $n$ & $\lambda_{n}^{(5)}$ & $\delta\lambda_{\text{I}}$ & $\lambda_{n}^{(\text{num I})}$ & $\delta\lambda_{\text{II}}$ & $\lambda_{n}^{(\text{num II})}$ & $\lambda_{n}^{(5)}$ & $\delta\lambda_{\text{I}}$ & $\lambda_{n}^{(\text{num I})}$ & $\lambda_{n}^{(5)}$ & $\delta\lambda_{\text{I}}$ & $\lambda_{n}^{(\text{num I})}$ & $\lambda_{n}^{(5)}$ & $\delta\lambda_{\text{I}}$ & $\lambda_{n}^{(\text{num I})}$ & $\lambda_{n}^{(5)}$ & $\delta\lambda_{\text{I}}$ & $\lambda_{n}^{(\text{num I})}$\\
    \hline
    $0$ & *.***** & $-$ & 0.12127 & $-$ & 0.12127 & *.***** & $-$ & 0.14683 & 0.42215 & 0.14 & 0.36988 & 0.60649 & $0.16$ & 0.52267 & 0.78038 & $7.9\cdot 10^{-2}$ & 0.66101 \\
    $1$ & 0.63856 & $1.2\cdot 10^{-2}$ & 0.64658 & $7.5\cdot10^{-6}$ & 0.63857 & 0.67199 & $5.9\cdot10^{-4}$ & 0.67239 & 0.96116 & $2.7\cdot 10^{-4}$ & 0.96142 & 1.15286 & $7.8\cdot 10^{-4}$ & 1.15196 & 1.32270 & $1.6\cdot 10^{-3}$ & 1.32056 \\
    $2$ & 1.11733 & $1.3\cdot 10^{-4}$ & 1.11748 & $1.3\cdot10^{-4}$ & 1.11748 & 1.15159 & $8.9\cdot 10^{-5}$ & 1.15169 & 1.46230 & $2.1\cdot 10^{-5}$ & 1.46227 & 1.67090 & $1.6\cdot 10^{-4}$ & 1.67063 & 1.85546 & $1.2\cdot 10^{-3}$ & 1.85481 \\
    $3$ & 1.63144 & $4.3\cdot 10^{-3}$ & 1.63951 & $4.2\cdot10^{-6}$ & 1.63143 & 1.66616 & $1.5\cdot10^{-5}$ & 1.66619 & 1.99252 & $2.8\cdot 10^{-5}$ & 1.99246 & 2.21446 & $1.3\cdot 10^{-4}$ & 2.21416 & 2.41097 & $2.9\cdot 10^{-4}$ & 2.41027 \\
    $4$ & 2.12033 & $1.9\cdot 10^{-6}$ & 2.12034 & $8.9\cdot10^{-7}$ & 2.12034 & 2.15530 & $9.7\cdot 10^{-6}$ & 2.15528 & 2.49304 & $2.7\cdot 10^{-5}$ & 2.49297 & 2.72512 & $1.0\cdot 10^{-4}$ & 2.72484 & 2.93084 & $1.9\cdot 10^{-3}$ & 2.93019 \\
    $5$ & 2.62923 & $3.1\cdot 10^{-3}$ & 2.63733 & $1.4\cdot10^{-6}$ & 2.62922 & 2.66444 & $9.2\cdot10^{-6}$ & 2.66446 & 3.01159 & $2.2\cdot 10^{-5}$ & 3.01153 & 3.25215 & $8.6\cdot 10^{-5}$ & 3.25187 & 3.46568 & $1.9\cdot 10^{-4}$ & 3.46503 \\
    $6$ & 3.12162 & $1.3\cdot 10^{-7}$ & 3.12162 & $8.7\cdot10^{-7}$& 3.12162 & 3.15699 & $7.9\cdot 10^{-6}$ & 3.15697 & 3.51192 & $2.0\cdot 10^{-5}$ & 3.51185 & 3.75965 & $7.4\cdot 10^{-5}$ & 3.75937 & 3.97982 & $2.5\cdot 10^{-3}$ & 3.97918 \\
    $7$ & 3.62815 & $2.2\cdot 10^{-3}$ & 3.63628 & $1.1\cdot10^{-6}$ & 3.62815 & 3.66369 & $6.9\cdot10^{-6}$ & 3.66371 & 4.02538 & $1.7\cdot 10^{-5}$ & 4.02531 & 4.27936 & $6.5\cdot 10^{-5}$ & 4.27908 & 4.50538 & $1.4\cdot 10^{-4}$ & 4.50475 \\
    $8$ & 4.12235 & $1.9\cdot 10^{-8}$ & 4.12235 & $9.8\cdot10^{-7}$ & 4.12234 & 4.15801 & $6.2\cdot 10^{-6}$ & 4.15798 & 4.52561 & $1.5\cdot 10^{-5}$ & 4.52554 & 4.78514 & $5.8\cdot 10^{-5}$ & 4.78487 & 5.01638 & $3.2\cdot 10^{-3}$ & 5.01574 \\
    $9$ & 4.62751 & $1.8\cdot 10^{-3}$ & 4.63566 & $1.0\cdot10^{-6}$ & 4.62751 & 4.66330 & $5.5\cdot10^{-6}$ & 4.66332 & 5.03618 & $1.4\cdot 10^{-5}$ & 5.03611 & 5.30069 & $5.2\cdot 10^{-5}$ & 5.30041 & 5.53662 & $1.1\cdot 10^{-4}$ & 5.53599 \\
    \hline
    \end{tabular}}
\end{center}
\caption{Values of the first $10$ eigenvalues of $\lambda_{n}$ from the large-$n$ WKB expansion for several values of $\alpha$. The $\lambda_{n}^{(5)}$ column is obtained from \eqref{WKB-normal-alpha-even}-\eqref{WKB-normal-alpha-odd} with truncation of the series behind the term $\propto\mathfrak{n}^{-5}$. The column $\lambda_{n}^{(\text{num I/II})}$ lists the eigenvalues obtained from Method I/II. The relative error is defined as $\delta\lambda_{\text{I/II}}=\frac{|\lambda_{n}^{(5)}-\lambda_{n}^{(\text{num I/II})}|}{\lambda_{n}^{(\text{num I/II})}}$, where the unrounded values of $\lambda_n^{(5)}$ and $\lambda_n^{(\text{num I/II})}$ are used in the estimation. 
}
\label{tab:WKB}
\end{table}

In \cite{Demeterfi:1993rs}, the ground state at $\alpha=0$ was reported in adjoint scalar QCD$_2$ as $M_0^2=2.38g_{\text{adj}}^2/\pi$, where the spectrum is governed by \eqref{tHooft-eq-bosons} with the identification $g^2\to g_{\text{adj}}^2=2g^2$. Our result, $M_0^2\overset{\eqref{mass-notation}}{=}2.39373g^2/\pi$, differs by only $0.6\%$. Given the high accuracy of our numerical method for the ground state, we believe that this provides a more accurate determination of the ground state mass.
\section{Analysis of results in limiting cases}\label{Limiting-cases-section}
In this section, we illustrate how our exact formulas can be applied and simplified in several physically relevant limiting regimes. These limits enable us to establish connections between our results and previously known findings reported in the literature for the model under study.
\subsection{Critical mass limit: \texorpdfstring{$\alpha\to0$}{}}
To obtain the asymptotic behavior of the spectral sums $\mathcal{G}^{(s)}_\pm$\eqref{Gpm1}-\eqref{Gpm3} and large-$n$ WKB expansion \eqref{WKB-normal-alpha-even}-\eqref{WKB-normal-alpha-odd}, we make use of the known asymptotics of the integrals $\mathtt{i}_{k}(\alpha)$ \eqref{i-def} and $\mathtt{u}_{2k-1}(\alpha)$ \eqref{u-def} provided in Appendix B.1 of \cite{Litvinov:2025geb}, together with the asymptotics of $\mathtt{v}(\alpha)$ \eqref{Proector-def}
\begin{equation}
    \mathtt{v}(\alpha)\Big|_{\alpha\to0}=\frac{\pi}{8\sqrt{\alpha}}-\frac{\log2}{2}+\frac{\pi\sqrt{\alpha}}{16}-\frac{\zeta(3)}{2\pi^2}\alpha-\frac{\pi}{576}\alpha^{3/2}+\frac{3\zeta(5)}{2\pi^4}\alpha^2-\frac{11\pi}{5760}\alpha^{5/2}+\left(\frac{\zeta(5)}{3\pi^4}-\frac{5\zeta(7)}{\pi^6}\right)\alpha^3+\mathcal{O}(\alpha^{7/2}).
\end{equation}

As a result, we have a series expansion of the large-$n$ WKB expansion in half-integer powers of $\alpha$:
\begin{equation}\label{WKB-small-alpha-even}
\resizebox{\textwidth}{!}{$
\begin{aligned}
    \lambda^{\text{even}}_{n}(\alpha)\Big|_{\alpha\to0}=&\;\left[n+\frac{1}{8}-\frac{1}{8\pi^2(n+\frac{1}{8})}-\frac{4-3\log{(8\pi e^{\gamma_E}(n+\frac{1}{8})})}{16\pi^4(n+\frac{1}{8})^2}+\mathcal{O}\left(\frac{\log^2n}{n^3}\right)\right]+
    \\&+\left[1-\frac{1}{16\pi^2(n+\frac{1}{8})^2}-\frac{1-\log{(8\pi e^{\gamma_E}(n+\frac{1}{8})})}{32\pi^4(n+\frac{1}{8})^3}+\mathcal{O}\left(\frac{\log^2n}{n^4}\right)\right]\frac{\sqrt{\alpha}}{\pi}+
    \\&+\left[\log\left(2\pi e^{\gamma_E}\left(n+\frac{1}{8}\right)\right)-\frac{9+6\log{2}-2\log{(2\pi e^{\gamma_E}(n+\frac{1}{8}))}}{16\pi^2(n+\frac{1}{8})^2}+\mathcal{O}\left(\frac{\log^2n}{n^3}\right)\right]\frac{\alpha}{\pi^2}+
    \\&-\left[\frac{1}{6}-\frac{1}{\pi^2(n+\frac{1}{8})}+\frac{5}{96\pi^2(n+\frac{1}{8})^2}+\frac{57-38\log2-49\log{(2\pi e^{\gamma_E}(n+\frac{1}{8}))}}{192\pi^4(n+\frac{1}{8})^3}+\mathcal{O}\left(\frac{\log^2n}{n^4}\right)\right]\frac{\alpha^{3/2}}{\pi}+
    \\&+\mathcal{O}(\alpha^2),\quad n=0,1,2,\dots
\end{aligned}
$}
\end{equation}
\begin{equation}\label{WKB-small-alpha-odd}
\resizebox{\textwidth}{!}{$
\begin{aligned}
    \lambda^{\text{odd}}_{n}(\alpha)\Big|_{\alpha\to0}=&\;\left[n-\frac{3}{8}+\frac{1}{8\pi^2(n-\frac{3}{8})}+\frac{3-2\log{(8\pi e^{\gamma_E}(n-\frac{3}{8})})}{16\pi^4(n-\frac{3}{8})^2}+\mathcal{O}\left(\frac{\log^2n}{n^3}\right)\right]+\left[1+\mathcal{O}\left(\frac{\log^3n}{n^5}\right)\right]\frac{\sqrt{\alpha}}{\pi}+
    \\&+\left[\log\left(2\pi e^{\gamma_E}\left(n-\frac{3}{8}\right)\right)+\frac{7+4\log{2}-2\log{(2\pi e^{\gamma_E}(n-\frac{3}{8}))}}{16\pi^2(n-\frac{3}{8})^2}+\mathcal{O}\left(\frac{\log^2n}{n^3}\right)\right]\frac{\alpha}{\pi^2}-
    \\&-\left[\frac{1}{6}-\frac{1}{\pi^2(n-\frac{3}{8})}-\frac{1}{24\pi^2(n-\frac{3}{8})^2}-\frac{1-2\log{(8\pi e^{\gamma_E}(n-\frac{3}{8}))}}{24\pi^4(n-\frac{3}{8})^3}+\mathcal{O}\left(\frac{\log^2n}{n^4}\right)\right]\frac{\alpha^{3/2}}{\pi}+\mathcal{O}(\alpha^2),\quad n=1,2,3,\dots
\end{aligned}
$}
\end{equation}
This expansion shows that $\alpha=0$ corresponds to a square-root branch point (see also discussion in Section \ref{BS-complex-Section}).

The spectral sums admit the following expansions:
\begin{equation}
    \begin{aligned}
        \mathcal{G}^{(1)}_+\Big|_{\alpha\to0}=&\;9+\log\left(\frac{\pi}{8}\right)-\frac{1}{3\pi}\left[312-\pi^2-48(5-\log2)\log2\right]\sqrt{\alpha}+
        \\&-\left[\frac{28}{3}-\log(16)-\frac{832-32\log2(33-(7-\log2)\log4)-5\zeta(3)}{\pi^2}\right]\alpha+\mathcal{O}(\alpha^{3/2}),
        \\
        \mathcal{G}^{(1)}_-\Big|_{\alpha\to0}=&\;\log(2\pi)-1-\frac{\pi}{3}\sqrt{\alpha}-\frac{3\zeta(3)}{\pi^2}\alpha+\mathcal{O}(\alpha^{3/2}),
    \end{aligned}
\end{equation}
\begin{equation}
    \begin{aligned}
        \mathcal{G}^{(2)}_+\Big|_{\alpha\to0}=&\;126-\frac{\pi^2}{3}-8\log2(11-\log4)-
        \\&-\frac{4}{3\pi}\left[1737-10\pi^2-4(489-\pi^2)\log2+744\log^22-96\log^32+3\zeta(3)\right]\sqrt{\alpha}+\mathcal{O}(\alpha),
        \\
        \mathcal{G}^{(2)}_-\Big|_{\alpha\to0}=&\;\frac{\pi^2}{3}-\frac{4}{3\pi}\left[\pi^2-3\zeta(3)\right]\sqrt{\alpha}+\left[\frac{8}{3}-\frac{7\pi^2}{45}-\frac{8\zeta(3)}{\pi^2}\right]\alpha+\mathcal{O}(\alpha^{3/2}),
    \end{aligned}
\end{equation}
\begin{equation}
    \begin{aligned}
        \mathcal{G}^{(3)}_+\Big|_{\alpha\to0}=&\;\frac{4214}{3}-1480\log2+528\log^22-64\log^32-\frac{8}{9}\pi^2(8-\log8)+2\zeta(3)+
        \\&+\Biggl[-\frac{23\pi^3}{45}+\frac{4\pi}{27}\left(2005+288\log^22-1500\log2\right)-\frac{4\left(87578+183\zeta(3)\right)}{9\pi}+
        \\&+\frac{4\log2}{9\pi}\left(12(10777+6\zeta(3))-72288\log2+18144\log^22-1728\log^32\right)\Biggl]\sqrt{\alpha}+\mathcal{O}(\alpha),
        \\
        \mathcal{G}^{(3)}_-\Big|_{\alpha\to0}=&\;\frac{2}{3}\left(\pi^2-3\zeta(3)\right)-\left[4\pi-\frac{\pi^3}{15}-\frac{12\zeta(3)}{\pi}\right]\sqrt{\alpha}+\left[4+\frac{2\pi^2}{15}-2\zeta(3)\left(1+\frac{6}{\pi^2}\right)+\frac{6\zeta(5)}{\pi^2}\right]\alpha+
        \\&+\mathcal{O}(\alpha^{3/2}).
    \end{aligned}
\end{equation}
First, these expansions satisfy a basic consistency check: by definition, the even spectral sums must exceed the odd ones (since $\lambda_{2n}<\lambda_{2n+1}$). Numerical values of $\mathcal{G}^{(s)}_\pm$ for $\alpha=0$ presented in Table \ref{tab:G}. Second, the expansions remain finite as $\alpha \to 0$, indicating that no massless states emerge in the spectrum. A proof of this statement was previously given in \cite{Aoki:1995dh}, while our formulas provide an independent confirmation. This situation stands in contrast to the 't Hooft model \cite{THOOFT1974461}, where in the chiral limit ($\alpha\to-1$) a massless pion appears. In that case, the spectral sums computed in \cite{Litvinov:2024riz,Artemev:2025cev} allow one to extract subleading contributions to the pion mass, whereas earlier only the leading term was known \cite{Brower:1979PhysRevD}, subsequent work \cite{Zhitnitsky:1985um} related it to the well-known relation between the pion mass and the chiral condensate.

It is worth noting, as already mentioned in the Introduction \ref{Introduction}, that the Bethe-Salpeter equation \eqref{BS-eq}, while providing an exact description of scalar QCD$_2$, can also be interpreted as a two-particle approximation to a system of two massive Ising models weakly coupled via the spin-spin interaction term \eqref{2IFT_action}. In the limit $\alpha\to0$, this coupled Ising system becomes integrable \cite{LeClair:1997gv} (we will refer to it as IIFT). The latter can be reformulated as a sine-Gordon–type model with a $\mathbb{Z}_2$ orbifold structure \cite{Zuber:1976aa,Boyanovsky:1989PhysRevB}, whose exact spectrum consists of two solitons  $(A_+,A_-)$, and six breathers $(B_1,\ldots,B_6)$ with mass relations \cite{Fateev:1993av}
\begin{equation}
    M_{B_1}=2\sin{\frac{\pi}{14}}M_{A_\pm}, \quad M_{B_n}=M_{B_1}\frac{\sin{\frac{\pi n}{14}}}{\sin{\frac{\pi}{14}}},\quad n=1,\ldots,6.
\end{equation}
From this perspective, the first six meson masses $M_n$ ($n=0,\ldots,5$) that we compute in the $\alpha\to0$ limit of scalar QCD$_2$ can simultaneously be regarded as approximations to the exact breather masses $M_{B_n}$ of the IIFT. This correspondence allows for a direct comparison, analogous to the analysis performed for the Ising Field Theory (IFT) in \cite{Litvinov:2025geb} (see Section 6.1), and the results are summarized in Table \ref{tab:mesons-near-alpha=0}.
\begin{table}[h!]
    \centering
    \begin{tabular}{|c||c|c|c|}
    \hline
    Ratio & IIFT & Scalar QCD$_2$ & $\delta_n$ \\
    \hline
    $M_{B_2}/M_{B_1}\; (M_1/M_0)$ & 1.94986 & 2.29473 & 0.176 \\
    \hline
    $M_{B_3}/M_{B_1}\; (M_2/M_0)$ & 2.80194 & 3.03562 & 0.083 \\
    \hline
    $M_{B_4}/M_{B_1}\; (M_3/M_0)$ & 3.51352 & 3.66786 & 0.044 \\
    \hline
    $M_{B_5}/M_{B_1}\; (M_4/M_0)$ & 4.04892 & 4.18148 & 0.033 \\
    \hline
    $M_{B_6}/M_{B_1}\; (M_5/M_0)$ & 4.38129 & 4.65631 & 0.063 \\
    \hline
    \end{tabular}
    \caption{Ratios of meson masses in IIFT (first column) and scalar QCD$_2$ (second column). The IIFT ratios are exact \cite{Fateev:1993av,LeClair:1997gv}. The scalar QCD$_2$ ratios were computed using numerical values of $\lambda_n \sim M_n^2$ from Table \ref{tab:WKB}: for even mesons we used numerical data, and for odd mesons we used WKB estimates (alternatively, they may be estimated from a system analogous to eq. 6.10 in \cite{Litvinov:2025geb}). The last column, $\delta_n$, shows the relative error 
    $\delta_n = \left|\frac{M_{n-1}/M_0 - M_{B_n}/M_{B_1}}{M_{B_n}/M_{B_1}}\right|$.}
    \label{tab:mesons-near-alpha=0}
\end{table}

However, in contrast to the Ising case studied in \cite{Litvinov:2025geb}, the level of accuracy obtained here is somewhat lower (especially for the first excited state). In a recent study \cite{Gao:2025mcg}, these ratios were also computed beyond the two-particle approximation using the Truncated Free Fermionic Space Approach (TFFSA) \cite{Yurov:1991my}, achieving even higher accuracy.
\subsection{Heavy mass limit: \texorpdfstring{$\alpha\to\infty$}{}}
\paragraph{Warm-up: low energy and semiclassical expansions.} In the regime of heavy bosons ($\alpha\approx\frac{\pi m^2}{g^2}\gg 1$), the Bethe-Salpeter equation \eqref{tHooft-eq-bosons} describing large-$N_c$ scalar QCD$_2$ reduces, at leading order, to a Schr\"{o}dinger equation with linear confinement \cite{Shei:1977ci}\footnote{Note that is a typo in \cite{Shei:1977ci}: the coefficient in front of $|x_1-x_2|$ is smaller by a factor of $2\pi$.}:
\begin{equation}\label{non-rel-Hamiltonian}
    H=2m+\frac{p_1^2+p_2^2}{2m}+\frac{g^2}{2}|x_1 - x_2|.
\end{equation}
Its spectrum is well known and takes the form
\begin{equation}\label{non-rel-spectrum}
    M_n=2m-m\left(\frac{g^2}{2m^2}\right)^{\frac{2}{3}}\cdot
    \begin{cases}
        \mu'_{k+1}, \quad n=2k;\\
        \mu_{k+1}, \quad n=2k+1,
    \end{cases}
    \quad \overset{\eqref{mass-notation}}{\Longrightarrow} \quad
    \lambda_n=\frac{2\alpha}{\pi^2}-\frac{2\alpha}{\pi^2}\left(\frac{\pi}{2\alpha}\right)^{\frac{2}{3}}\cdot
    \begin{cases}
        \mu'_{k+1}, \quad n=2k;\\
        \mu_{k+1}, \quad n=2k+1,
    \end{cases}
\end{equation}
where $\mu_n$ and $\mu'_n$ denote consecutive zeros of the Airy function $\text{Ai}(z)$ and its derivative $\text{Ai}'(z)$, respectively. This result precisely coincides with the non-relativistic bound-state spectrum of the 't Hooft model, reflecting the fact that the Schr\"{o}dinger formulation does not distinguish between bosonic and fermionic constituents. Recently, following the approach of \cite{Fonseca:2006au}, in \cite{Gao:2025mcg} the Bethe-Salpeter equation governing interchain mesons in a system of two massive Ising chains coupled via the spin-spin interaction term \eqref{2IFT_action}. Remarkably, it turns out to coincide with the Bethe-Salpeter equation of large-$N_c$ scalar QCD$_2$, and their analysis provides systematic corrections to the interchain meson spectrum, i.e. refinements of the leading-order result \eqref{non-rel-spectrum}.

The approximation \eqref{non-rel-spectrum} provides an accurate description for low-lying states with $n\ll\alpha$. However, for higher excitations with $n\sim\alpha$, relativistic corrections become essential, and the appropriate framework is semi-classical quantization. To analyze this regime, let us consider a relativistic meson composed of two bosons\footnote{The argument is insensitive to the particle statistics, but for definiteness we take them to be bosons.}. In the center-of-mass frame, each constituent follows a periodic back-and-forth motion, tracing out segments of hyperbolic trajectories. The resulting classical orbits form closed ``lentil-shaped’’ curves, with the boson rapidities parameterized as $\pm w$. In these variables, the classical trajectories take the form:
\begin{equation}
    p=\frac{p_2-p_1}{2}=m\sinh w, \quad x=x_2-x_1=\frac{4m}{g^2}(\cosh w-\cosh\vartheta), 
\end{equation}
where $\vartheta$ denotes the maximal rapidity attained when the two particles cross. The classical trajectory carries a finite action over one oscillation cycle,
\begin{equation}
    \oint p\;dx=\frac{4m^2}{g^2}\int_{-\vartheta}^{\vartheta}\limits \sinh^2w\;dw=\frac{2\alpha}{\pi}(\sinh2\vartheta-2\vartheta),\quad\text{with}\quad\alpha\approx\frac{\pi m^2}{g^2},
\end{equation}
and this action must be quantized. Applying the Bohr-Sommerfeld prescription, we obtain
\begin{equation}
    2\oint p\;dx=2\pi\left(N+\frac{1}{2}\right),
\end{equation}
with $N$ restricted to even integers in the even sector and odd integers in the odd sector. This leads to a discrete set of rapidities $\vartheta_n$ and, consequently, to the meson spectrum,
\begin{equation}\label{semi-classical-spectrum}
    \frac{\sinh2\vartheta_n-2\vartheta_n}{2}=\frac{\pi^2}{2\alpha}\left(n-\left[\frac{1}{2}\pm\frac{1}{4}\right]\right)+\mathcal{O}(\alpha^{-2}), \quad M_n=2m\cosh\vartheta_n,\quad n=1,2,3,\dots
\end{equation}
The ``$+$'' sign corresponds to even energy levels, and ``$-$'' to odd energy levels. A more systematic analysis, which allowed one to extract additional terms in the expansion on the right-hand side of the equation, was first carried out in \cite{Fonseca:2006au} (more terms are given in \cite{Rutkevich:2009zz}) within the Ising Field Theory (where, due to the Majorana nature of the fermions, only the odd series appears). Later, the same technique was extended to the 't Hooft model \cite{ZIYATDINOV:2010ModPhysA}, where the fermions are Dirac, and consequently both series of levels are present. For the two-flavor case of the 't Hooft model, this technique was used in \cite{Artemev:2025cev}, albeit only to extract the leading term.
\paragraph{WKB expansion.}
As shown in Section \ref{WKB-section}, the spectrum in the two-particle approximation of IFT exhibits linear growth for $n\to\infty$ at fixed $\alpha$, see \eqref{WKB-normal-alpha-even}-\eqref{WKB-normal-alpha-odd}. We now study the opposite limit: $\alpha\to\infty$ first, followed by $n\gg1$. These two limits do not commute. Following the method of \cite{Artemev:2025cev}, the regime $\alpha\to\infty$ requires careful treatment of levels with $n\lesssim\alpha\log{\alpha}$. In this case, the WKB parameter $\mathfrak{n}$ \eqref{WKB-parameter} can become negative, making \eqref{WKB-normal-alpha-even}-\eqref{WKB-normal-alpha-odd} invalid. One expects the corresponding eigenvalues to remain of order $\alpha$, see \eqref{heavy-lambda-estimate}.

Moreover, higher-order phase contributions $\Phi_\pm^{(k)}$ are no longer suppressed, since they grow as $\alpha^{k+1}/\lambda^k$ (for $k\geq1$), while the $k=0$ term contributes as $\alpha\log\frac{\alpha}{\lambda}$; subleading terms are suppressed by $\alpha^{-2}$ or smaller. Resumming the dominant terms in \eqref{quantisation-condition-lambda}, and using the asymptotics of $\mathtt{i}_k(\alpha)$ \eqref{i-def}, $\mathtt{u}_{2k-1}(\alpha)$ \eqref{u-def} (see Appendix B.2 in \cite{Litvinov:2025geb}), the quantization condition takes the form
\begin{equation}
   \frac{2\alpha}{\pi^2}\left(\frac{2}{A}+\log{A}-1-\log{4}+\frac{A}{4}\left(1+\frac{A}{4}+\frac{5A^2}{48}+\frac{7A^3}{128}+\frac{21A^4}{640}+\dots\right)\right)+\mathcal{O}\left(\frac{1}{\alpha}\right)=n+\frac{1}{2},
\end{equation}
with $A=\frac{2\alpha}{\pi^2\lambda}$, $n=0,1,\dots$. The bracketed series matches the Maclaurin expansion of ${}_3F_2(1,1,\tfrac{3}{2};2,3;A)$. The coefficient at $\alpha$ vanishes at $A=1$, giving for small $n$: $\lambda_n=\tfrac{2\alpha}{\pi^2}+\dots$, $M_n^2=4m^2+\dots,$
consistent with \eqref{non-rel-spectrum}. Expanding around $A=1$ yields (here $n=0,1,2,\dots$)
\begin{equation}\label{A-to-1-quantization-cond}
    \frac{2\alpha}{\pi^2}\left(A^{-1}-1\right)^{3/2}\left(\frac{4}{3}-\frac{2(A^{-1}-1)}{5}+\frac{3}{14}(A^{-1}-1)^2+\dots\right)+\mathcal{O}\left(\frac{1}{\alpha} \right)=n+\frac{1}{2},
\end{equation}
leading to
\begin{equation} \label{large-alpha-wkb}
    \frac{\lambda_n}{2\alpha/\pi^2}=1+\left(\frac{3\pi^2}{8\alpha}(n+1/2)\right)^{2/3}+\dots
\end{equation}
which coincides with good accuracy\footnote{The leading terms that we have summed up give only the asymptotics of the zeros of the Airy function and its derivative. A systematic resummation of the $\mathcal{O}(1/\alpha)$ corrections to the phases $\Phi^{(k)}_{\pm}$ would be required to rigorously justify the asymptotic formula \eqref{non-rel-spectrum} (see also the end of Section 6.2 in \cite{Litvinov:2025geb}).} the subleading term \eqref{non-rel-spectrum}, and is valid for $1\ll n\ll \alpha$.
\paragraph{Spectral sums.} 
The low-lying bound-state spectrum $M_n$ approaches $2m$ from above \eqref{non-rel-spectrum}. Although all spectral sums except $\mathcal{G}^{(1)}_\pm$ are expected to vanish as $\alpha\to\infty$, this behavior is not immediately apparent from the explicit expressions \eqref{Gpm1}-\eqref{Gpm3}. By employing the large-$\alpha$ asymptotics of the integrals $\mathtt{i}_k(\alpha)$ \eqref{i-def}, $\mathtt{u}_{2k-1}(\alpha)$ \eqref{u-def} (see Appendix B.2 in \cite{Litvinov:2025geb}), as well as the asymptotic expansion of $\mathtt{v}(\alpha)$ \eqref{Proector-def}
\begin{equation}
    \mathtt{v}(\alpha)\Big|_{\alpha\to\infty}=\frac{1}{4\alpha}-\frac{1}{8\alpha^2}+\frac{12+\pi^2}{144\alpha^3}-\frac{3+\pi^2}{48\alpha^4}+\frac{6+5\pi^2+\frac{7\pi^4}{40}}{120\alpha^5}-\frac{3+5\pi^2+\frac{161\pi^4}{240}}{72\alpha^6}+\mathcal{O}(\alpha^{-7}),
\end{equation}
one can systematically derive the corresponding expansions of the spectral sums in the limit $\alpha \to \infty$, providing a controlled framework for analyzing the non-relativistic regime of the theory:
\begin{equation}\label{G1pm-heavy}
    \begin{aligned}
        \mathcal{G}^{(1)}_+\Big|_{\alpha\to\infty}=&\;\log\left(\frac{2\pi^2e^{-2-\gamma_E}}{\alpha}\right)+\frac{\pi^2}{8\alpha}-\frac{\pi^2}{8\alpha^2}+\frac{17\pi^2-32}{48\alpha^3}-\frac{7\pi^2\left(21+\pi^2\right)}{576\alpha^4}+\frac{\pi^2(2271+424\pi^2)}{6912\alpha^5}+
        \\&-\frac{\pi^2(56325+25852\pi^2+744\pi^4)}{138240\alpha^6}+\mathcal{O}\left(\frac{1}{\alpha^7}\right),
        \\
        \mathcal{G}^{(1)}_-\Big|_{\alpha\to\infty}=&\;\log\left(\frac{2\pi^2e^{-2-\gamma_E}}{\alpha}\right)-\frac{\pi^2}{8\alpha}+\frac{\pi^2}{8\alpha^2}-\frac{16+\pi^2}{24\alpha^3}+\frac{7\pi^2\left(120+7\pi^2\right)}{2880\alpha^4}-\frac{\pi^2(30+7\pi^2)}{80\alpha^5}+
        \\&+\frac{11\pi^2\left(1680+980\pi^2+31\pi^4\right)}{40320\alpha^6}+\mathcal{O}\left(\frac{1}{\alpha^7}\right),
    \end{aligned}
\end{equation}
\begin{equation}\label{G2pm-heavy}
    \begin{aligned}
        \mathcal{G}^{(2)}_+\Big|_{\alpha\to\infty}=&\;\frac{\pi^2}{3\alpha}+\frac{\pi^4}{16\alpha^2}-\frac{11\pi^4}{60\alpha^3}+\frac{367\pi^4}{720\alpha^4}-\frac{\pi^4\left(1563+74\pi^2\right)}{1440\alpha^5}+\frac{\pi^4\left(480459+90344\pi^2\right)}{241920\alpha^6}+\mathcal{O}\left(\frac{1}{\alpha^7}\right),
        \\
        \mathcal{G}^{(2)}_-\Big|_{\alpha\to\infty}=&\;\frac{\pi^2}{3\alpha}-\frac{\pi^4}{16\alpha^2}+\frac{3\pi^4}{20\alpha^3}-\frac{31\pi^4}{72\alpha^4}+\frac{\pi^4\left(1540+87\pi^2\right)}{1680\alpha^5}-\frac{\pi^4\left(3585+811\pi^2\right)}{2160\alpha^6}+\mathcal{O}\left(\frac{1}{\alpha^7}\right),
    \end{aligned}
\end{equation}
\begin{equation}\label{G3pm-heavy}
    \begin{aligned}
        \mathcal{G}^{(3)}_+\Big|_{\alpha\to\infty}=&\;\frac{\pi^4}{15\alpha^2}+\frac{\pi^6}{32\alpha^3}-\frac{71\pi^6}{420\alpha^4}+\frac{151\pi^6}{210\alpha^5}-\frac{13\pi^6\left(9979+468\pi^2\right)}{60480\alpha^6}+\mathcal{O}\left(\frac{1}{\alpha^7}\right),
        \\
        \mathcal{G}^{(3)}_-\Big|_{\alpha\to\infty}=&\;\frac{\pi^4}{15\alpha^2}-\frac{\pi^6}{32\alpha^3}+\frac{11\pi^6}{84\alpha^4}-\frac{17\pi^6}{30\alpha^5}+\frac{\pi^6\left(25522+1419\pi^2\right)}{15120\alpha^6}+\mathcal{O}\left(\frac{1}{\alpha^7}\right),
    \end{aligned}
\end{equation}
These asymptotic results confirm that $\mathcal{G}_+^{(s)}>\mathcal{G}_-^{(s)}$, in agreement with the expected ordering of the spectral sums, although this inequality is not immediately transparent from the explicit expressions \eqref{Gpm1}-\eqref{Gpm3}. Furthermore, it is important to note that the lowest-lying energy levels contribute only subdominantly to the spectral sums. This follows from
\begin{equation}\label{heavy-lambda-estimate}
    \lambda_n \sim M_{n}^2\sim m^2 \sim \alpha\quad \Rightarrow\quad (\lambda_n)^{-s}\sim\alpha^{-s}.   
\end{equation}
The dominant contribution arises from relativistic mesons \eqref{semi-classical-spectrum}, whose masses are well above the two–quark threshold. Strikingly, the first two terms in the spectral sums \eqref{G1pm-heavy}–\eqref{G3pm-heavy} exactly match the results previously obtained both in the 't Hooft model \cite{Litvinov:2024riz} and in the two-particle approximation of the Ising Field Theory \cite{Litvinov:2025geb}. As demonstrated in \cite{Litvinov:2024riz}, this agreement is a direct consequence of the semiclassical spectrum \eqref{semi-classical-spectrum}, which provides a systematic framework for deriving these leading contributions.
\section{Large-\texorpdfstring{$N_c$}{} scalar QCD\texorpdfstring{$_2$}{} beyond real masses}\label{BS-complex-Section}
In this section, we initiate a study of the analytic continuation of meson masses $\lambda_n(\alpha)$ in large-$N_c$ scalar QCD$_2$ with respect to the scaling parameter $\alpha$, extending it into the complex plane $\alpha\in\mathbb{C}$. Our motivation stems from the crucial role played by complex singularities of spectral sums in shaping the analytic structure of meson spectra and their relation to critical phenomena. 
\subsection{Analytical continuation}
To investigate the analytic structure of the eigenvalues $\lambda_n$, we analytically continue the Bethe-Salpeter equation \eqref{BS-eq}, originally defined for real $\alpha\geq 0$, into the complex $\alpha$-plane. The idea of probing complex values of $\alpha$ originates in the two-particle approximation of the Ising Field Theory, first suggested in \cite{Fonseca:2006au} and further developed in \cite{Litvinov:2025geb}. In the large-$N_c$ scalar QCD$_2$ case, the pole structure of $\Psi(\nu)$ \eqref{Psi-all-poles} turns out to be identical to that of the Ising model, so we only summarize the main steps here. Related questions were also raised in the context of the 't Hooft model \cite{Zamolodchikov:2009pres}\cite{Ambrosino:2023dik}.

\begin{figure}[h!]
    \centering
    \includegraphics[width=0.49\linewidth]{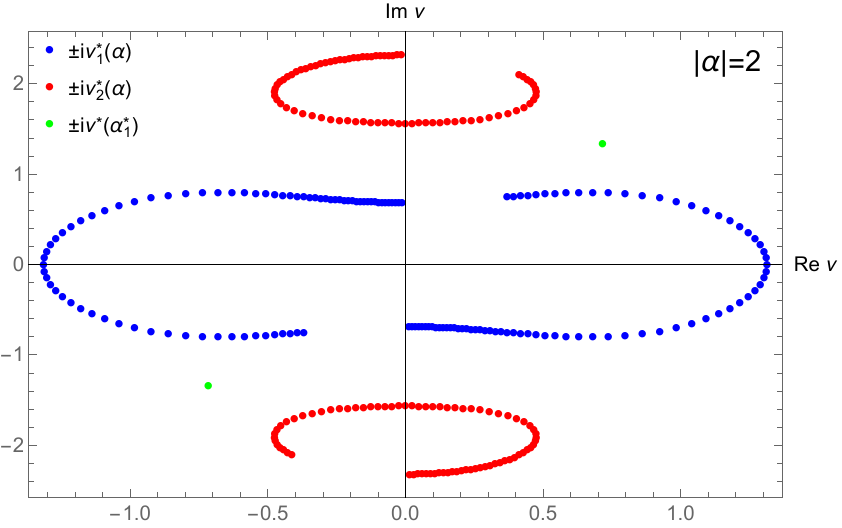}
    \includegraphics[width=0.49\linewidth]{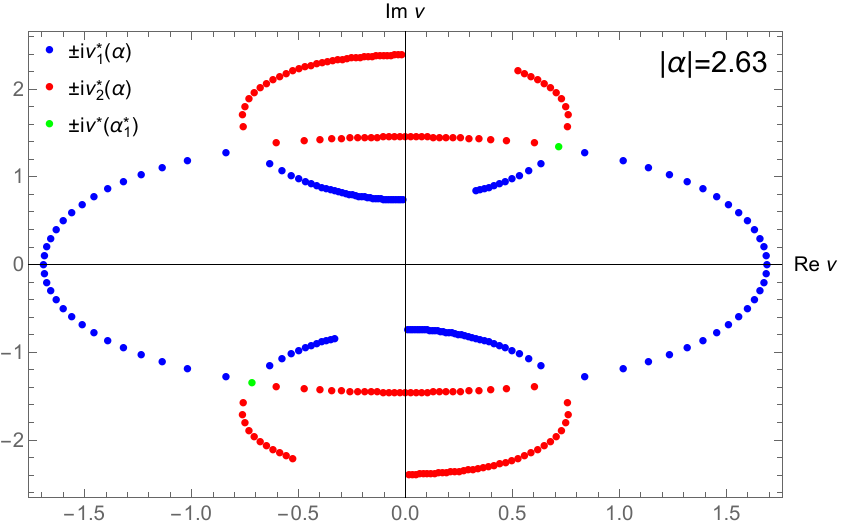}
    \includegraphics[width=0.49\linewidth]{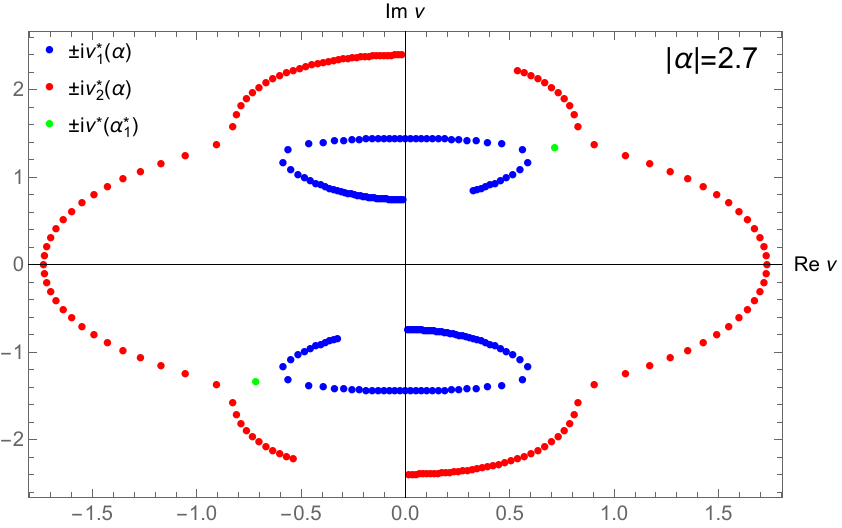}
    \includegraphics[width=0.49\linewidth]{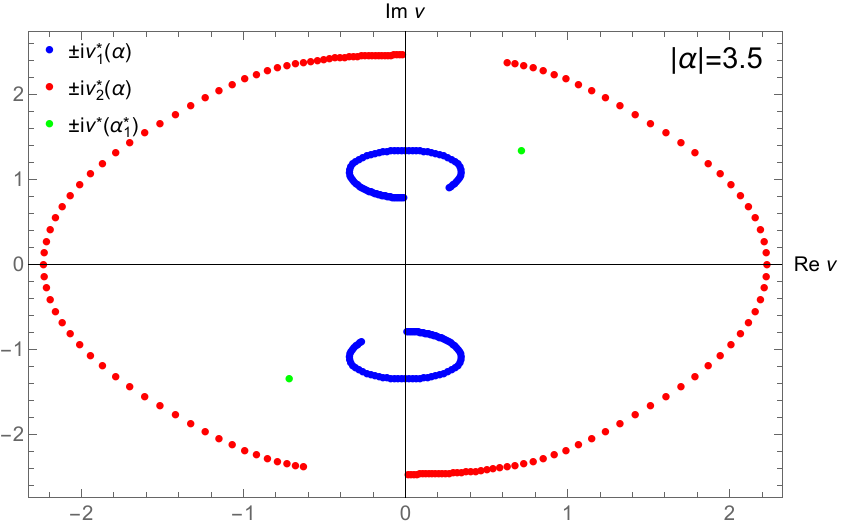}
    \caption{Trajectories of the first two pole pairs of $\Psi(\nu)$ under the continuation \eqref{analyt-cont}, shown for $|\alpha|=\{2,\; 2.63,\; 2.7,\; 3.5\}$. Blue and red points denote the positions $\pm i\nu^*_1(\alpha)$ and $\pm i\nu^*_2(\alpha)$ as $\arg \alpha$ varies from $0$ to $\tfrac{3\pi}{2}$. Green points mark the double zero of \eqref{main-transcendental-equation} (collision of two poles of $\Psi(\nu)$), located at $\pm i\nu^*(\alpha^*_1)=\pm\frac{2}{\pi}\sqrt{\alpha^*_1(1+\alpha^*_1)}$, corresponding to the critical value $\alpha^*_1\approx -1.65061-2.05998i$.}
    \label{fig:poles-evolution}
\end{figure}

For real $\alpha>0$, the continuation is performed along a counterclockwise circular contour\footnote{A clockwise continuation is equally possible; it simply reverses the pole trajectories, flipping the signs of the residue contributions and mapping the critical points of Fig. \ref{fig:crit-points} to their complex conjugates. More general contours can also be considered, see Section 7 of \cite{Litvinov:2025geb}.}:
\begin{equation}\label{analyt-cont}
\mathbb{R}\ni\alpha\quad\longmapsto\quad \alpha=e^{i\phi}|\alpha|\in\mathbb{C}.
\end{equation}
The procedure is straightforward for $|\arg\alpha|<\pi$. However, as $\arg\alpha$ approaches $\pm\pi$, the two poles of $\Psi(\nu)$ at $\nu=\pm i\nu^*_k(\alpha)$ reach the real $\nu$-axis (see Fig. \ref{fig:poles-evolution}, reproduced from \cite{Litvinov:2025geb}). The ordinal number $k$ of the pole pair crossing the axis is determined by the modulus of $\alpha$ through the condition $|\alpha^*_{k-1}|<|\alpha|<|\alpha^*_{k}|$, with the first few critical values $\alpha^*_{k}$ listed in Table \ref{tab:crit-points}.

Crossing into the second Riemann sheet ($|\arg\alpha|>\pi$), these poles pass through the real axis, generating additional residue contributions. The Bethe-Salpeter equation thus acquires an extra term, leading to
\begin{multline}\label{BS_second_worldsheet}
    \left(\frac{2\alpha}{\pi}+\nu\tanh{\frac{\pi\nu}{2}}\right)\Psi(\nu)+\frac{1}{8}\frac{1}{\cosh{\frac{\pi\nu}{2}}}\int_{-\infty}^{\infty}\limits\limits d\nu'\frac{1}{\cosh{\frac{\pi\nu'}{2}}}\Psi(\nu')-\lambda(\alpha)\int_{-\infty}^{\infty}\limits d\nu'\frac{\pi(\nu-\nu')}{2\sinh{\frac{\pi(\nu-\nu')}{2}}}\Psi(\nu')+
    \\+2\pi i\left(\underset{\nu'=i\nu^*_k}{\textrm{Res }}-\underset{\nu'=-i\nu^*_k}{\textrm{Res }}\right)\left[\frac{1}{8}\frac{1}{\cosh{\frac{\pi\nu}{2}}}\frac{1}{\cosh{\frac{\pi\nu'}{2}}}\Psi(\nu')-\lambda(\alpha)\frac{\pi(\nu-\nu')}{2\sinh{\frac{\pi(\nu-\nu')}{2}}}\Psi(\nu')\right]=0.
\end{multline}
Equation \eqref{BS_second_worldsheet} continues to hold at $\arg\alpha=2\pi$, i.e. after one full rotation around $\alpha=0$. A second rotation generates an additional residue contribution, which precisely cancels the first\footnote{In the 't Hooft model \cite{THOOFT1974461} the cancellation does not always occur, since the singularities $\alpha_k^*$ ($k\ne0$) on the second sheet are quartic-root rather than square-root \cite{Litvinov:2025geb} (see formula 7.8). Consequently, after two large-radius rotations---or any contour encircling $\alpha=0$ while passing an odd number of times around some $\alpha_k^*$---the cancellation fail.}. This cancellation establishes $\alpha=0$ as a square-root branch point of the solution, in agreement with the expansions \eqref{WKB-small-alpha-even}-\eqref{WKB-small-alpha-odd}.
\subsection{Critical points on the second sheet}
\begin{figure}[h!]
    \centering
    \includegraphics[width=0.65\linewidth]{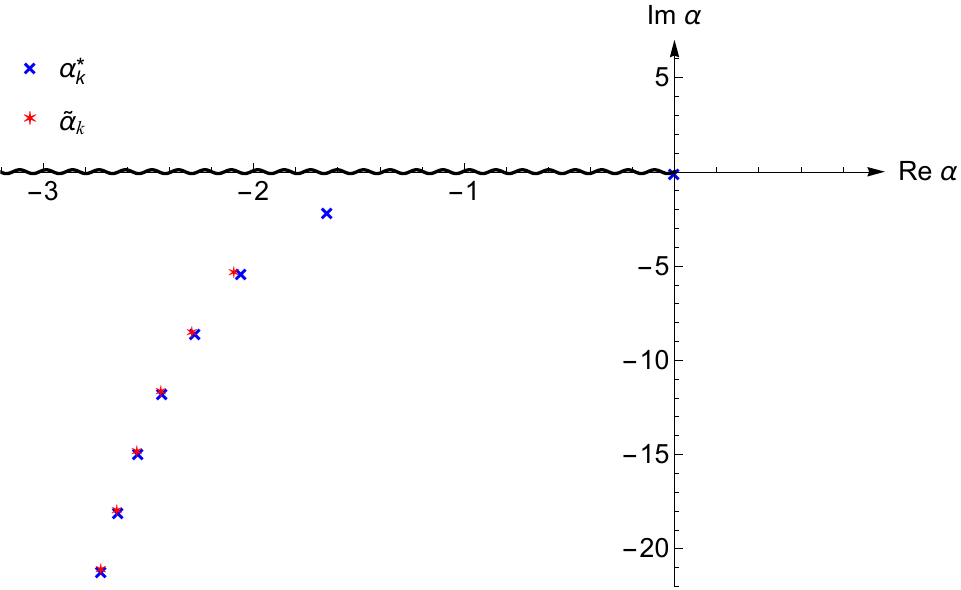} 
    \caption{Critical points of $\lambda_n$ on the second sheet of the $\alpha$-plane: $\alpha^*_k$ \eqref{critical-points1-approx} (blue) and $\widetilde{\alpha}_k$ \eqref{critical-points2-approx} (red). The point $\alpha^*_0=0$ is a square-root branch point, while all other $\alpha^*_k$ are square-root branch points on the second sheet, as indicated by \eqref{WKB-complex-1}. Numerical values for $\alpha^*_k$ and $\widetilde{\alpha}_k$ are listed in Tables \ref{tab:crit-points} and \ref{tab:crit-points-2}, respectively.}
    \label{fig:crit-points}
\end{figure}

The spectral sums $\mathcal{G}^{(s)}_\pm$ \eqref{Gpm1}-\eqref{Gpm3}—polynomials in the integrals $\mathtt{u}_{2k-1}(\alpha)$, with additional factors $(1+\mathtt{v}(\alpha))^{-l}$ in the even sector from the matrix elements $\langle p|K^n|p\rangle$ \eqref{matr-el-K}—expose a rich analytic structure in complex $\alpha$. They do not reconstruct the full meson spectrum; rather, they act as diagnostics for the emergence of massless states. On the second Riemann sheet, their singularities form two families that accumulate toward $|\alpha|\to\infty$ (see Fig. \ref{fig:crit-points}):
\begin{itemize}
    \item \textbf{Pole-collision singularities (odd-sector).} At the critical points $\alpha=\alpha_k^*$, where two poles of $\Psi(\nu)$ collide, the integrals $\mathtt{u}_{2k-1}(\alpha)$ diverge. As a result, the odd sums $\mathcal{G}^{(s)}_-$ blow up, producing a massless odd state and generating square-root branch points in $\alpha_k^*$.
    \item \textbf{Zero-induced singularities (even-sector).} At the critical points $\alpha=\widetilde{\alpha}_k$, where $1+\mathtt{v}(\widetilde{\alpha}_k)=0$, the even sums $\mathcal{G}^{(s)}_+$ diverge. As a result, the spectrum contains a massless even state, but without square-root branching of the meson levels.
\end{itemize}
\paragraph{Pole-collision singularities $\alpha=\alpha_k^*$.} We begin with the first family of singularities, $\alpha_k^*$. Their origin lies in the pole-collision mechanism analyzed in detail in \cite{Litvinov:2025geb} (see also Fig.~\ref{fig:poles-evolution}). The positions of these critical points are determined by a coupled system consisting of the transcendental equation \eqref{main-transcendental-equation} together with its derivative with respect to $t=\frac{\pi\nu}{2}$\footnote{In the 't Hooft model, the critical points are determined by a system of equations analogous to \eqref{crit-points-system}, a fact first observed in \cite{Zamolodchikov:2009pres}
\begin{equation*}
    \begin{cases}
        \alpha+t\coth{t}&=0;\\
        \partial_t(\alpha+t\coth{t})&=0.
    \end{cases}
\end{equation*}
}:
\begin{equation}\label{crit-points-system}
    \begin{cases}
        \alpha+t\tanh{t}&=0;\\
        \partial_t(\alpha+t\tanh{t})&=0,
    \end{cases}\quad\Rightarrow\quad
    \begin{cases}
        it_k^*&=\pm\sqrt{\alpha_k^*(1+\alpha_k^*)};\\
        2it_k^*+\sinh{(2it_k^*)}&=0.
    \end{cases}
\end{equation}
Solving the system \eqref{crit-points-system} produces an infinite sequence of critical values $\alpha^*_k$, each residing on the second Riemann sheet of the $\alpha$-plane, beyond the branch cut along $(-\infty,0)$. Since our analytic continuation \eqref{analyt-cont} is performed counterclockwise, we focus on solutions with $\arg\alpha>\pi$; continuation in the opposite direction would instead produce their complex conjugates. The first few of these points are listed in Table \ref{tab:crit-points} and illustrated in Fig. \ref{fig:crit-points}. Each nontrivial critical point $\alpha^*_k$ (with the exception of $\alpha=\alpha^*_0=0$, discussed earlier in Section \ref{Limiting-cases-section}) corresponds to a pair of values $\pm i\nu^*_k=\pm\frac{2}{\pi}i t^*_k$, one lying in the upper and the other in the lower half-plane. This structure reflects the pole dynamics: one pole descends from the upper half-plane to collide with a partner, while the other ascends from the lower half-plane and undergoes an analogous collision.
\begin{table}[h!]
    \begin{center}
    \resizebox{\textwidth}{!}{$
        \begin{tabular}{| c | c | c | c | c |}
            \hline 
            $k$ & $-it^*_k$ & $-\alpha^*_k$ & $|\alpha^*_k|$ & $\phi^*_k$ \\
            \hline
            0 & 0 & 0 & 0 & $-$\\
            \hline
            1 & $1.12536430580 + 2.10619611525i$ & $1.6506112935 + 2.0599814572i$ & 2.6397047650 & 0.28497600762 \\
            \hline
            2 & $1.55157437291 + 5.35626869864i$ & $2.0578451096 + 5.3347083072i$ & 5.7178526754 & 0.38281140736 \\
            \hline 
            3 & $1.77554367351 + 8.53668242658i$ & $2.2784697373 + 8.5226372750i$ & 8.8219482239 & 0.41684664443 \\
            \hline
            4 & $1.92940449655 + 11.69917761283i$ & $2.4311221188 + 11.6887718663i$ & 11.9389171410 & 0.43472603131 \\
            \hline
            5 & $2.04685246238 + 14.85405991264i$ & $2.5479913751 + 14.8457993910i$ & 15.0628689035 & 0.44589547061 \\
            \hline
            \end{tabular}
            $}
        \end{center}
        \caption{Numerical values for critical points $\alpha^*_k=|\alpha^*_k|e^{i\pi(1+\phi^*_k)}$ and corresponding points $it^*_k$ in the lower-left quadrant of the complex plane.
        \label{tab:crit-points}}
\end{table}

By applying the iterative method, one derives an explicit asymptotic expression for the roots $it^*_k$ and $\alpha^*_k$ of the system \eqref{crit-points-system}:
\begin{equation}\label{critical-points1-approx}
    \resizebox{\textwidth}{!}{$
    \begin{aligned}
        -it^*_k&=\left[\frac{1}{2}\log\kappa+\frac{2\log^2\kappa-4\log\kappa-1}{2\kappa^2}+\mathcal{O}\left(\frac{\log^4\kappa}{\kappa^4}\right)\right]+i\left[\frac{\kappa}{4}-\frac{\log \kappa}{\kappa}+\frac{4\log^3\kappa-18\log^2\kappa+6\log\kappa+3}{3\kappa^3}+\mathcal{O}\left(\frac{\log^5\kappa}{\kappa^5}\right)\right],
        \\
        -\alpha^*_k&=\left[\frac{1}{2}(\log\kappa+1)+\frac{2\log^2\kappa-2\log\kappa-1}{2\kappa^2}+\mathcal{O}\left(\frac{\log^4\kappa}{\kappa^4}\right)\right]+i\left[\frac{\kappa}{4}-\frac{2\log\kappa+1}{2\kappa}+\frac{8\log^3\kappa-24\log^2\kappa+3}{6\kappa^3}+\mathcal{O}\left(\frac{\log^5\kappa}{\kappa^5}\right)\right],
        \end{aligned}
    $}
\end{equation}
with $\kappa=4\pi\left(k-\tfrac{1}{4}\right),\; k=1,2,3,\ldots$. Remarkably, the approximation is already highly accurate: for the first nontrivial root ($k=1$), the relative error does not exceed $0.2\%$.

Using the expansions of the integrals $\mathtt{u}_{2k-1}(\alpha)$ near the critical points $\alpha^*_k$ (see Appendix B.3 in \cite{Litvinov:2025geb}), we obtain the corresponding expansions of the spectral sums in the vicinity of these potential singularities. Here, we present only the essential terms of these expansions, sufficient to draw conclusions about the leading behavior of the massless state:
\begin{equation}\label{spec-sums-expansion-in-crit-points}
    \begin{aligned}
        \mathcal{G}^{(s)}_+\Big|_{\alpha\to\alpha_k^*}=&\;\text{const}(s)+\mathcal{O}\left(\sqrt{\alpha-\alpha_k^*}\right),
        \\
        \mathcal{G}^{(s)}_-\Big|_{\alpha\to\alpha_k^*}=&\;\left(\frac{2\pi}{\sqrt{\alpha-\alpha_k^*}}\right)^{s}+\mathcal{O}\left((\alpha-\alpha_k^*)^{-\frac{s-1}{2}}\right),
    \end{aligned}
\end{equation}
The expansions show that even spectral sums remain finite, while odd ones diverge with increasing order. This implies that, under $\alpha\to\alpha^*_k$, the spectrum contains a meson $\lambda^{(2k+1)}$ with vanishing mass. The index $2k+1$ is assigned to indicate correspondence with $\alpha^*_k$ and its origin as an odd state, though its exact label (before analytic continuation) cannot be fixed from the expansions. The mass vanishes as
\begin{equation}\label{lambda-odd-critical}
    \resizebox{\textwidth}{!}{$
    \begin{gathered}
        \lambda^{(2k+1)}(\alpha)\Big|_{\alpha\to\alpha_k^*}=\frac{1}{2\pi}\sqrt{\alpha-\alpha_k^*}+\left(\frac{(\alpha_k^*)^2c_{3}^{(0)}}{4\pi^2}-\frac{2+\alpha_k^*}{2\pi^2}+\frac{\alpha_k^*\zeta(3)}{2\pi^4}-\frac{i(3+2\alpha_k^*)}{3\pi\sqrt{\alpha_k^*(1+\alpha_k^*)}}\right)(\alpha-\alpha_k^*)+\mathcal{O}\left((\alpha-\alpha)^{\frac{3}{2}}\right),
        \\
        c^{(l)}_{2n-1}=(-1)^l\int_{-\infty+i\epsilon}^{\infty+i\epsilon}\limits dt\frac{\cosh^{2+l}{t}}{t\sinh^{2n-1}{t}(\alpha_k^*\cosh{t}+t\sinh{t})^{l+1}},
    \end{gathered}
    $}
\end{equation}
with subleading terms inherited from the expansion of the odd spectral sums $\mathcal{G}^{(s)}_-$.

At each critical point $\alpha_k^*$, in addition to the vanishing of the odd meson mass \eqref{lambda-odd-critical}, the meson spectrum develops square-root branch behavior: the masses become multivalued and admit expansions in half-integer powers of $(\alpha-\alpha_k^*)^{1/2}$. For low-lying states, this behavior indirectly follows from the expansions of spectral sums \eqref{spec-sums-expansion-in-crit-points} into half-integer powers, while for highly excited states it can be verified directly by substituting the expansions of the integrals $\mathtt{i}_k(\alpha)$ \eqref{i-def} near $\alpha_k^*$ (Appendix B.3 of \cite{Litvinov:2025geb}) into the WKB formulas \eqref{WKB-normal-alpha-even}-\eqref{WKB-normal-alpha-odd}
\begin{equation}\label{WKB-complex-1}
    \resizebox{\textwidth}{!}{$
    \begin{gathered}
        \lambda_n(\alpha)\Big|_{\alpha\to\alpha_k^*}=\left[\frac{N_n}{2}+\frac{\alpha_k^*\log\left(\pi e^{\gamma_E}N_n\right)}{\pi^2}+\mathcal{O}\left(\frac{\log N_n}{N_n}\right)\right]+\frac{2\sqrt{\alpha-\alpha_k^*}}{\pi}\left[1+\frac{2\alpha_k^*}{\pi^2N_n}+\mathcal{O}\left(\frac{\log N_n}{N_n^2}\right)\right]+\mathcal{O}(\alpha-\alpha_k^*),\\ N_n=n+\frac{1}{4}+\frac{(\alpha_k^*)^2}{\pi^2}\left(\frac{2\zeta(3)}{\pi^2}+\alpha_k^*c^{(0)}_{3}-c^{(0)}_{1}-\frac{4\pi i}{\alpha_k^*}\sqrt{\frac{1+\alpha_k^*}{\alpha_k^*}}\right).
    \end{gathered}
    $}
\end{equation}
This expression holds identically for both even and odd mesons and coincides with the Ising Field Theory result \cite{Litvinov:2025geb} at leading order, with distinctions arising only in the subleading corrections.
\paragraph{Zero-induced singularities $\alpha=\widetilde{\alpha}_k$.} We now turn to the second class of potential singularities, which arise when the condition $1+\mathtt{v}(\widetilde{\alpha})=0$ is satisfied. This relation originates from the behavior of even spectral sums $\mathcal{G}^{(s)}_\pm$ \eqref{Gp-with-matr-el} or alternatively from the Bethe-Salpeter equation \eqref{BS-eq} by setting $\lambda=0$ and projecting onto the kernel $\frac{1}{\cosh{\frac{\pi\nu}{2}}}$. A closer examination of integral $\mathtt{v}(\alpha)$ \eqref{Proector-def} shows that, on the first sheet, its modulus attains its maximum at $\alpha=0$, with a value of $-\frac{\log2}{2}$. Consequently, any solutions must lie on the second sheet.

Moreover, since the integrand diverges at the infinite set of points $\alpha^*_k$ (the denominator contains the same transcendental function as in \eqref{main-transcendental-equation}), the equation $1+\mathtt{v}(\widetilde{\alpha})=0$ necessarily admits infinitely many solutions, which are expected to be located near the $\alpha^*_k$. Upon analytic continuation to the second sheet, the integral $\mathtt{v}(\alpha)$ acquires residue contributions, yielding the following defining equation for the critical points $\widetilde{\alpha}_k$:
\begin{equation}\label{1+v=0}
    1+\frac{\pi i}{2\widetilde{t}_k+\sinh{2\widetilde{t}_k}}+\frac{1}{8}\int_{-\infty}^{\infty}\limits dt\frac{1}{\cosh^2{t}(\widetilde{\alpha}_{k}+t\tanh{t})}=0,\quad \widetilde{\alpha}_{k}+\widetilde{t}_k\tanh{\widetilde{t}_k}=0,
\end{equation}
where the first term arises from the residues at $\pm\widetilde{t}_k$, with $\widetilde{t}_k$ located in the lower-left quadrant of the complex plane, where they cross the real axis. Finding solutions to this equation is considerably more challenging than for \eqref{crit-points-system}. Table \ref{tab:crit-points-2} presents the first few roots in the lower-left quadrant, and their location is also illustrated in Fig. \ref{fig:crit-points}.
\begin{table}[h!]
    \begin{center}
    \resizebox{\textwidth}{!}{$
        \begin{tabular}{| c || c | c | c | c | c |c |}
        \hline 
        $k$ & $-\widetilde{t}_k$ & $-\widetilde{\alpha}_k$ & $|\widetilde{\alpha}_k|$ & $\phi_k$ & $1+\mathtt{v}(\widetilde{\alpha}_k)$ & error $=|1+\mathtt{v}|$ \\
        \hline
        1 & $1.380161693+5.327877602i$ & $2.090075377+5.345971862i$ & $5.740020055$ & 0.3813686521 & $-1.8\cdot10^{-10}+4.6\cdot10^{-11}i$ & $1.9\cdot10^{-10}$ \\
        \hline
        2 & $1.675025408+8.523800071i$ & $2.289110212+8.525462921i$ & $8.827431313$ & $0.4165022909$ & $7.2\cdot10^{-11}+8.2\cdot10^{-10}i$ & $8.3\cdot10^{-10}$ \\
        \hline
        3 & $1.857942214+11.691620693i$ & $2.436420834+11.689921817i$ & $11.94112300$ & $0.4345939869$ & $-4.7\cdot10^{-10}+5.5\cdot10^{-10}i$ & $7.2\cdot10^{-10}$ \\
        \hline
        4 & $1.991345496+14.849025726i$ & $2.551163008+14.846386387i$ & $15.06398425$ & $0.4458315163$ & $-3.9\cdot10^{-10}-1.3\cdot10^{-10}i$ & $4.2\cdot10^{-10}$ \\
        \hline
        5 & $2.096494558+18.001310501i$ & $2.644816531+17.998428580i$ & $18.19171476$ & $0.4535576714$ & $3.0\cdot10^{-10}-3.5\cdot10^{-10}i$ & $4.6\cdot10^{-10}$ \\
        \hline
        6 & $2.183314153+21.150667361i$ & $2.723842932+21.147786167i$ & $21.32248063$ & $0.4592260499$ & $-4.4\cdot10^{-10}-5.3\cdot10^{-10}i$ & $6.9\cdot10^{-10}$ \\
        \hline
        \end{tabular}
    $}
    \end{center}
    \caption{Numerical values for critical points $\widetilde{\alpha}_k=|\widetilde{\alpha}_k|e^{i\pi(1+\phi_k)}$ and corresponding points $\widetilde{t}_k$ in the lower-left quadrant of the complex plane.}
     \label{tab:crit-points-2}
\end{table}

By ignoring the integral term, we can obtain an estimate for the roots of the equation \ref{1+v=0}, which already gives an accuracy better than $0.2\%$ even for the first solution
\begin{equation}\label{critical-points2-approx}
    \resizebox{\textwidth}{!}{$
    \begin{aligned}
        -\widetilde{t}_k&\approx\left[\frac{1}{2}\log\kappa+\frac{2\log^2\kappa-4\log\kappa-1}{2\kappa^2}\right]+i\left[\frac{\kappa}{4}+\frac{\pi}{2}-\frac{\log \kappa}{\kappa}+\frac{4\log^3\kappa-18\log^2\kappa+6\log\kappa+3}{3\kappa^3}\right],
        \\
        -\widetilde{\alpha}_k&\approx\left[\frac{1}{2}(\log\kappa+1)+\frac{\pi}{\kappa}+\frac{2\log^2\kappa-2\log\kappa-1}{2\kappa^2}\right]+i\left[\frac{\kappa}{4}+\frac{\pi}{2}-\frac{2\log\kappa+1}{2\kappa}+\frac{\pi(2\log\kappa-1)}{\kappa^2}+\frac{8\log^3\kappa-24\log^2\kappa+3}{6\kappa^3}\right],
    \end{aligned}
    $}
\end{equation}
where $\kappa=4\pi\left(k+\frac{1}{4}\right),$ $k=1,2,3,\ldots$. 

Since the roots $\widetilde{\alpha}_k$ of \eqref{1+v=0} are simple zeros, the integral behaves linearly in their vicinity, $1+\mathtt{v}(\alpha)\sim\alpha-\widetilde{\alpha}_k$. Furthermore, the maximal power of $\mathcal{V}(\alpha)\sim\frac{1}{1+\mathtt{v}(\alpha)}$ in the even spectral sums \eqref{Gp-with-matr-el} matches the index of the sum. Consequently, the spectrum in this limit contains a single meson whose mass vanishes linearly:
\begin{equation}\label{lambda-even-critical}
    \lambda^{(2k)}(\alpha)\Big|_{\alpha\to\widetilde{\alpha}_k}\sim\alpha-\widetilde{\alpha}_k.
\end{equation}
Here, as in the case above, we cannot determine the ordinal number of the state from the spectral sums before the analytical continuation, and we use the notation with the superscript $2k$ to emphasize that this state is associated with the critical point $\widetilde{\alpha}_k$ and that it was originally even. The analyticity of $1+\mathtt{v}(\alpha)$ at $\widetilde{\alpha}_k$ implies that, in contrast to the points $\alpha_k^*$, the meson spectrum does not develop square-root branchings at $\widetilde{\alpha}_k$.

Notably, the vanishing behavior of mesons \eqref{lambda-odd-critical} and \eqref{lambda-even-critical}, as extracted from the spectral sums, is fully consistent with the analytic formula for the ratio of spectral determinant constants, $d_-/d_+$ \eqref{dm/dp}. In particular, at $\alpha=0$, this ratio remains finite, confirming the absence of massless states in large-$N_c$ scalar QCD$_2$. Near the pole-collision points $\alpha^*_k$ (excluding $\alpha^*_0=0$), the ratio diverges as $(\alpha-\alpha^*_k)^{-1/2}$, while in the vicinity of $\widetilde{\alpha}_k$, it vanishes linearly as $\alpha-\widetilde{\alpha}_k$
\begin{equation}
    \frac{d_-}{d_+}\Bigg|_{\alpha\to0}=\frac{\sqrt{2}}{8}, \qquad \frac{d_-}{d_+}\Bigg|_{\alpha\to\alpha_k^*}=\frac{\sqrt{2\alpha_k^*}}{4(1+\alpha_k^*)}\frac{1}{\sqrt{\alpha-\alpha_k^*}}, \qquad \frac{d_-}{d_+}\Bigg|_{\alpha\to\widetilde{\alpha}_k}\sim(\alpha-\widetilde{\alpha}_k)\Big|_{\alpha\to\widetilde{\alpha}_k}=0.
\end{equation}
This agreement provides further evidence of the internal consistency of our results and demonstrates that the formulas \eqref{DD-integral} (appropriately modified according to \eqref{newDet-oldDet}), \eqref{DD-QQ-relation-new}, and \eqref{2.17-new} capture genuinely non-perturbative physics, remaining valid in both the large- and small-$\lambda$ regimes. A similar correspondence between the vanishing behavior of mesons and the ratio $d_-/d_+$ (see formula 4.25 in \cite{Litvinov:2024riz}) has also been observed in the 't Hooft model and in the two-particle approximation of the Ising Field Theory \cite{Litvinov:2025geb}.
\paragraph{Physical meaning of the critical points.} The full physical significance of the critical points lying on the second sheet of the complex $\alpha$–plane remains unresolved in all cases considered: large-$N_c$ scalar QCD$_2$, the ’t Hooft model, and the two-particle approximations of IFT (as well as spin-field coupled double IFT). A definitive explanation of their origin requires going beyond strict asymptotic regimes and developing a framework that consistently accounts for finite-$N_c$ corrections in two-dimensional QCD together with genuine multiparticle dynamics in IFT. Accounting for corrections by $1/N_c$ will most likely lead to $1/N_c$ corrections in critical exponents
\begin{equation}\label{massless-mesons}
    \begin{aligned}
        \text{scalar QCD$_2$: }&\lambda^{(2k)}\sim(\alpha-\widetilde{\alpha}_k)^{1+\mathcal{O}\left(\frac{1}{N_c}\right)}, \quad \lambda^{(2k+1)}\sim(\alpha-\alpha_k^*)^{\frac{1}{2}+\mathcal{O}\left(\frac{1}{N_c}\right)},\\
        \text{'t Hooft model: }&\lambda^{(2k)}\sim(\alpha-\alpha_k^*)^{\frac{1}{2}+\mathcal{O}\left(\frac{1}{N_c}\right)},\quad \lambda^{(2k+1)}_\pm\sim(\alpha-\alpha_k^*)^{\frac{1}{4}+\mathcal{O}\left(\frac{1}{N_c}\right)}.
    \end{aligned}
\end{equation}

At present, the systematic study of these corrections is still at a very early stage in both the QCD$_2$ \cite{Barbon:1994NucPhysB,Ji:2018waw}\cite{Kochergin:2024quv} and IFT \cite{Fonseca:2001dc}\cite{Fonseca:2006au,Rutkevich:2009zz}, and remains an important open problem. Yet, there exists one well-understood benchmark case: in the full IFT, the first spectral singularity---where the lightest meson mass vanishes---has a precise physical interpretation as the Yang–Lee edge singularity, realized at a purely imaginary magnetic field. The associated critical theory is the non-unitary minimal model $\mathcal{M}_{2,5}$ \cite{Fisher:1978pf,Cardy:1985yy} (see also the discussion in Section 7 of \cite{Litvinov:2025geb}, where the two-particle IFT approximation is compared with the exact result of the full theory).

This observation motivates a broader conjecture: as first suggested in the IFT context \cite{Fonseca:2006au} and later extended to the 't Hooft model \cite{Fateev:2009jf}, the critical points lying on the second sheet of the complex plane (at least some of them) may universally correspond to specific families of non-unitary Conformal Field Theories labeled by integer $N_c$ that dictate the infrared behavior in the vicinity of these singularities. Establishing such correspondences remains an open and compelling challenge. Its resolution would not only clarify the physical content of these critical points but also deepen our understanding of non-perturbative phenomena in confining Quantum Field Theories.
\section{Discussion}\label{Discussion}
In this work, we have investigated the analytic structure of mesons in large-$N_c$ scalar QCD$_2$, and—as discussed in Section \ref{BS-and-Integr}—the same analysis also applies to the interchain mesons in the two-particle approximation of the doubled Ising model coupled via the spin-spin interaction term. The central step is the reformulation of the Bethe-Salpeter equation \eqref{tHooft-eq-bosons} as a Baxter-type TQ relation \eqref{TQ-equation}, which situates the problem within the framework of integrability and makes it amenable to the Fateev–Lukyanov–Zamolodchikov method \cite{Fateev:2009jf}, originally developed for the 't Hooft model and subsequently extended in \cite{Litvinov:2024riz,Artemev:2025cev,Litvinov:2025geb}.

Within this framework, we establish a set of nontrivial functional relations for one flavor case $\alpha_1=\alpha_2=\alpha$—\eqref{DD-integral} (modified by \eqref{newDet-oldDet}), \eqref{DD-QQ-relation-new}, and \eqref{2.17-new}—that allow direct extraction of spectral data from the TQ-equation solutions described in \cite{Litvinov:2025geb}. These relations provide explicit formulas for the spectral sums $\mathcal{G}^{(s)}_\pm(\alpha)$ \eqref{Gpm1}-\eqref{Gpm3} and yield a systematic large-$n$ WKB expansion \eqref{WKB-normal-alpha-even}-\eqref{WKB-normal-alpha-odd}, both of which coincide with numerical results and reproduce known behavior in the appropriate limits. In addition, the analytic continuation of these expressions into the complex $\alpha$-plane reveals a striking structure: two infinite sequences of critical points where individual mesons become massless, together with a precise characterization of their behavior near these singularities \eqref{massless-mesons}. The points from one of these sequences are square root branching points.

Although in this work the problem is reformulated as a TQ-equation \eqref{TQ-equation} and the necessary analytic properties of its solutions are established for the two-flavor case \eqref{Q-is-bounded}-\eqref{Q_quantization_cond}, the extraction of spectral data in full generality remains unresolved. Achieving this would require constructing explicit solutions to the TQ-equation with the correct analytic behavior and extending the relations between the spectral determinants $\mathcal{D}_\pm$ and the $Q$-functions. Nevertheless, by analogy with the WKB analysis of the 't Hooft model \cite{Artemev:2025cev} in the most general case of two flavors, one may anticipate that the leading terms of the WKB expansion in the present theory take the form:
\begin{equation}
    \lambda_n(\alpha_1,\alpha_2)=\frac{\mathfrak{n}}{2}+\frac{\alpha_1+\alpha_2}{2\pi^2}\log(\pi e^{\gamma_E}\mathfrak{n})+\mathcal{O}\left(\frac{\log{\mathfrak{n}}}{\mathfrak{n}}\right)\quad\text{with}\quad\mathfrak{n}=n+\frac{1}{4}-\frac{\alpha_1^2\mathtt{i}_2(\alpha_1)+\alpha_2^2\mathtt{i}_2(\alpha_2)}{4\pi^2}.
\end{equation}

The analytic methods developed in this work are expected to have applications in other theories where Bethe-Salpeter equations with integrable kernels \cite{Its:1980,Its:1990MPhysB} naturally appear. Notable examples include:
\begin{itemize}
    \item Large-$N_c$ QCD$_2$ with both bosonic and fermionic quarks: in this case, mesons composed of constituents with different statistics are described by equations of the form \cite{Aoki:1993ma}
    \begin{equation}\label{tHooft-eq-bf-regularized}
        2\pi^2\lambda\;\phi(x)=\left(\frac{\alpha_1}{x}+\frac{\alpha_2}{1-x}\right)\phi(x)-\fint_0^1\limits dy\frac{\phi(y)}{(x-y)^2}\frac{x+y}{\sqrt{xy}}.
    \end{equation}
    The meson spectrum for constituents with different statistics is expected to lie between those of identical-statistics cases \cite{Aoki:1995dh}. Guided by the WKB expansions in the 't Hooft model and scalar QCD$_2$, one can conjecture that the subleading constant term is approximately the average of the corresponding terms in the two theories. For small $\alpha$, this suggests the spectrum behaves as $\lambda_n\approx\frac{1}{2}\left(n + \frac{1}{2}\right)$. Numerical verification using Method II from Section \ref{Numerics} extended to this theory confirms this hypothesis\footnote{In \cite{Aoki:1995dh}, when applying Multhopp's method, a typo appears in formula (3.21) for boson-fermion case: an extra factor of $\cos\theta$ is mistakenly included. For this theory, in Method II only the part of the Hamiltonian $H^{(2)}_{mn}$ \eqref{H2-Chebyshev} is modified:
    \begin{equation}
        H^{(2)}_{mn}=\frac{n\pi^2}{2}\delta_{n,m}-\frac{\pi^2}{4}\delta_{n,m}-\frac{\pi^2}{2}\Theta(m-n-1).
    \end{equation}
    }. 
    \item The Schwinger model: in the two-particle (valence) sector, its dynamics can also be cast into a similar Bethe-Salpeter form \cite{Bergknoff:1976xr} (see also \cite{Hornbostel:1988fb}) \begin{equation}\label{tHooft-eq-Schwinger}
        2\pi^2\lambda\;\phi(x)=\frac{\alpha}{x(1-x)}\phi(x)-\fint_0^1\limits dy\frac{\phi(y)}{(x-y)^2}+\int_0^1\limits dx\;\phi(x).
    \end{equation}
    We have obtained preliminary results, and a more detailed investigation of this case will be published elsewhere.
    \item Adjoint QCD$_2$: theories with fermionic or bosonic quarks in the adjoint representation provide another natural arena. Here, one can analyze the spectrum of glueballs and confining flux tubes. A system of infinitely many coupled eigenvalue equations was derived in \cite{Bhanot:1993xp,Demeterfi:1993rs}, representing a far-reaching generalization of \eqref{tHooft-eq-fermions}, \eqref{tHooft-eq-bosons}. Developing analytic approaches in this direction remains a challenging but important problem.  
\end{itemize}
\subsection*{Supplemental material}
Along with this submission, we provide three supplementary Wolfram Mathematica notebooks: \texttt{Spectral-sums-scalar-QCD.nb}, \texttt{Phi.nb}, and \texttt{WKB.nb}. The first notebook contains closed-form expressions for the initial five spectral sums $\mathcal{G}^{(s)}_\pm$ together with the corresponding matrix elements $\bra{p}\hat{K}^n\ket{p}$ for $n=1,\ldots,5$. The second notebook collects the phase functions $\Phi_\pm^{(k)}(l)$ for $k=0,\ldots,5$. The third file presents higher-order contributions to the large-$n$ WKB expansion \eqref{WKB-normal-alpha-even}-\eqref{WKB-normal-alpha-odd}, including terms up to order $\mathfrak{n}^{-5}$. All three notebooks rely on the fundamental integrals $\mathtt{i}_{k}(\alpha)$, $\mathtt{u}_{2k-1}(\alpha)$, and $\mathtt{v}(\alpha)$, defined in \eqref{i-def}, \eqref{u-def}, and \eqref{Proector-def}, respectively.
\section*{Acknowledgments}
We acknowledge discussions with Alexander Artemev, Igor Klebanov, Ilia Kochergin, Alexey Litvinov, and Alexander Zamolodchikov. We would like to thank Igor Klebanov for highlighting the relevance of the problem addressed in this work, as well as Igor Klebanov and Alexey Litvinov for their helpful advice during the preparation of this manuscript. This work was supported in part by the Simons Collaboration on Confinement and QCD Strings through the Simons Foundation Grant $917464$.
\appendix
\section{Matrix elements of spectral operator \texorpdfstring{$\hat{K}$}{}}\label{matrix-el-of-K}
Here we describe the computation of the matrix elements $\bra{p}\hat{K}^n\ket{p}$ and derive equation \eqref{g-for-appendix}.
Consider first the Liouville-Neumann series for $\Psi_+(\nu|\lambda)$,
\begin{equation}\label{Liouville-Neumann_for_Psi_m}
    f(\nu)\Psi_+(\nu|\lambda)=\frac{1}{\cosh{\frac{\pi\nu}{2}}}+\sum\limits_{k=1}^\infty\lambda^k\int_{-\infty}^\infty\limits\frac{1}{\cosh{\frac{\pi \nu_k}{2}}}\prod\limits_{j=1}^k\frac{d\nu_j}{f(\nu_j)}S(\nu_j-\nu_{j-1}),\quad \nu_0\equiv \nu,
\end{equation}
where $f(\nu)=\frac{2\alpha}{\pi}+\nu\tanh{\frac{\pi\nu}{2}},\; S(\nu)=\frac{\pi\nu}{2\sinh{\frac{\pi\nu}{2}}}$. By definition $Q_+(\nu|\lambda)=\cosh{\frac{\pi\nu}{2}} f(\nu)\Psi_+(\nu|\lambda)$, so that equation \eqref{Liouville-Neumann_for_Psi_m} can be rewritten as
\begin{equation}\label{Liouville-Neumann_for_Qp}
    \frac{Q_+(\nu)}{\cosh{\frac{\pi\nu}{2}}}=\frac{1}{\cosh{\frac{\pi\nu}{2}}}+\sum\limits_{k=1}^\infty\lambda^k\int_{-\infty}^\infty\limits\frac{1}{\cosh{\frac{\pi\nu_k}{2}}}\prod\limits_{j=1}^k \frac{d\nu_j}{f(\nu_j)}S(\nu_j-\nu_{j-1}).
\end{equation}

Next, let us examine the explicit form of the matrix element,
\begin{equation}
    \bra{p}\hat{K}^n\ket{p}=\frac{1}{8\mathtt{v}(\alpha)} \int_{-\infty}^\infty\limits \frac{d \nu_0}{f(\nu_0)} \prod_{j=1}^n \frac{d\nu_j}{f(\nu_j)} S(\nu_j-\nu_{j-1}) \frac{1}{\cosh{\frac{\pi \nu_0}{2}}} \cdot \frac{1}{\cosh{\frac{\pi \nu_n}{2}}},
\end{equation}
from which it follows that $\bra{p}\hat{K}^n\ket{p}$ can be obtained by expanding the left-hand side of \eqref{Liouville-Neumann_for_Qp} at the point $\nu=i$ in powers of $\lambda$:
\begin{equation}\label{matrix-elements-series}
    \resizebox{\textwidth}{!}{$
    \begin{aligned}
        \frac{1}{4\pi\mathtt{v}(\alpha)}\frac{Q_+(\nu)-1}{\cosh{\frac{\pi\nu}{2}}}\Biggl|_{\nu=i}=-\frac{iQ'_+(i)}{2\pi^2\mathtt{v}(\alpha)}=\lambda+\sum\limits_{k=1}^\infty \lambda^{k+1}\bra{p}\hat{K}^k\ket{p}\quad\Rightarrow\quad -1-\frac{iQ'_+(i)}{2\pi^2\mathtt{v}(\alpha)\lambda}=\sum\limits_{k=1}^\infty \lambda^{k}\bra{p}\hat{K}^k\ket{p}.
    \end{aligned}
    $}
\end{equation}
For example,
\begin{equation}\label{matr-el-K}
    \begin{aligned}
        \bra{p}\hat{K}\ket{p}=&\;\frac{1}{8\pi^2\mathtt{v}(\alpha)}\left[\frac{\pi^4}{\alpha}+8\pi^2-8\alpha\zeta(3)-4\pi^2\alpha^2\mathtt{u}_3(\alpha)\right],\\
        \bra{p}\hat{K}^2\ket{p}=&\;\frac{1}{8\pi^4\mathtt{v}(\alpha)}\Bigl[-\pi^4\left(\frac{4\pi^2}{\alpha}+10-4\alpha\right)+2\alpha^2\left(6\zeta(5)-\pi^4\alpha\left(\mathtt{u}_3(\alpha)+3\mathtt{u}_5(\alpha)\right)\right)+
        \\
        &+\left(2\zeta(3)+\pi^2\alpha\mathtt{u}_3(\alpha)\right)\left(\pi^4+4\pi^2(1-\alpha)\alpha-2\alpha^2\zeta(3)-\pi^2\alpha^3\mathtt{u}_3(\alpha)\right)\Bigl],
        \\
        \bra{p}\hat{K}^3\ket{p}=&\;\frac{1}{324\pi^6\mathtt{v}(\alpha)}\Biggl[\frac{27\pi^{10}}{\alpha^2}+\pi^8\left(\frac{405}{\alpha}-98\right)+552\pi^6-24\pi^6\alpha(17-3\alpha)+
        \\&+\frac{9\pi^2\alpha}{2}\left(3\pi^2-32\alpha^2\right)\left(2\zeta(3)+\pi^2\alpha\mathtt{u}_3(\alpha)\right)^2-
        \\&-2\left(2\zeta(3)+\pi^2\alpha\mathtt{u}_3(\alpha)\right)\left(\pi^6(63-39\alpha)+2\pi^4\alpha(54-(108-23\alpha)\alpha)-216\alpha^3\zeta(5)\right)-
        \\&-\pi^2\alpha\left(39\pi^2+16(9-5\alpha)\alpha\right)\left(6\zeta(5)-\pi^4\alpha\left(\mathtt{u}_3(\alpha)+3\mathtt{u}_5(\alpha)\right)\right)-
        \\&-72\pi^4\alpha^4\left(2\zeta(3)+\pi^2\alpha\mathtt{u}_3(\alpha)\right)\left(\mathtt{u}_3(\alpha)+3\mathtt{u}_5(\alpha)\right)-  \\&-4\alpha^3\left(90\zeta(7)+\pi^6\alpha\left(2\mathtt{u}_3(\alpha)+30\mathtt{u}_5(\alpha)+45\mathtt{u}_7(\alpha)\right)\right)\Biggl].
    \end{aligned}
\end{equation}
This procedure yields the explicit representation for $g(\lambda)$,
\begin{equation}\label{g-func-explicit}
    g(\lambda)=\sum_{k=1}^{\infty}\limits\mathcal{V}(\alpha)\bra{p}\hat{K}^k\ket{p}\lambda^k=\mathcal{V}(\alpha)\left(-1-\frac{iQ'_+(i)}{2\pi^2\mathtt{v}(\alpha)\lambda}\right).
\end{equation}

We also prove here that the quantization condition \eqref{Q_quantization_cond} coincides with \eqref{Quant_cond2}. Indeed, starting from \eqref{Q_quantization_cond} and using the TQ-equation \eqref{TQ-equation}, one finds
\begin{equation}
    Q_+(i)=-\frac{1}{8}\int_{-\infty}^\infty\limits d \nu'\frac{1}{\cosh^2{\frac{\pi\nu'}{2}}}\frac{(Q_+(\nu'+2i)-Q_+(\nu'))+(Q_+(\nu'-2i)-Q_+(\nu'))}{-4\pi\lambda_n}.
\end{equation}
Analytically continuing this expression, we obtain
\begin{equation}
    Q_+(i)=\frac{1}{32\pi\lambda_n}\left[-\oint_{\mathcal{C}_i}\limits d\nu\;\frac{Q_+(\nu)}{\cosh^2{\frac{\pi\nu}{2}}}+\oint_{\mathcal{C}_{-i}}\limits d\nu\;\frac{Q_+(\nu)}{\cosh^2{\frac{\pi\nu}{2}}}\right]=\frac{iQ'_+(i)}{2\pi^2\lambda_n}=1,
\end{equation}
where $\mathcal{C}_i$ and $\mathcal{C}_{-i}$ are small contours surrounding the points $i$ and $-i$, respectively. Here we have used the normalization conditions \eqref{Qpm-normalization-conditions}. This confirms the equivalence with the condition \eqref{Quant_cond2}.
\bibliographystyle{MyStyle}
\bibliography{MyBib}
\end{document}